\documentclass{article}

\usepackage{amsmath}
\usepackage{amssymb}
\usepackage{setspace}
\usepackage{tikz-cd}
\usepackage{graphicx}
\usepackage{authblk}
\usepackage{hyperref}
\usepackage{physics}
\usepackage{xcolor}
\usepackage{fullpage}

\newtheorem{definition}{Definition}

\title{ZX-Calculus and Extended Hypergraph Rewriting Systems I:\\
A Multiway Approach to Categorical Quantum Information Theory}
\author[1]{Jonathan Gorard\footnote{Corresponding Author: \url{jg865@cam.ac.uk}}}
\author[2]{Manojna Namuduri\footnote{\url{manon@wolfram.com}}}
\author[3]{Xerxes D. Arsiwalla\footnote{\url{x.d.arsiwalla@gmail.com}}}
\affil[1]{\small University of Cambridge, Cambridge, UK}
\affil[2]{Wolfram Research, USA}
\affil[3]{Pompeu Fabra University, Barcelona, Spain}

\begin{document}
\maketitle

\begin{abstract}
Categorical quantum mechanics and the Wolfram model offer distinct but complementary approaches to studying the relationship between diagrammatic rewriting systems over combinatorial structures and the foundations of physics; the objective of the present article is to begin elucidating the formal correspondence between the two methodologies in the context of the ZX-calculus formalism of Coecke and Duncan for reasoning diagrammatically about linear maps between qubits. After briefly summarizing the relevant formalisms, and presenting a categorical formulation of the Wolfram model in terms of adhesive categories and double-pushout rewriting systems, we illustrate how the diagrammatic rewritings of the ZX-calculus can be embedded and realized within the broader context of Wolfram model multiway systems, and illustrate some of the capabilities of the software framework (\textit{ZXMultiwaySystem}) that we have developed specifically for this purpose. Finally, we present a proof (along with an explicitly computed example) based on the methods of Dixon and Kissinger that the multiway evolution graphs and branchial graphs of the Wolfram model are naturally endowed with a monoidal structure based on \textit{rulial composition} that is, furthermore, compatible with the monoidal product of ZX-diagrams.
\end{abstract}

\clearpage

\tableofcontents

\clearpage

\section{Introduction}

The \textit{ZX-calculus}, as first outlined by Coecke and Duncan in  2008\cite{coecke}\cite{coecke2}, is a natural outgrowth of the field of \textit{categorical quantum mechanics}, as pioneered by Abramsky and Coecke\cite{abramsky}\cite{abramsky2} (whose primary objective is to describe the foundations of quantum mechanics within the language of \textit{monoidal category theory}\cite{baez}), and provides a novel method of reasoning diagrammatically about linear maps between qubits in quantum information theory\cite{backens}\cite{backens2} that is provably both complete and sound\cite{jeandel}\cite{hadzihasanovic}\cite{jeandel2}, such that two ZX-diagrams represent the same linear map if and only if they can be transformed into one another via the rules of the ZX-calculus. Philosophically, categorical quantum mechanics differs from the standard Hilbert space formalism of Dirac and von Neumann in that it treats quantum processes and their compositions as being the fundamental objects of study, as opposed to quantum states; within the mathematical context of a ``dagger symmetric monoidal category''\cite{selinger}, sequential compositions of quantum processes are captured in terms of compositions of morphisms, whilst parallel compositions are captured in terms of monoidal products of those morphisms\cite{coecke8}.

The conventional \textit{quantum circuit} model of quantum computation ultimately treats every linear map between qubits as being a large unitary matrix that is applied to some initial quantum state, where this matrix is derived via the sequential composition of \textit{quantum gates} (which are themselves unitary matrices)\cite{nielsen}. The ZX-calculus model differs in at least two fundamental ways. Firstly, whereas quantum circuits exhibit a rigid topological structure, in which the distinction between the inputs and outputs of gates, and indeed of the circuit as a whole, is crucial, the linear map described by a ZX-diagram is invariant under arbitrary topological deformations - in some sense, within a ZX-diagram, there is ``only'' topology. Secondly, whereas quantum circuits distinguish between the specification of the circuit/matrix and its actualization (i.e. its application to a particular quantum state), ZX-diagrams enact their own computations, via their own diagrammatic transformation rules. In this regard, ZX-diagrams are highly analogous to \textit{combinators} in mathematical logic: in both ZX-diagrams and combinator expressions, no fundamental distinction is made between the specification of a program and the state of its execution, because the program is executed by simply applying symbolic transformation rules to its own specification. Thus, ZX-calculus may be viewed as an attempt to ``break apart'' the formalism of the quantum circuit model, in such a way as to lay bare its underlying computational structure.

In much the same way, the \textit{Wolfram model}\cite{wolfram}\cite{wolfram2}\cite{gorard} is an attempt to ``break apart'' the fundamental structure of spacetime, and to make manifest the computational structure that (potentially) underlies it. The Wolfram model describes an idealized class of ``laws of physics'' in terms of symbolic transformation rules applied to hypergraphs (generalizations of graphs in which edges can connect arbitrary non-empty subsets of vertices); such transformation rules naturally yield combinatorial structures known as \textit{causal graphs}, \textit{multiway systems} and \textit{branchial graphs}, amongst many others, with surprising formal analogies to Lorentzian manifolds, path integrals and projective Hilbert spaces, respectively\cite{gorard2}. Thus, both the Wolfram model and the ZX-calculus ultimately find their foundations in diagrammatic rewriting systems. Indeed, just as ZX-diagrams eschew the rigid structure of quantum circuits in favor of a pure description of the topology of linear maps, the Wolfram model eschews the rigid structure of spacetime in favor of a pure description of the topology of causal relationships. Moreover, just as ZX-diagrams do not distinguish between the specification of a quantum process and the state of its execution, the Wolfram model does not distinguish between the ``background'' structure of space and the ``foreground'' structure of physical processes: everything is described entirely in terms of symbolic transformations on hypergraphs. Therefore, at least at first glance, these two formalisms appear as though they may end up being deeply related.

Indeed, at a more fundamental level, the recent treatment of symmetric monoidal categories as a general language for reasoning about physical systems (with morphisms between objects playing the role of physical transformations between states, and with morphism composition and monoidal composition playing the role of sequential and parallel combination of such processes, respectively) is very much in the same spirit as the goal of the Wolfram Physics Project to describe physical processes in terms of multiway systems defined via abstract rewriting rules over arbitrary symbolic expressions. The principal objective of the present article is to begin the process of making the correspondence mathematically precise.

Since the number of people who are intimately familiar with both categorical quantum mechanics and the Wolfram model is (presumably) still relatively small, we will begin this article with a brief but hopefully gentle introduction to both formalisms, including a categorical formalization of (a restricted case of) the Wolfram model in terms of adhesive categories and double-pushout rewriting (DPO) systems. We will then proceed to describe how the diagrammatic rewritings of the ZX-calculus can be compiled and embedded within Wolfram model multiway systems, including a brief illustration of the Wolfram Language software packages (such as \textit{MakeZXDiagram}\cite{wfr1} and \textit{ZXMultiwaySystem}\cite{wfr2}) that we have developed specifically for this purpose, which may be thought of as constituting the beginnings of Wolfram model-based variants of frameworks like \textit{Quantomatic}\cite{kissinger}. We will also briefly discuss some potential applications of this embedding, including a new approach to parallelizing the automatic rewriting of ZX-diagrams, a new method for performing lemma selection in automated reasoning algorithms over ZX-diagrams, and potentially even a new technique for proving consistency, completeness and soundness results for the ZX-calculus (and related formalisms) using combinatorial methods. Finally, we will begin to make manifest the connection between the present conjectural formulation of quantum mechanics in the Wolfram model (described in terms of multiway systems and branchial graphs) and the categorical formulation of quantum mechanics inherent to the ZX-calculus. In particular, we will show that the categories of branchial graphs and of multiway systems are both naturally endowed with a (compatible) monoidal structure, given in terms of \textit{rulial} composition, and moreover that this monoidal structure is also compatible with the natural monoidal product of ZX-diagrams. We will first illustrate this compatibility empirically, via explicit computation, before presenting a general proof using the methods of Dixon and Kissinger\cite{dixon}.

As a consequence of this choice of structure, Section \ref{sec:section1} of this article is partially expository, albeit involving a somewhat novel description of the Wolfram model in terms of double-pushout rewritings, and an original presentation of the formalism of multiway operator systems. The majority of the novel content of the article is contained within Sections \ref{sec:section2} and \ref{sec:section3}, in which the two primary original contributions that we wish to emphasize are:

\begin{enumerate}
\item
An explicit demonstration that the diagrammatic rewritings of formalisms such as the ZX-calculus may be recast cleanly and consistently into the more general framework of a Wolfram model multiway operator system, illustrating that, in a precise sense, such multiway systems form an embedding space for the collection of all possible diagrammatic rewritings.

\item
A demonstration (and subsequent proof) that the categories of branchial graphs and multiway evolution graphs in the Wolfram model formalism are naturally endowed, by the properties of the \textit{rulial multiway system}, with a monoidal structure that is compatible with the monoidal structure of ZX-diagrams.
\end{enumerate}

We also attach an appendix, containing in Section \ref{sec:section5} a glossary of basic terminology and concepts commonly encountered in the formalism of the Wolfram model, as well as in Section \ref{sec:section6} an overview of the theory of monoidal categories, as commonly employed in category-theoretic approaches to quantum mechanics in general, and in the ZX-calculus approach to quantum information theory in particular.

\clearpage

\section{The Wolfram Model, Multiway Systems and Term Rewriting}
\label{sec:section1}

\begin{figure}[ht]
\centering
\includegraphics[width=0.295\textwidth]{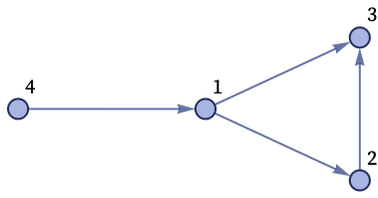}\hspace{0.25\textwidth}
\includegraphics[width=0.295\textwidth]{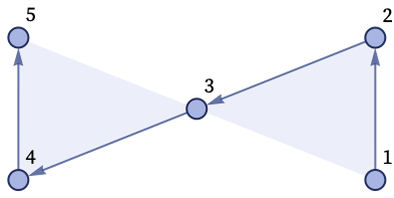}
\caption{Spatial hypergraphs corresponding to finite collections of ordered relations between elements, namely ${\left\lbrace \left\lbrace 1, 2 \right\rbrace, \left\lbrace 1, 3 \right\rbrace, \left\lbrace 2, 3 \right\rbrace, \left\lbrace 4, 1 \right\rbrace \right\rbrace}$ and ${\left\lbrace \left\lbrace 1, 2, 3 \right\rbrace, \left\lbrace 3, 4, 5 \right\rbrace \right\rbrace}$, respectively.}
\label{fig:Figure1}
\end{figure}

We begin this section with a novel reformulation of the Wolfram model in terms of double-pushout rewriting systems and adhesive categories. The Wolfram model is a discrete spacetime formalism in which apparently continuous structures such as space, time and (projective) Hilbert space emerge as large-scale limits of underlying discrete structures such as hypergraphs, causal networks and so-called \textit{branchial graphs}. At its most basic level, the Wolfram model is based upon diagrammatic rewriting rules acting on hypergraphs\cite{gorard}:

\begin{definition}
A  ``spatial hypergraph", denoted ${H = \left( V, E \right)}$, is a finite, undirected hypergraph:

\begin{equation}
E \subset \mathcal{P} \left( V \right) \setminus \left\lbrace \emptyset \right\rbrace,
\end{equation}
where ${\mathcal{P} \left( V \right)}$ denotes the power set of $V$.
\end{definition}
A crucial observation is that (directed) spatial hypergraphs can therefore be represented purely abstractly as finite collections of ordered relations (i.e. hyperedges) between elements (i.e. hypernodes), as shown in Figure \ref{fig:Figure1}.

\begin{figure}[ht]
\centering
\includegraphics[width=0.395\textwidth]{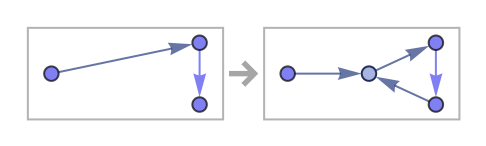}
\caption{A hypergraph transformation rule corresponding to the set substitution system ${\left\lbrace \left\lbrace x, y \right\rbrace, \left\lbrace y, z \right\rbrace \right\rbrace \to \left\lbrace \left\lbrace w, y \right\rbrace, \left\lbrace y, z \right\rbrace, \left\lbrace z, w \right\rbrace, \left\lbrace x, w \right\rbrace \right\rbrace}$.}
\label{fig:Figure2}
\end{figure}

One can then define the dynamics of a Wolfram model system in terms of hypergraph rewriting rules:

\begin{definition}
An ``update rule'', denoted $R$, for a spatial hypergraph ${H = \left( V, E \right)}$ is an abstract rewrite rule of the form ${H_1 \to H_2}$, in which a subhypergraph matching pattern ${H_1}$ is replaced by a distinct subhypergraph matching pattern ${H_2}$.
\end{definition}
Each such rewriting rule is formally equivalent to a set substitution system (one in which a subset of ordered relations matching a particular pattern is replaced with a distinct subset of ordered relations matching a particular pattern), as shown in Figure \ref{fig:Figure2}.

\begin{figure}[ht]
\centering
\includegraphics[width=0.595\textwidth]{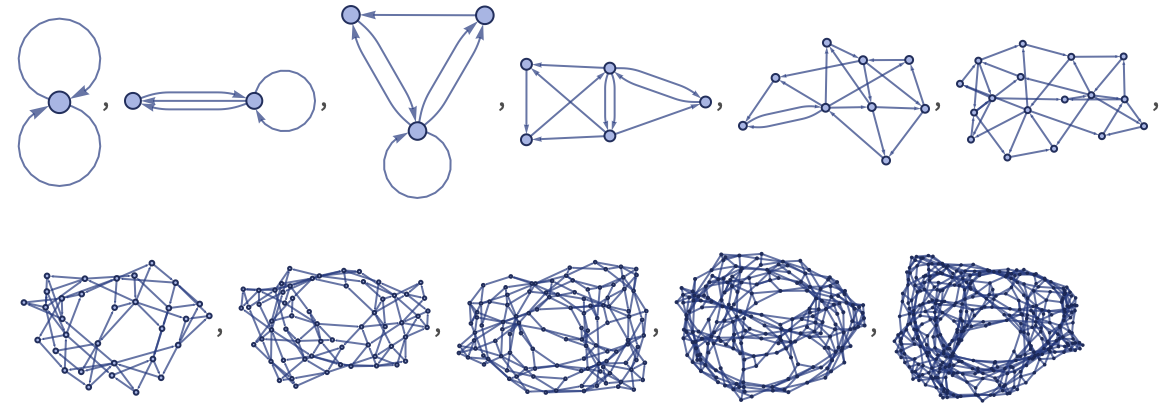}
\caption{The results of the first 10 steps in the evolution history of the set substitution system ${\left\lbrace \left\lbrace x, y \right\rbrace, \left\lbrace y, z \right\rbrace \right\rbrace \to \left\lbrace \left\lbrace w, y \right\rbrace, \left\lbrace y, z \right\rbrace, \left\lbrace z, w \right\rbrace, \left\lbrace x, w \right\rbrace \right\rbrace}$, starting from a double self-loop initial condition.}
\label{fig:Figure3}
\end{figure}

\begin{figure}[ht]
\centering
\includegraphics[width=0.495\textwidth]{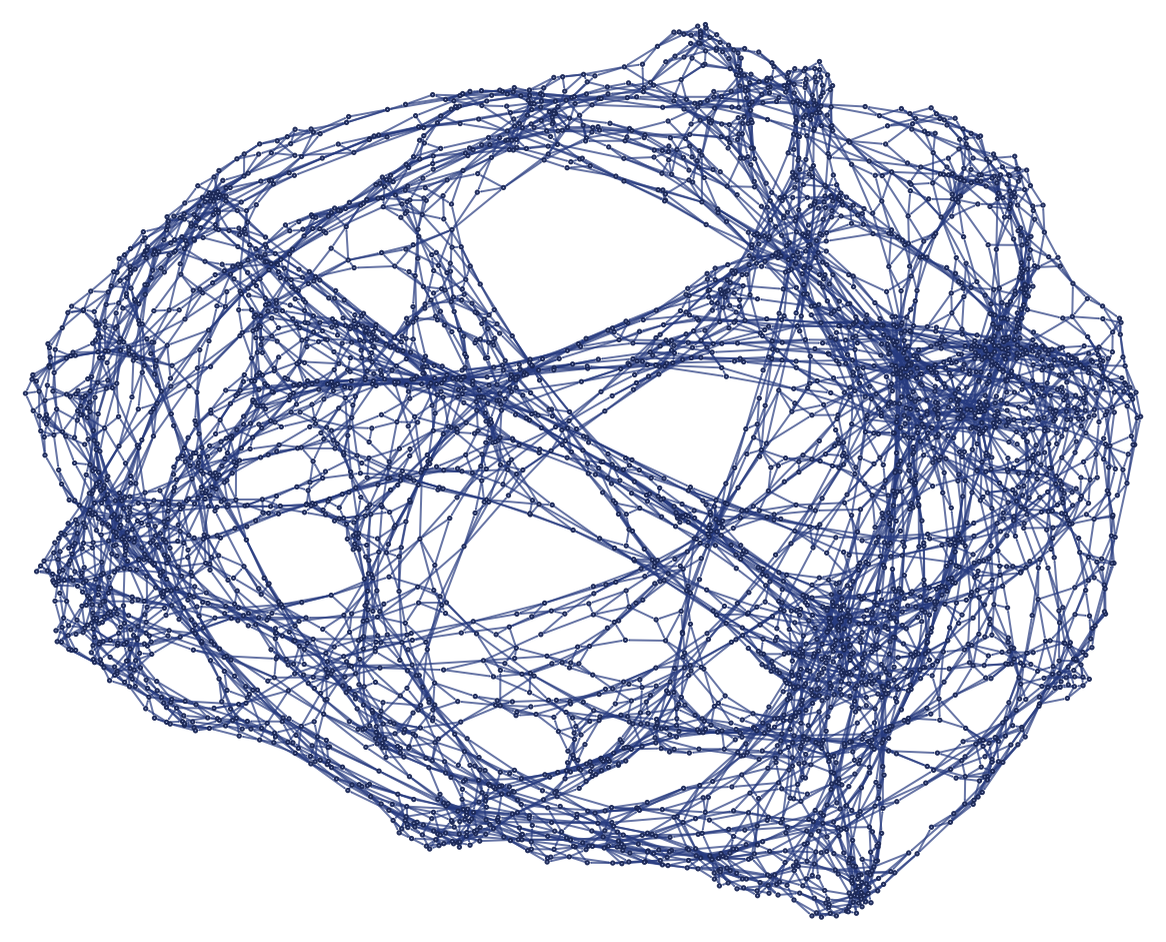}
\caption{The result after 14 steps of evolution of the set substitution system ${\left\lbrace \left\lbrace x, y \right\rbrace, \left\lbrace y, z \right\rbrace \right\rbrace \to \left\lbrace \left\lbrace w, y \right\rbrace, \left\lbrace y, z \right\rbrace, \left\lbrace z, w \right\rbrace, \left\lbrace x, w \right\rbrace \right\rbrace}$, starting from a double self-loop initial condition.}
\label{fig:Figure4}
\end{figure}

Note that, in general, the order in which to apply the transformation rules is not well-defined; in the simplest case, we could simply apply the rule to every possible matching (and non-overlapping) subhypergraph, as illustrated in Figures \ref{fig:Figure3} and \ref{fig:Figure4}. However, even in this simplified case, the initial choice of the subhypergraph to which to apply the first transformation is still ambiguous, and different such choices will in general yield non-isomorphic sequences of hypergraphs in the evolution.

Therefore, the evolution of any given spatial hypergraph will, generically, be   non-deterministic, due to this lack of any canonical updating order; we can parametrize this non-determinism by treating the Wolfram model as an abstract rewriting system\cite{baader}\cite{bezem}:

\begin{definition}
An ``abstract rewriting system'' (or ``ARS'') is a set, denoted $A$ (with each element known as an ``object''), equipped with some binary relation, denoted ${\to}$, known as the ``rewrite relation''.
\end{definition}
  
\begin{figure}[ht]
\centering
\includegraphics[width=0.595\textwidth]{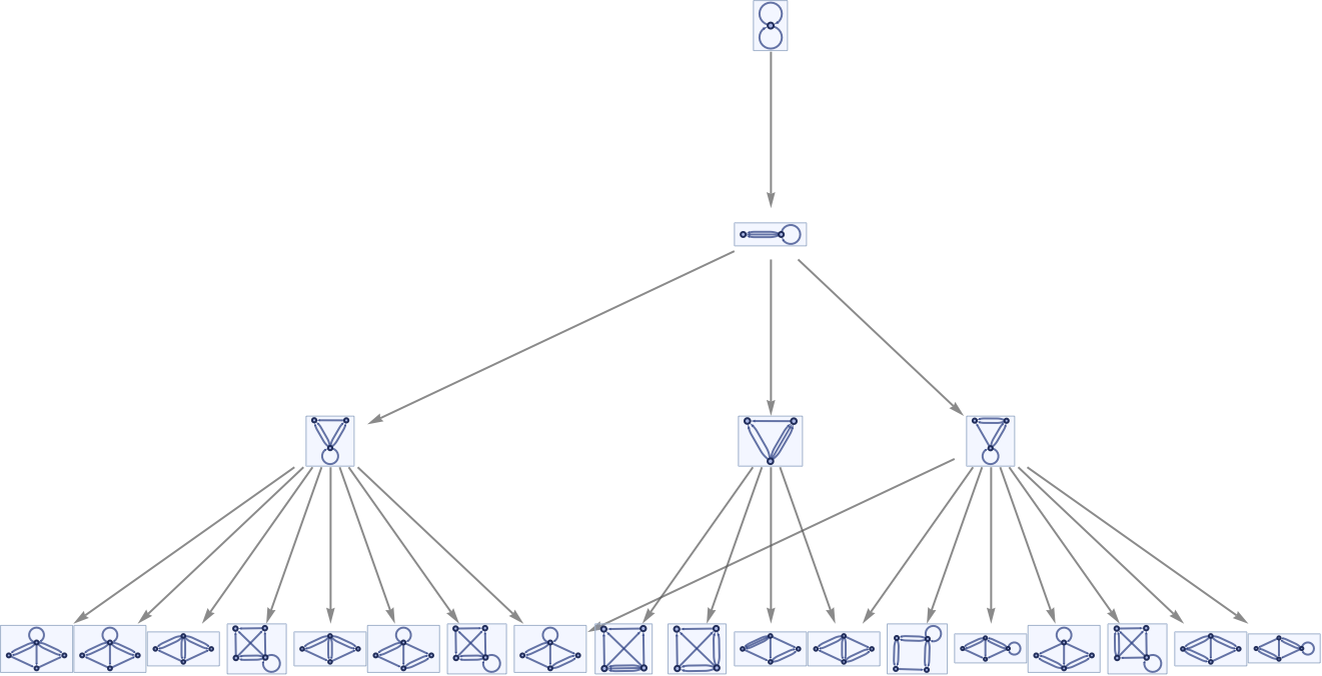}
\caption{The multiway evolution graph corresponding to the non-deterministic evolution of the set substitution system ${\left\lbrace \left\lbrace x, y \right\rbrace, \left\lbrace y, z \right\rbrace \right\rbrace \to \left\lbrace \left\lbrace w, y \right\rbrace, \left\lbrace y, z \right\rbrace, \left\lbrace z, w \right\rbrace, \left\lbrace x, w \right\rbrace \right\rbrace}$.}
\label{fig:Figure5}
\end{figure}

\begin{figure}[ht]
\centering
\includegraphics[width=0.495\textwidth]{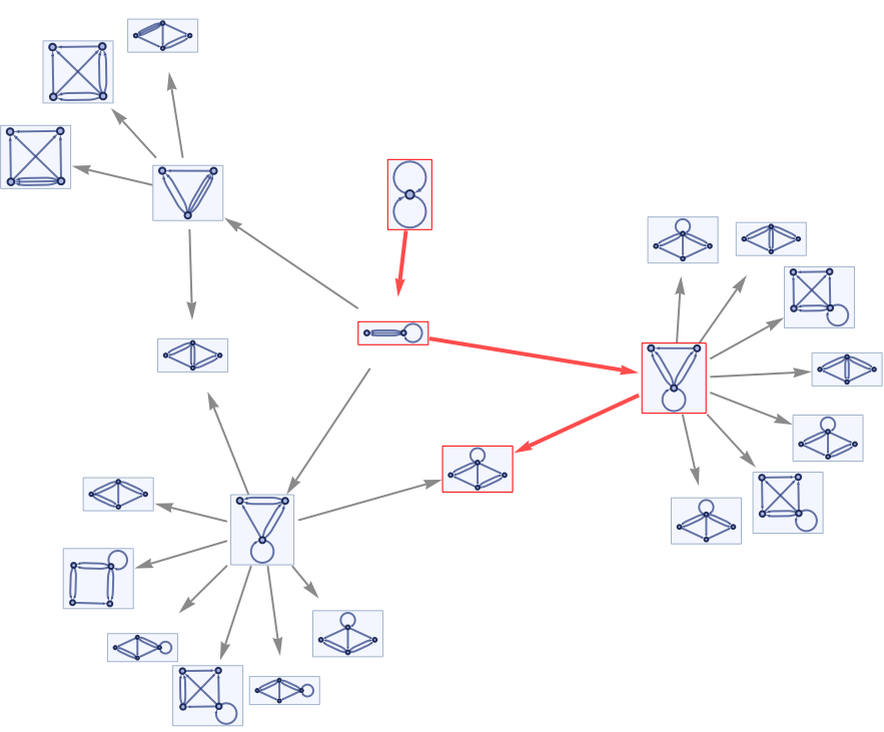}
\caption{The standard updating order for the evolution of the set substitution system ${\left\lbrace \left\lbrace x, y \right\rbrace, \left\lbrace y, z \right\rbrace \right\rbrace \to \left\lbrace \left\lbrace w, y \right\rbrace, \left\lbrace y, z \right\rbrace, \left\lbrace z, w \right\rbrace, \left\lbrace x, w \right\rbrace \right\rbrace}$, highlighted as a single path in the associated multiway evolution graph.}
\label{fig:Figure6}
\end{figure}

\begin{definition}
${\to^{*}}$ is the reflexive transitive closure of ${\to}$, i.e. the transitive closure of ${\to \cup =}$, where $=$ denotes the identity relation.
\end{definition}
In other words, ${\to^{*}}$ is the smallest preorder containing ${\to}$, i.e. the smallest binary relation containing ${\to}$ and also satisfying the axioms of reflexivity and transitivity:

\begin{equation}
a \to^{*} a, \qquad \text{ and } \qquad a \to^{*} b, b \to^{*} c \implies a \to^{*} c.
\end{equation}
A concrete way of representing the abstract rewriting structure of a Wolfram model system is through the use of a general combinatorial structure known as a \textit{multiway system}:

\begin{definition}
A ``multiway system'' (or, more strictly, a ``multiway evolution graph''), denoted ${G_{multiway}}$, is a directed, acyclic graph corresponding to the evolution of a (generically non-confluent) abstract rewriting system, in which each vertex corresponds to an object, and the directed edge ${A \to B}$ exists if and only if there exists a rewrite rule application that transforms object $A$ to object $B$.
\end{definition}
More specifically, directed edges connect vertices $A$ and $B$ if and only if ${A \to B}$ in the associated rewriting system, and a directed path connects $A$ and $B$ if and only if ${A \to^{*} B}$, i.e. there exists a finite rewrite sequence of the form:

\begin{equation}
a \to a^{\prime} \to a^{\prime \prime} \to \cdots \to b^{\prime} \to b.
\end{equation}
Thus, the evolution of a generic Wolfram model system will correspond to a multiway evolution graph, within which the ``standard'' updating order shown above will correspond to a single path, as illustrated in Figures \ref{fig:Figure5} and \ref{fig:Figure6}.

In the above definitions, the notion of \textit{confluence} is invoked in order to formalize the condition in which, within certain classes of multiway systems, all bifurcations in the evolution history will eventually converge\cite{dershowitz}\cite{huet}:
 
\begin{definition}
An object ${a \in A}$ is ``confluent'' if and only if:

\begin{equation}
\forall b, c \in A, \text{ such that } a \to^{*} b \text{ and } a \to^{*} c, \qquad \exists d \in A \text{ such that } b \to^{*} d \text{ and } c \to^{*} d.
\end{equation}
\end{definition}

\begin{figure}[ht]
\centering
\includegraphics[width=0.595\textwidth]{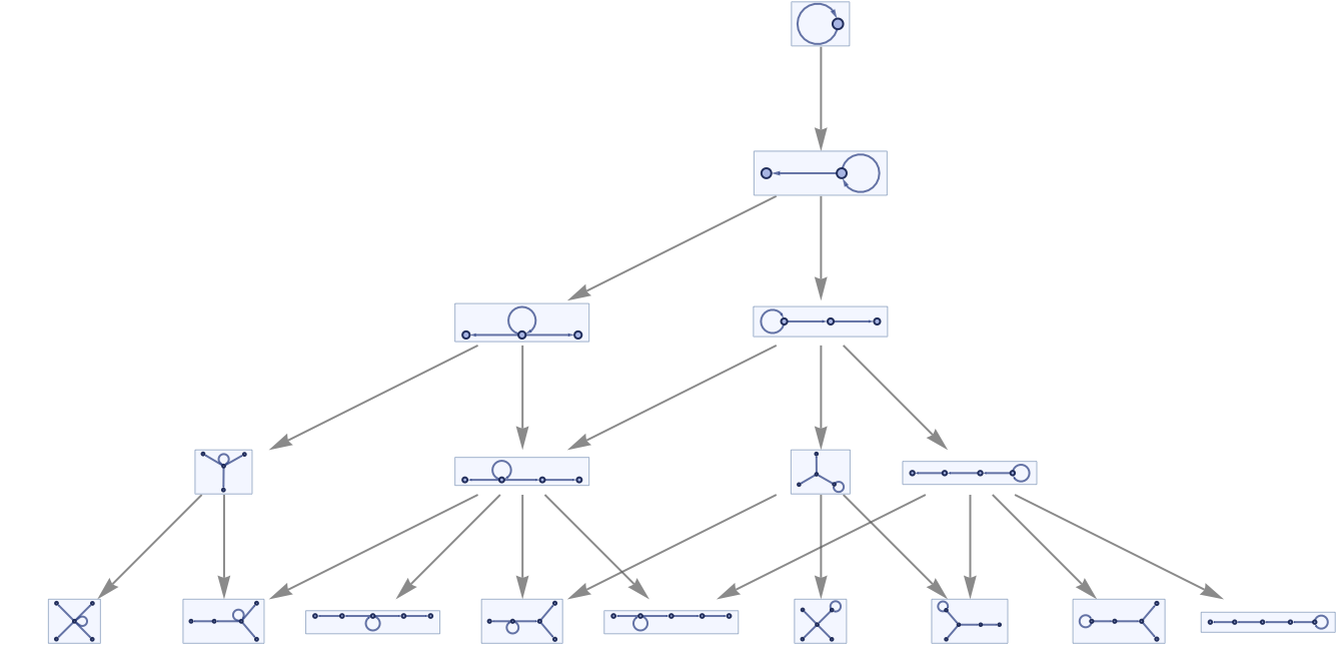}
\caption{An example evolution of a (globally) confluent set substitution system, namely ${\left\lbrace \left\lbrace x, y \right\rbrace \right\rbrace \to \left\lbrace \left\lbrace x, y \right\rbrace, \left\lbrace y, z \right\rbrace \right\rbrace}$, in which all bifurcations converge after a single step.}
\label{fig:Figure7}
\end{figure}

\begin{definition}
An abstract rewriting system $A$ is (globally) ``confluent'' (or exhibits the ``Church-Rosser property'') if and only if every object ${a \in A}$ is confluent.
\end{definition}
An example of such a (globally) confluent multiway evolution for a Wolfram model system is shown in Figure \ref{fig:Figure7}.

\begin{figure}[ht]
\centering
\includegraphics[width=0.395\textwidth]{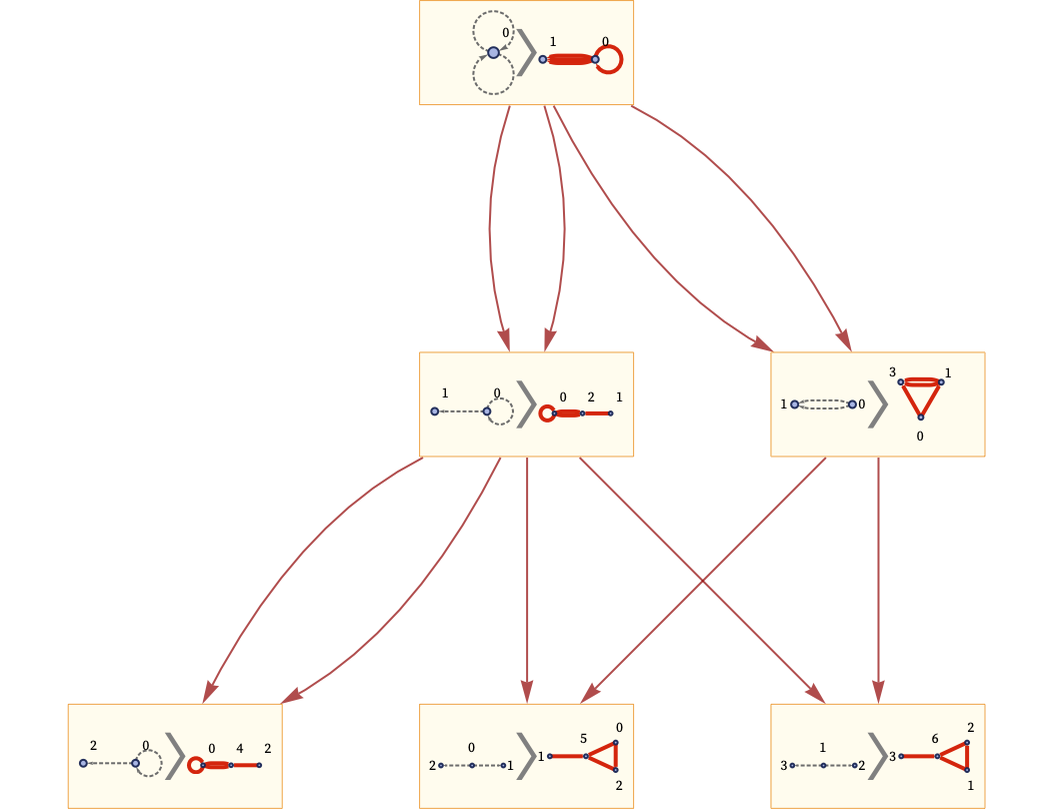}
\caption{The causal network for the set substitution system ${\left\lbrace \left\lbrace x, y \right\rbrace, \left\lbrace x, z \right\rbrace \right\rbrace \to \left\lbrace \left\lbrace x, y \right\rbrace, \left\lbrace x, w \right\rbrace, \left\lbrace y, w \right\rbrace, \left\lbrace z, w \right\rbrace \right\rbrace}$.}
\label{fig:Figure34}
\end{figure}

Furthermore, future applications of transformation rules may have dependencies upon prior applications, such that updating event $B$ could only have been applied if event $A$ had previously been applied; such dependencies may be captured by means of a causal network:

\begin{definition}
A ``causal network'', denoted ${G_{causal}}$, is a directed, acyclic graph in which every vertex corresponds to an application of an update rule (i.e. an update ``event''), and in which the directed edge ${A \to B}$ exists if and only if:

\begin{equation}
\mathrm{In} \left( B \right) \cap \mathrm{Out} \left( A \right) \neq \emptyset,
\end{equation}
i.e. the input for event $B$ makes use of hyperedges that were produced by the output of event $A$.
\end{definition}
In the context of the Wolfram model, the transitive reduction of a causal network is presumed to correspond to the Hasse diagram of the causal partial order for some discrete approximation to a Lorentzian manifold. More specifically, we interpret the partial order relation ${\prec}$ defined by the causal network as corresponding to a statement of causal precedence\cite{penrose2} for points on an associated Lorentzian manifold ${\mathcal{M}}$, such that, for instance, the sets:

\begin{equation}
J^{+} \left( x \right) = \left\lbrace y \in \mathcal{M} \mid x \prec y \right\rbrace,
\end{equation}
and:

\begin{equation}
J^{-} \left( x \right) = \left\lbrace y \in \mathcal{M} \mid y \prec x \right\rbrace,
\end{equation}
which correspond combinatorially to the out and in-components of vertex $x$ in the causal network, respectively, are interpreted as being the future and past light cones of the associated event in ${\mathcal{M}}$, respectively. In this way, one can think of a single path through a Wolfram model multiway system as representing a deterministic algorithmic method for generating causal sets\cite{malament}\cite{dowker} (albeit ones that are equipped with additional topological structure arising from the hypergraph). An example of a causal network for a simple Wolfram model system is shown in Figure \ref{fig:Figure34}.

\begin{figure}[ht]
\centering
\includegraphics[width=0.395\textwidth]{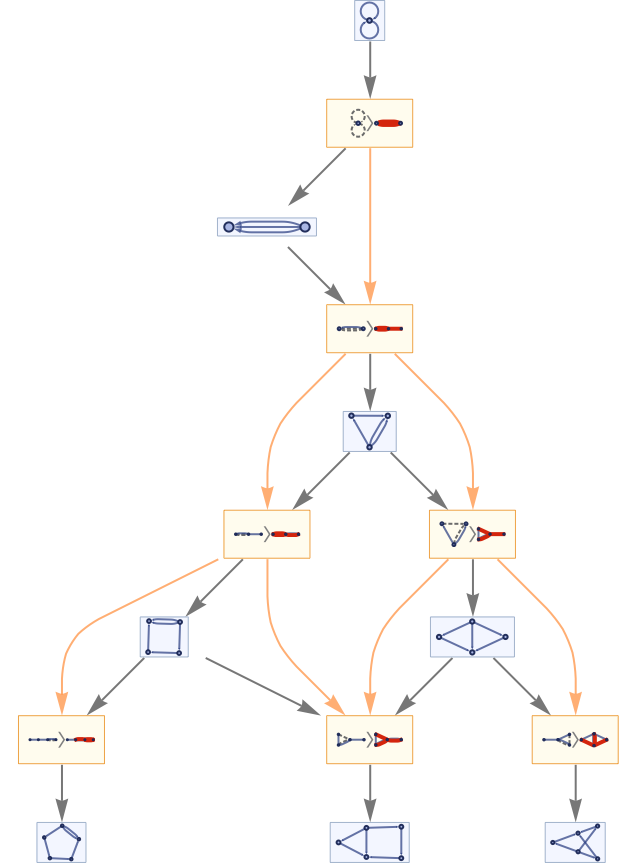}
\caption{The multiway evolution causal graph (with evolution edges shown in gray, and causal edges shown in orange) for the set substitution system ${\left\lbrace \left\lbrace x, y \right\rbrace, \left\lbrace z, y \right\rbrace \right\rbrace \to \left\lbrace \left\lbrace x, w \right\rbrace, \left\lbrace y, w \right\rbrace, \left\lbrace z, w \right\rbrace \right\rbrace}$, illustrating trivial causal invariance.}
\label{fig:Figure35}
\end{figure}

The notion of confluence in abstract rewriting theory is deeply related to (and, indeed, is a necessary but not sufficient condition for) the criterion of \textit{causal invariance} in multiway evolution:

\begin{definition}
A multiway system is ``causal invariant'' if and only if the causal networks generated by following all paths through the multiway system are (eventually) isomorphic as directed, acyclic graphs.
\end{definition}
An example of a multiway evolution causal graph (in which updating events are shown in yellow, state vertices are shown in blue, evolution edges are shown in gray and causal edges are shown in orange) for a system exhibiting trivial causal invariance is featured in Figure \ref{fig:Figure35}.

Although our description so far has treated hypergraph transformation rules in terms of elementary operations on sets, it is important to note that there also exists a purely categorical description of the same class of transformations. Specifically, a generic abstract rewriting system can be represented category-theoretically by considering the rewrite relation ${\to}$ in the system ${\left( A, \to \right)}$ to be an indexed union of subrelations, such as:

\begin{equation}
\to_1 \cup \to_2 = \to,
\end{equation}
which will be the case in general, since there can exist multiple transformation rules within any given system. Thus, we have a labeled state transition system ${\left( A, \Lambda, \to \right)}$, with the index set given by ${\Lambda}$. This system is simply a bijective function from the set $A$ to a subset of the power set of $A$ indexed by ${\Lambda}$, i.e. ${\mathcal{P} \left( \Lambda \times A \right)}$:

\begin{equation}
p \mapsto \left\lbrace \left( \alpha, q \right) \in \Lambda \times A : p \to^{\alpha} q \right\rbrace.
\end{equation}

\begin{definition}
An ``endofunctor'', denoted $F$, is a functor that maps from a category ${\mathbf{C}}$ to itself:

\begin{equation}
F : \mathbf{C} \to \mathbf{C}.
\end{equation}
\end{definition}

\begin{definition}
The ``F-coalgebra'' for the endofunctor:

\begin{equation}
F : \mathbf{C} \to \mathbf{C},
\end{equation}
is an object $A$ in ${\mathrm{ob} \left( \mathbf{C} \right)}$, equipped with a morphism:

\begin{equation}
\alpha : A \to F A,
\end{equation}
in ${\hom \left( \mathbf{C} \right)}$, and which is consequently denoted ${\left( A, \alpha \right)}$.
\end{definition}
We see, therefore, that this state transition system is simply an F-coalgebra for the power set functor ${\mathcal{P} \left( \Lambda \times \left( - \right) \right)}$, since the power set construction on the category of sets (denoted ${\mathbf{Set}}$) can be represented as a covariant endofunctor:

\begin{equation}
\mathcal{P} : \mathbf{Set} \to \mathbf{Set},
\end{equation}
such that the abstract rewriting system ${\left( A, \to \right)}$ consists of the object $A$ equipped with a morphism of ${\mathbf{Set}}$, denoted ${\to}$ (i.e. the rewrite relation):

\begin{equation}
\to : A \to \mathcal{P} A.
\end{equation}

In the particular case of multiway Wolfram model systems, we construct the following category-theoretic formulation of hypergraph rewriting in terms of adhesive categories and double-pushout (DPO) rewriting  \cite{ehrig}\cite{habel}:

\begin{definition}
A ``span'' is any diagram that consists of two maps with a common domain:

\begin{equation}
B \leftarrow A \rightarrow C.
\end{equation}
More specifically, a span generalizes the binary relation between two objects in a category ${\mathbf{C}}$, by considering instead three objects, $A$, $B$ and $C$ in ${\mathrm{ob} \left( \mathbf{C} \right)}$, and the pair of morphisms:

\begin{equation}
f : A \to B, \qquad { and } \qquad g : A \to C,
\end{equation}
in ${\hom \left( \mathbf{C} \right)}$.
\end{definition}

\begin{definition}
A ``monomorphism'' is a morphism that is left-cancellative under composition. More specifically, it is a morphism:

\begin{equation}
f : A \to B,
\end{equation}
in ${\hom \left( \mathbf{C} \right)}$ for some category ${\mathbf{C}}$, such that, for every object $C$ in ${\mathrm{ob} \left( \mathbf{C} \right)}$, and every pair of morphisms:

\begin{equation}
g_1, g_2 : C \to A,
\end{equation}
in ${\hom \left( \mathbf{C} \right)}$, one has:

\begin{equation}
f \circ g_1 = f \circ g_2 \implies g_1 = g_2,
\end{equation}
which generalizes the notion of $f$ being an injective function.
\end{definition}

\begin{definition}
The ``pullback'' of the pair of morphisms:

\begin{equation}
f : A \to C, \qquad \text{ and } \qquad g : B \to C,
\end{equation}
in ${\hom \left( \mathbf{C} \right)}$ for some category ${\mathbf{C}}$ (i.e. a pair of morphisms with a common codomain), denoted ${P = A \times_{C} B}$, is defined by an object $P$ in ${\mathrm{ob} \left( \mathbf{C} \right)}$ and a pair of morphisms:

\begin{equation}
p_1 : P \to A, \qquad \text{ and } p_2 : P \to B,
\end{equation}
in ${\hom \left( \mathbf{C} \right)}$ such that the following diagram commutes:

\begin{equation}
\begin{tikzcd}
P \arrow[r, "p_2"] \arrow[d, "p_1"] & B \arrow[d, "g"]\\
A \arrow[r, "f"] & C
\end{tikzcd},
\end{equation}
and such that the pullback ${\left( P, p_1, p_2 \right)}$ is universal with respect to this diagram. More specifically, for any other triple ${\left( Q, q_1, q_2 \right)}$ with morphisms:

\begin{equation}
q_1 : Q \to A, \qquad \text{ and } \qquad q_2: Q \to B,
\end{equation}
in ${\hom \left( \mathbf{C} \right)}$ such that:

\begin{equation}
f \circ q_1 = g \circ q_2,
\end{equation}
there must exist a unique morphism:

\begin{equation}
u : Q \to P,
\end{equation}
in ${\hom \left( \mathbf{C} \right)}$, such that the following compositional equations are satisfied:

\begin{equation}
p_2 \circ u = q_2, \qquad \text{ and } \qquad p_1 \circ u = q_1.
\end{equation}
\end{definition}
These compositional equations for the universal property are equivalent to stating that, for any triple ${\left( Q, q_1, q_2 \right)}$ for which the following diagram commutes, there much exist a unique morphism ${u : Q \to P}$ for which the diagram also commutes:

\begin{equation}
\begin{tikzcd}
Q \arrow[rrd, "q2", bend left = 20] \arrow[rd, dashed, "u"] \arrow[rdd, "q_1", bend right = 20]\\
& P \arrow[r, "p_2"] \arrow[d, "p_1"] & B \arrow[d, "g"]\\
& A \arrow[r, "f"] & C
\end{tikzcd}.
\end{equation}

\begin{definition}
The ``pushout'' of the pair of morphisms:

\begin{equation}
f : C \to A \qquad \text{ and } \qquad g : C \to B,
\end{equation}
in ${\hom \left( \mathbf{C} \right)}$ for some category ${\mathbf{C}}$ (i.e. a pair of morphisms with a common domain), denoted ${P = A +_{C} B}$, is the dual notion to a pullback, and is defined by an object $P$ and a pair of morphisms:

\begin{equation}
p_1 : A \to P, \qquad \text{ and } \qquad p_2 : B \to P,
\end{equation}
in ${\hom \left( \mathbf{C} \right)}$ such that the following diagram commutes:

\begin{equation}
\begin{tikzcd}
P & B \arrow[l, "p_2"]\\
A \arrow[u, "p_1"] & C \arrow[l, "f"] \arrow[u, "g"]
\end{tikzcd},
\end{equation}
and such that the pushout ${\left( P, p_1, p_2 \right)}$ is universal with respect to this diagram. More specifically, for any other triple ${Q \left( q_1, q_2 \right)}$ with morphisms:

\begin{equation}
q_1 : A \to Q, \qquad \text{ and } \qquad q_2 : B \to Q,
\end{equation}
in ${\hom \left( \mathbf{C} \right)}$ such that:

\begin{equation}
q_1 \circ f = q_2 \circ g,
\end{equation}
there must exist a unique morphism:

\begin{equation}
u : P \to Q,
\end{equation}
in ${\hom \left( \mathbf{C} \right)}$, such that the following compositional equations are satisfied:

\begin{equation}
u \circ p_2 = q_2, \qquad \text{ and } \qquad u \circ p_1 = q_1.
\end{equation}
\end{definition}
These compositional equations for the universal property, much as in the pullback case above, are equivalent to stating that, for any triple ${\left( Q, q_1, q_2 \right)}$ for which the following diagram commutes, there must exist a unique morphism ${u : P \to Q}$ for which the diagram also commutes:

\begin{equation}
\begin{tikzcd}
Q\\
& P \arrow[ul, "u", dashed] & B \arrow[ull, "q_2", bend right = 20] \arrow[l, "p_2"]\\
& A \arrow[u, "p_1"] \arrow[uul, "q_1", bend left = 20] & C \arrow[l, "f"] \arrow[u, "g"]
\end{tikzcd}.
\end{equation}

\begin{definition}
A pushout of morphisms:

\begin{equation}
f^{\prime} : B \to D, \qquad \text{ and } g^{\prime} : C \to D,
\end{equation}
in ${\hom \left( \mathbf{C} \right)}$ for some category ${\mathbf{C}}$, of a span:

\begin{equation}
g : A \to B, \qquad \text{ and } \qquad f : A \to C,
\end{equation}
is known as a ``van-Kampen square'' if and only if, for every commutative diagram:

\begin{equation}
\begin{tikzcd}
B^{\prime} \arrow[rd, "h_{B}"] \arrow[ddd, "f_{h}^{\prime}"] & & & A^{\prime} \arrow[lll, "g_h"] \arrow[ld, "h_{A}"] \arrow[ddd, "f_{h}"]\\
& B \arrow[d, "f^{\prime}"] & A \arrow[d, "f"] \arrow[l, "g"]\\
& D & C \arrow[l, "g^{\prime}"]\\
D^{\prime} \arrow[ur, "h_{D}"] & & & C^{\prime} \arrow[ul, "h_{C}"] \arrow[lll, "g_{h}^{\prime}"]
\end{tikzcd},
\end{equation}
for which the sub-diagrams:

\begin{equation}
\begin{tikzcd}
B^{\prime} \arrow[d, "h_{B}"] & A^{\prime} \arrow[l, "g_{H}"] \arrow[d, "h_{A}"]\\
B & A \arrow[l, "g"]
\end{tikzcd},
\end{equation}
and:

\begin{equation}
\begin{tikzcd}
A \arrow[d, "f"] & A^{\prime} \arrow[l, "h_{A}"] \arrow[d, "f_{h}"]\\
C & C^{\prime} \arrow[l, "h_{C}"]
\end{tikzcd},
\end{equation}
are pullbacks, a certain compatibility condition between pushouts and pullbacks is satisfied. More specifically, the pair of morphisms:

\begin{equation}
f_{h}^{\prime} : B^{\prime} \to D^{\prime}, \qquad \text{ and } \qquad g_{h}^{\prime} : C^{\prime} \to D^{\prime},
\end{equation}
is a pushout of the span:

\begin{equation}
g_{h} : A^{\prime} \to B^{\prime}, \qquad \text{ and } \qquad f_{h} : A^{\prime} \to C^{\prime},
\end{equation}
if and only if the sub-diagrams:

\begin{equation}
\begin{tikzcd}
B^{\prime} \arrow[r, "h_{B}"] \arrow[d, "f_{h}^{\prime}"] & B \arrow[d, "f^{\prime}"]\\
D^{\prime} \arrow[r, "h_{D}"] & D
\end{tikzcd},
\end{equation}
and:

\begin{equation}
\begin{tikzcd}
D & C \arrow[l, "g^{\prime}"]\\
D^{\prime} \arrow [u, "h_{D}"] & C^{\prime} \arrow[u, "h_{C}"] \arrow[l, "g_{h}^{\prime}"]
\end{tikzcd},
\end{equation}
are pullbacks.
\end{definition}

\begin{definition}
A category ${\mathbf{C}}$ is known as an ``adhesive category''\cite{lack} if and only if it has pullbacks, and all pushouts along monomorphisms are van-Kampen squares.
\end{definition}
In the context of an adhesive category, hypergraph transformation rules can thus be defined in terms of double-pushout rewrites and direct derivations as follows\cite{ehrig2}:

\begin{definition}
A ``transformation rule'':

\begin{equation}
\rho = \left( l : K \to L, r : K \to R \right),
\end{equation}
is a span of monomorphisms, where the left- and right-hand-sides of the rule are given by the objects $L$ and $R$, respectively.
\end{definition}

\begin{definition}
A ``rule match'':

\begin{equation}
m : L \to G,
\end{equation}
for a transformation rule ${\rho}$ within an object (such as a hypergraph) $G$ is a morphism from the left-hand-side of the rule ${\rho}$ to $G$.
\end{definition}

\begin{definition}
A transformation rule ${\rho}$ is ``applicable'' at match $m$ if there exists a pair of pushout diagrams of the form:

\begin{equation}
\begin{tikzcd}
L \arrow[d, "m"] & K \arrow[l, "l"] \arrow[d, "n"] \arrow[r, "r"] & R \arrow[d, "p"]\\
G & D \arrow[l, "g"] \arrow[r, "h"] & H
\end{tikzcd},
\end{equation}
i.e. if the pairs of morphisms:

\begin{equation}
m : L \to G, \qquad \text{ and } \qquad g : D \to G,
\end{equation}
and:

\begin{equation}
p : R \to H, \qquad \text{ and } \qquad h : D \to H,
\end{equation}
constitute pushouts of the pairs of morphisms:

\begin{equation}
l : K \to L, \qquad \text{ and } \qquad n : K \to D,
\end{equation}
and:

\begin{equation}
r : K \to R, \qquad \text{ and } \qquad n : K \to D,
\end{equation}
respectively.
\end{definition}

\begin{definition}
A ``direct derivation'' refers to the pair of pushouts that appears in the application of rule ${\rho}$ at match $m$.
\end{definition}
Clearly, by these definitions, only reversible transformations can be described by such a scheme; if there exists a direct derivation from $G$ to $H$ using rule ${\rho}$, then there must exist a direct derivation from $H$ to $G$ using the inverse rule ${\rho^{-1}}$. This constitutes an obstruction to obtaining a full description of the Wolfram model using double-pushout rewritings (since the Wolfram model scheme is sufficiently general that it also considers rewritings that are not strictly reversible). However, since the equational rewriting rules for the ZX-calculus that the present article considers \textit{are} necessarily reversible, this restriction will not be of concern to us at present.

\begin{figure}[ht]
\centering
\includegraphics[width=0.295\textwidth]{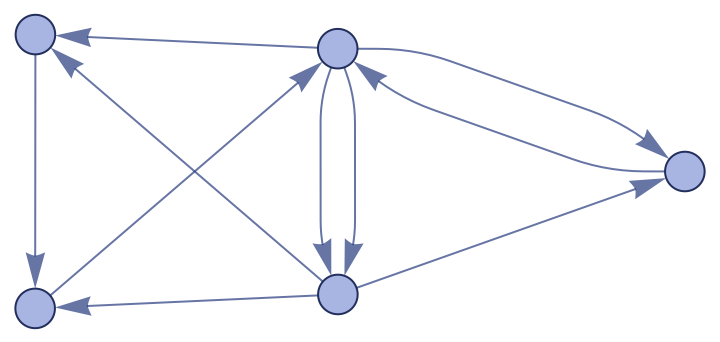}
\caption{The set system ${\left\lbrace \left\lbrace 0, 2 \right\rbrace, \left\lbrace 1, 2 \right\rbrace, \left\lbrace 3, 0 \right\rbrace, \left\lbrace 0, 1 \right\rbrace, \left\lbrace 1, 3 \right\rbrace, \left\lbrace 2, 3 \right\rbrace, \left\lbrace 4, 0 \right\rbrace, \left\lbrace 0, 1 \right\rbrace, \left\lbrace 1, 4 \right\rbrace, \left\lbrace 0, 4 \right\rbrace \right\rbrace}$, represented as a directed hypergraph.}
\label{fig:Figure8}
\end{figure}

\begin{figure}[ht]
\centering
\includegraphics[width=0.595\textwidth]{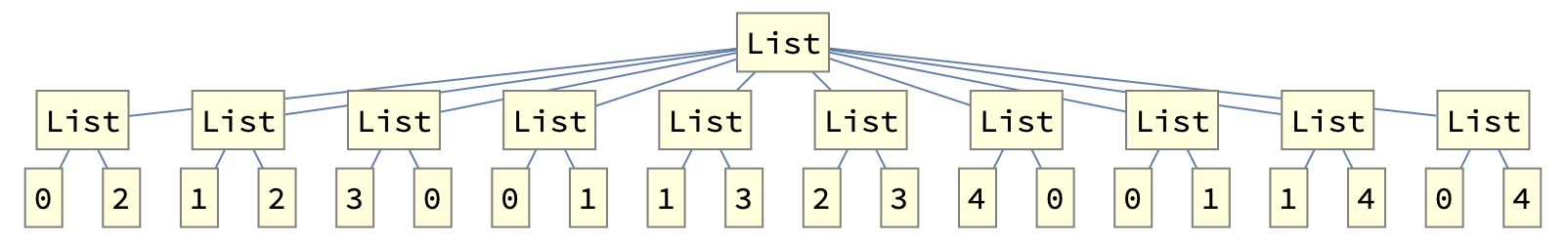}
\caption{The set system ${\left\lbrace \left\lbrace 0, 2 \right\rbrace, \left\lbrace 1, 2 \right\rbrace, \left\lbrace 3, 0 \right\rbrace, \left\lbrace 0, 1 \right\rbrace, \left\lbrace 1, 3 \right\rbrace, \left\lbrace 2, 3 \right\rbrace, \left\lbrace 4, 0 \right\rbrace, \left\lbrace 0, 1 \right\rbrace, \left\lbrace 1, 4 \right\rbrace, \left\lbrace 0, 4 \right\rbrace \right\rbrace}$, represented as a rooted tree denoting the hierarchical collection of ordered relations between elements.}
\label{fig:Figure9}
\end{figure}

\begin{figure}[ht]
\centering
\includegraphics[width=0.495\textwidth]{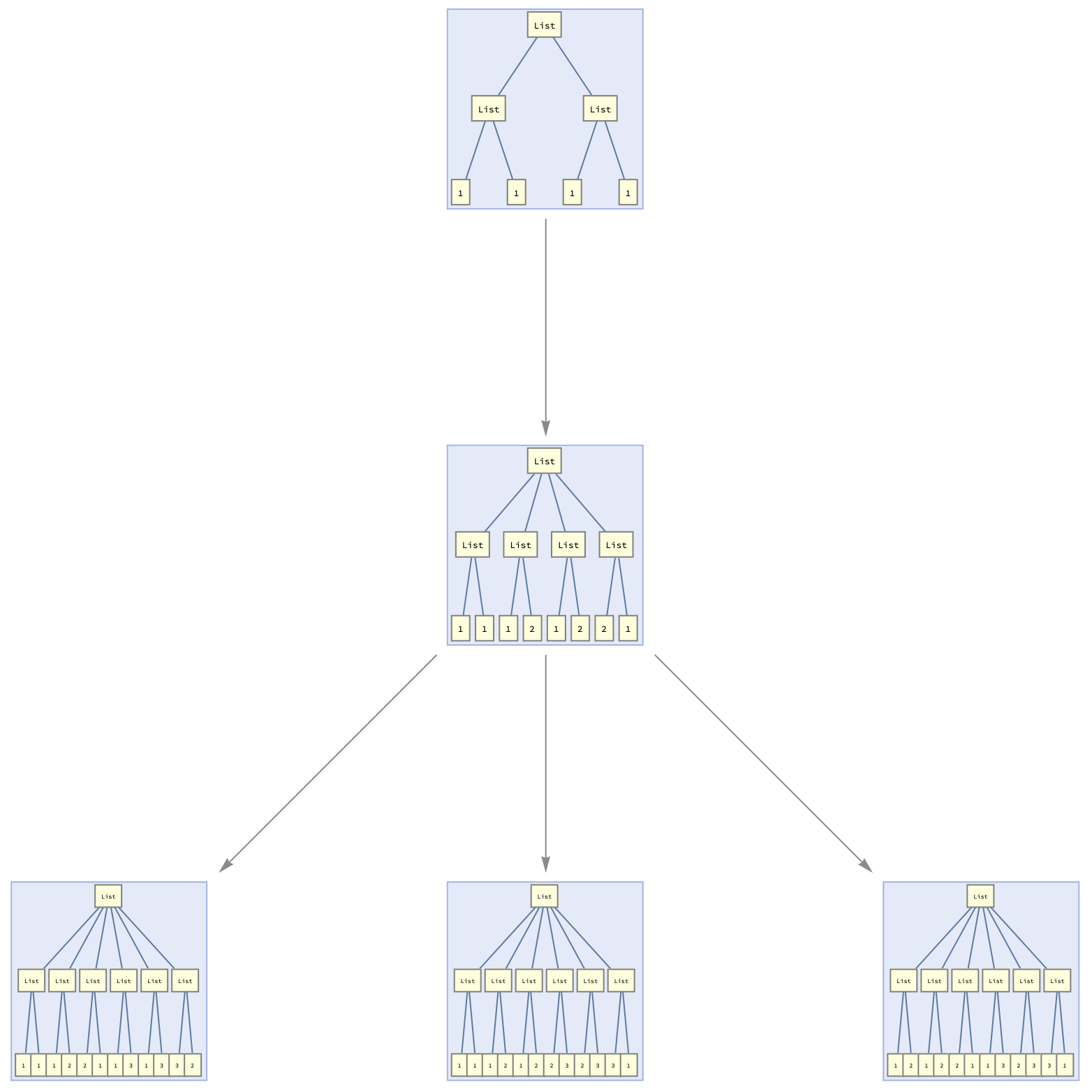}
\caption{The multiway evolution graph corresponding to the non-deterministic evolution of the set substitution system ${\left\lbrace \left\lbrace x, y \right\rbrace, \left\lbrace y, z \right\rbrace \right\rbrace \to \left\lbrace \left\lbrace w, y \right\rbrace, \left\lbrace y, z \right\rbrace, \left\lbrace z, w \right\rbrace, \left\lbrace x, w \right\rbrace \right\rbrace}$, represented in terms of graph transformations being applied to rooted trees.}
\label{fig:Figure10}
\end{figure}

As mentioned towards the beginning of this section, hypergraph transformation rules and set substitution rules are trivially interconvertible, such that each hypergraph can be represented alternately as a rooted tree representing the hierarchical collection of ordered relations between abstract elements, as shown in Figures \ref{fig:Figure8} and \ref{fig:Figure9}. The same multiway evolution shown above for the set substitution rule:

\begin{equation}
\left\lbrace \left\lbrace x, y \right\rbrace, \left\lbrace y, z \right\rbrace \right\rbrace \to \left\lbrace \left\lbrace w, y \right\rbrace, \left\lbrace y, z \right\rbrace, \left\lbrace z, w \right\rbrace, \left\lbrace x, w \right\rbrace \right\rbrace,
\end{equation}
can thus be recast in terms of graph transformations being applied to such rooted trees, as shown in Figure \ref{fig:Figure10}.

This realization allows us to consider certain generalizations of Wolfram model multiway systems described by term rewriting systems:

\begin{definition}
A ``term'' is an expression which contains nested sub-expressions.
\end{definition}

\begin{figure}[ht]
\centering
\includegraphics[width=0.595\textwidth]{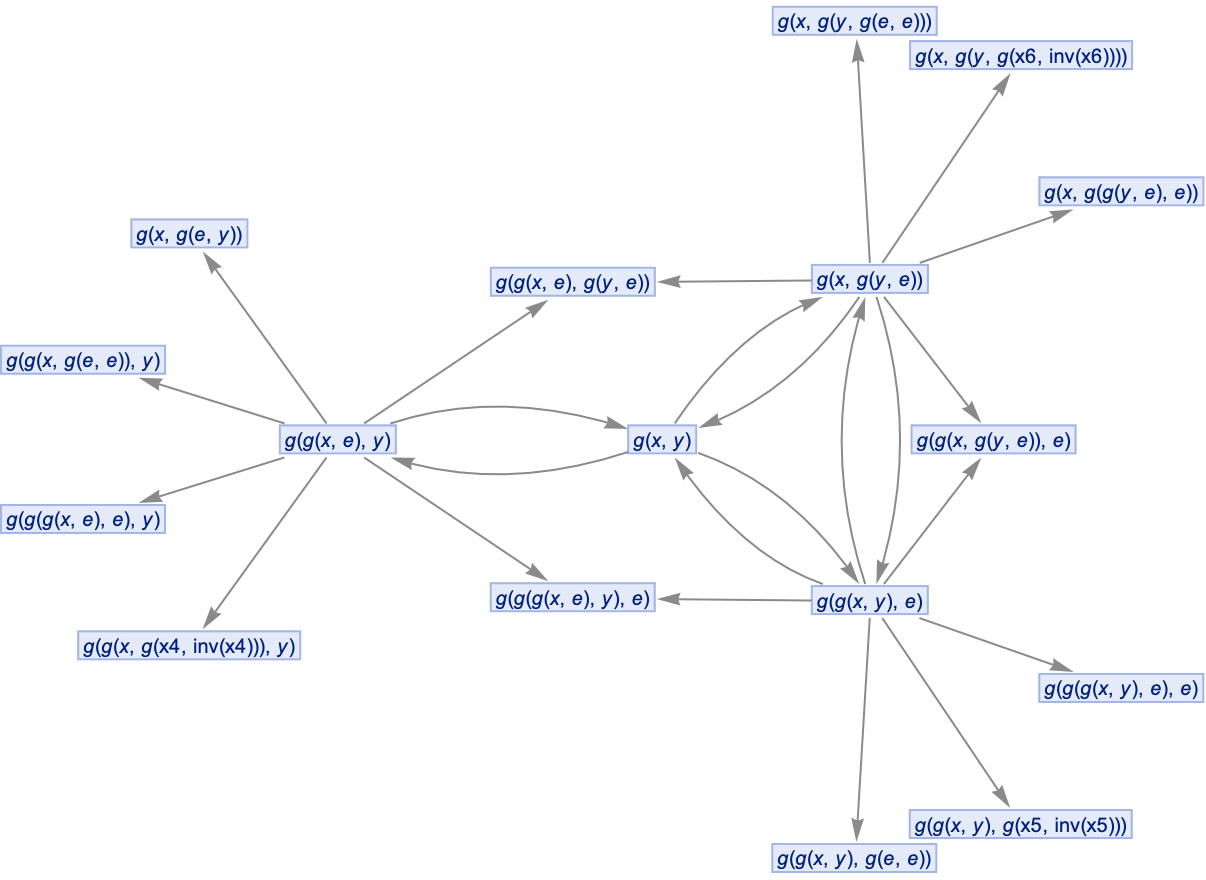}
\caption{The multiway states graph (i.e. a variant of a multiway evolution graph in which cycles are permitted) corresponding to the evolution of the multiway operator system for the axioms of group theory, defined by the term rewriting rules ${\left\lbrace g[x\_, g[y\_, z\_]] :> g[g[x, y], z], g[g[x\_, y\_], z\_] :> g[x, g[y, z]] \right\rbrace}$ (associativity), ${\left\lbrace g[a\_, e] :> a, a\_ :> g[a, e] \right\rbrace}$ (right identity) and ${\left\lbrace g[a\_, inv[a\_]] :> e, e :> g[a, inv[a]] \right\rbrace}$ (right inverse).}
\label{fig:Figure11}
\end{figure}

\begin{figure}[ht]
\centering
\includegraphics[width=0.595\textwidth]{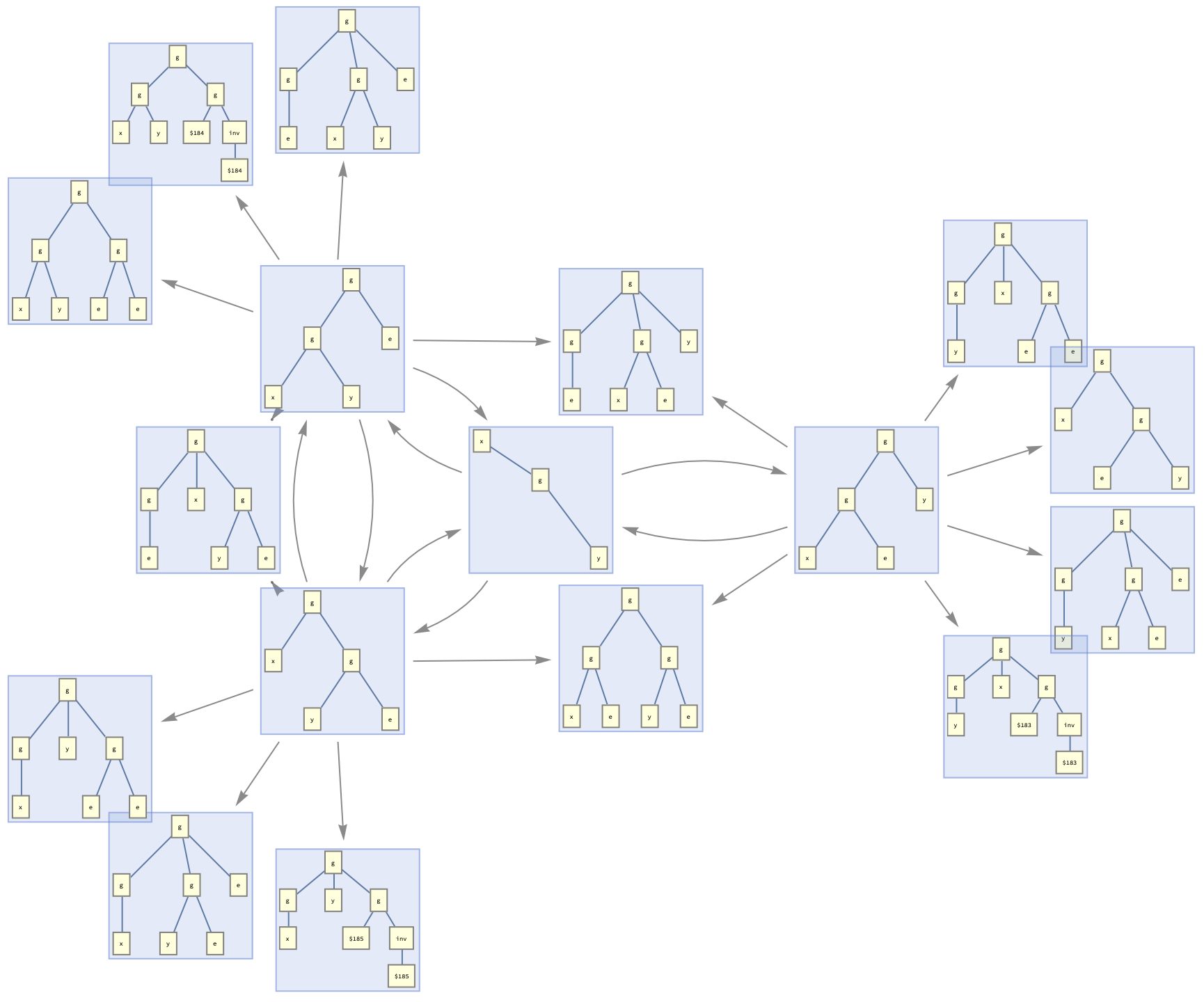}
\caption{The multiway states graph (i.e. a variant of a multiway evolution graph in which cycles are permitted) corresponding to the evolution of the multiway operator system for the axioms of group theory, defined by the term rewriting rules ${\left\lbrace g[ x\_, g[y\_, z\_]] :> g[g[x, y], z], g[g[x\_, y\_], z\_] :> g[x, g[y,z]] \right\rbrace}$ (associativity), ${\left\lbrace g[a\_, e] :> a, a\_ :> g[a, e] \right\rbrace}$ (right identity) and ${\left\lbrace g[a\_, inv[a\_]] :> e, e :> g[a, inv[a]] \right\rbrace}$ (right inverse), represented in terms of graph transformations being applied to rooted trees.}
\label{fig:Figure12}
\end{figure}

\begin{definition}
A ``term rewriting system'' (TRS) is an abstract rewriting system whose objects are all terms.
\end{definition}
In other words, whereas Wolfram model multiway systems consider only nestings of expressions at a single level (hence why all of the rooted trees shown in the multiway system described above have a depth of exactly two), we can instead consider multiway systems which involve arbitrary nestings of expressions, in which the associated rooted trees can therefore have arbitrary depth. For instance, the axioms of group theory can be represented as a multiway operator system, with term rewriting rules ${\left\lbrace g[x\_, g[y\_, z\_]] :> g[g[x, y], z], \right.}$ ${\left. g[g[x\_, y\_], z\_] :> g[x, g[y, z]] \right\rbrace}$ (associativity), ${\left\lbrace g[a\_, e] :> a, a\_ :> g[a, e] \right\rbrace}$ (right identity) and ${\left\lbrace g[a\_, inv[a\_]] :> e, \right.}$ ${\left. e:> g[a, inv[a]] \right\rbrace}$ (right inverse), as shown in Figure \ref{fig:Figure11}, or, in terms of transformations being applied to rooted trees, as in Figure \ref{fig:Figure12}. Crucially, by enabling the manipulation of rooted trees of arbitrary depth, we are greatly expanding the class of possible symbolic rewriting systems that can be considered using this framework, including (in particular) arbitrary diagrammatic rewriting systems, with the ZX-calculus being a notable special case.

\clearpage

\section{The Multiway System as an Embedding Space of ZX-Diagrams}
\label{sec:section2}

The multiway representation of an abstract rewriting system serves to provide a discrete \textit{embedding space} for the collection of all possible deductions in a given diagrammatic reasoning system. In what follows, we will explicitly illustrate this embedding for the case of the ZX-calculus.

\subsection{ZX Generators within the Wolfram Model}

\begin{figure}[ht]
\centering
\includegraphics[width=0.395\textwidth]{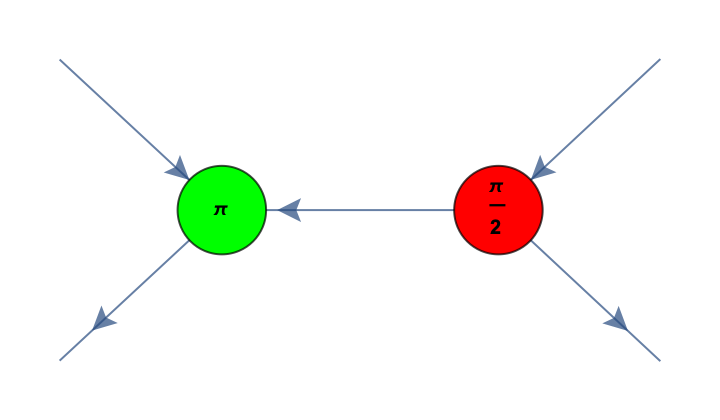}
\includegraphics[width=0.595\textwidth]{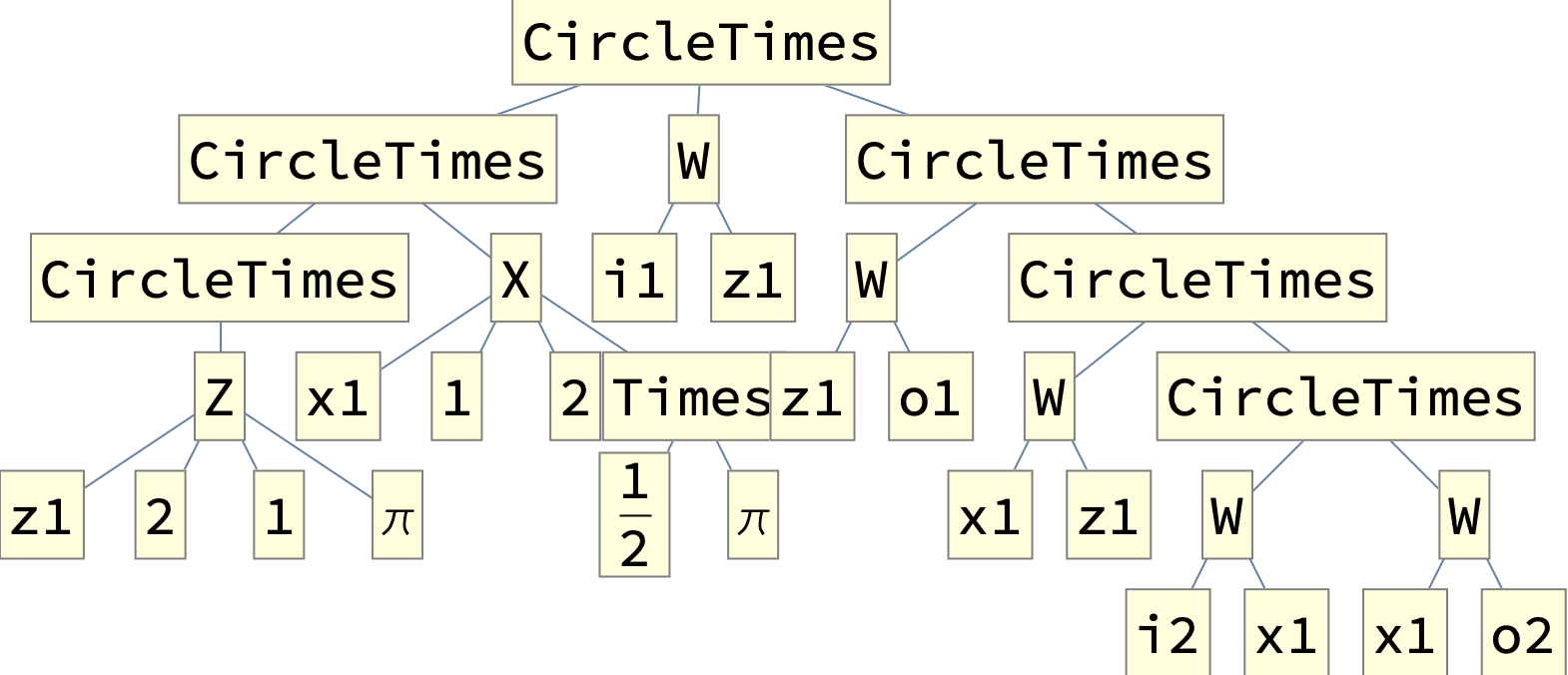}
\caption{On the left, a ZX-diagram corresponding to the nested operator expression ${Z \left[ z_1, 2, 1, \pi \right] \otimes \left( X \left[ x_1, 1, 2, \frac{\pi}{2} \right] \otimes \left( W \left[ i_1, z_1 \right] \otimes \left( W \left[ z_1, o_1 \right] \otimes \left( W \left[ x_1, z_1 \right] \otimes \left( W \left[ i_2, x_1 \right] \otimes W \left[ x_1, o_2 \right] \right) \right) \right) \right) \right)}$. On the right, a representation of the same expression as a rooted tree.}
\label{fig:Figure13}
\end{figure}

\begin{figure}[ht]
\centering
\includegraphics[width=0.595\textwidth]{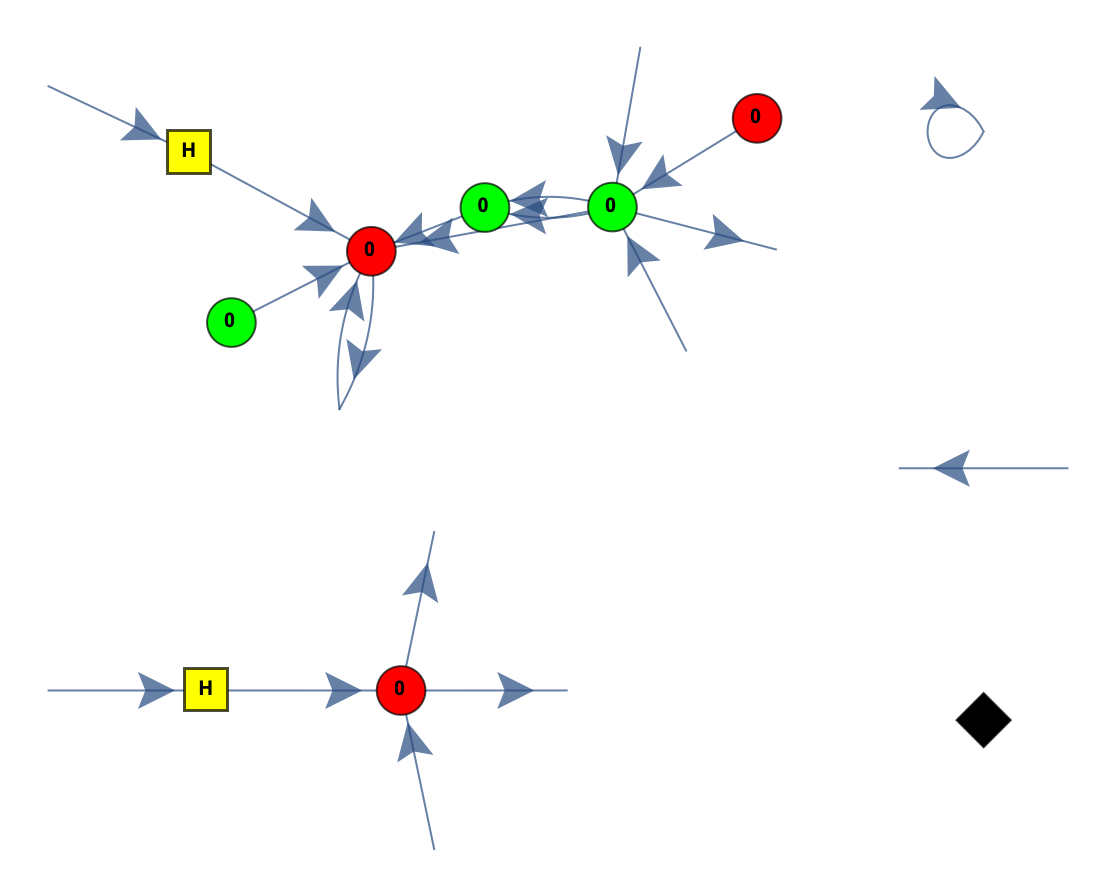}
\includegraphics[width=0.895\textwidth]{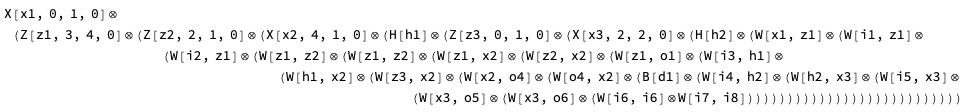}
\caption{A rendering of a more complicated ZX-diagram, again represented as a nested operator expression (shown below).}
\label{fig:Figure14}
\end{figure}

\begin{figure}[ht]
\centering
\includegraphics[width=0.695\textwidth]{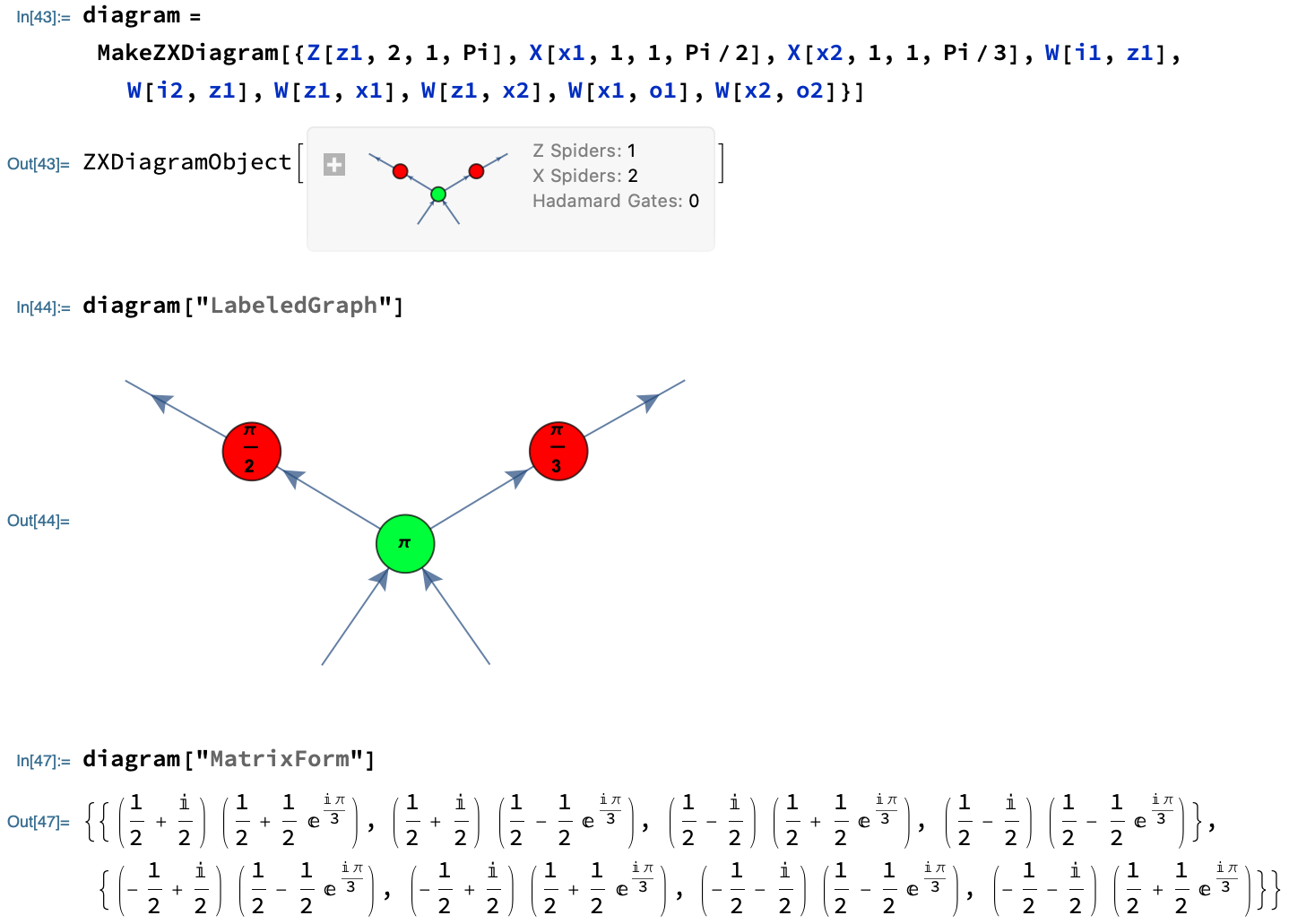}
\caption{An illustrative example of the mutual inter-convertibility between symbolic ZX-diagrams and their associated explicit matrix forms, as demonstrated using the \textit{ZXMultiwaySystem} software package in the Wolfram Language.}
\label{fig:Figure64}
\end{figure}

For the purpose of compiling and realizing diagrammatic rewritings of the ZX-calculus as Wolfram model multiway systems, we consider now a particular class of nested operator expressions corresponding to ZX-diagrams, as shown in Figure \ref{fig:Figure13} for the case of the expression:

\begin{equation}
Z \left[ z_1, 2, 1, \pi \right] \otimes \left( X \left[ x_1, 1, 2, \frac{\pi}{2} \right] \otimes \left( W \left[ i_1, z_1 \right] \otimes \left( W \left[ z_1, o_1 \right] \otimes \left( W \left[ x_1, z_1 \right] \otimes \left( W \left[ i_2, x_1 \right] \otimes W \left[ x_1, o_2 \right] \right) \right) \right) \right) \right),
\end{equation}
along with a much more complicated diagram, represented using the same basic scheme, in Figure \ref{fig:Figure14}. To this end, we have developed the Wolfram Language software packages \textit{MakeZXDiagram}\cite{wfr1}, for easily generating, manipulating, computing and displaying these diagrams (using the nested operator form as the internal representation), and \textit{ZXMultiwaySystem}\cite{wfr2}, for actually simulating the diagrammatic rewrites as Wolfram model multiway systems. This functionality is built upon a previous (and more general) software framework that we developed known as \textit{MultiwayOperatorSystem}\cite{wfr3}, which we also make extensive use of here. Amongst several other algorithmic and visualization features, the \textit{ZXMultiwaySystem} software package offers the ability to convert between symbolic ZX-diagrams and their explicit matrix forms, as shown in Figure \ref{fig:Figure64}.

The two principal generators of the ZX-diagrams shown here are the ``Z-spiders'' (colored in green) and the ``X-spiders'' (colored in red) - so-named because they correspond to generalized variants of rotations around the Z- and X-axes of the Bloch sphere, respectively. The Z-spiders are used to denote states/unitary operators/linear isometries/projections/etc. with respect to the computational basis:

\begin{equation}
\left\lbrace \ket{0}, \ket{1} \right\rbrace = \left\lbrace \begin{bmatrix}
1\\
0
\end{bmatrix}, \begin{bmatrix}
0\\
1
\end{bmatrix} \right\rbrace,
\end{equation}
whilst the X-spiders are used to denote states/unitary operators/linear isometries/projections/etc. with respect to the Hadamard-transformed basis:

\begin{equation}
\left\lbrace \ket{+}, \ket{-} \right\rbrace = \left\lbrace \frac{1}{\sqrt{2}} \left( \ket{0} + \ket{1} \right), \frac{1}{\sqrt{2}} \left( \ket{0} - \ket{1} \right) \right\rbrace.
\end{equation}
The Z- and X-spiders are represented internally by operators of the form ${Z \left[ n, i, o, p \right]}$ and ${X \left[ n, i, o, p \right]}$, respectively, where $n$ designates the ``name'' of the spider (used as an unique identifier when specifying wire configurations), $i$ designates the input arity of the spider (i.e. the number of incoming wires), $o$ designates the output arity of the spider (i.e.  the number of outgoing wires) and $p$ designates the phase of the spider (taken to be an element of the closed interval ${\left[ -2 \pi, 2 \pi \right]}$). The wires are specified by binary operators of the form ${W \left[ s_1, s_2 \right]}$, where ${s_1}$ and ${s_2}$ designate the names of the spiders lying at the start and end points of the wire, respectively. Both spiders and wires are composed using the ${\otimes}$ operator, indicating the fact that all diagram compositions are assumed to be monoidal (i.e. parallel) unless they are explicitly specified to be compositional (i.e. sequential) by the choice of wire configurations.

\begin{figure}[ht]
\centering
\includegraphics[width=0.205\textwidth]{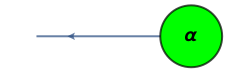}\hspace{0.25\textwidth}
\includegraphics[width=0.205\textwidth]{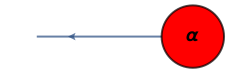}
\caption{Z- and X-spiders with input arities of 0 and output arities of 1, corresponding to the expressions ${Z \left[ z_1, 0, 1, \alpha \right] \otimes W \left[ z_1, o_1 \right]}$ and ${X \left[ x_1, 0, 1, \alpha \right] \otimes W \left[ x_1, o_1 \right]}$, respectively, are interpreted as pure states.}
\label{fig:Figure15}
\end{figure}

Spiders with input arities of 0 and output arities of 1, i.e. expressions of the general form:

\begin{equation}
Z \left[ z_1, 0, 1, \alpha \right] \otimes W \left[ z_1, o_1 \right], \qquad \text{ or } \qquad X \left[ x_1, 0, 1, \alpha \right] \otimes W \left[ x_1, o_1 \right],
\end{equation}
can be interpreted as pure states, as shown in Figure \ref{fig:Figure15}. More specifically, one has the pure states:

\begin{equation}
\ket{0} + e^{i \alpha} \ket{1}, \qquad \text{ and } \qquad \ket{+} + e^{i \alpha} \ket{-},
\end{equation}
for the Z- and X-spiders, respectively.

\begin{figure}[ht]
\centering
\includegraphics[width=0.255\textwidth]{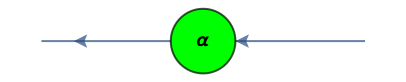}\hspace{0.25\textwidth}
\includegraphics[width=0.255\textwidth]{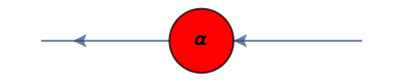}
\caption{Z- and X-spiders with input arities of 1 and output arities of 1, corresponding to the expressions ${Z \left[ z_1, 1, 1, \alpha \right] \otimes \left( W \left[ i_1, z_1 \right] \otimes W \left[ z_1, o_1 \right] \right)}$ and ${X \left[ x_1, 1, 1, \alpha \right] \otimes \left( W \left[ i_1, x_1 \right] \otimes W \left[ x_1, o_1 \right] \right)}$, respectively, are interpreted as unitary maps.}
\label{fig:Figure16}
\end{figure}

Spiders with input arities of 1 and output arities of 1, i.e. expressions of the general form:

\begin{equation}
Z \left[ z_1, 1, 1, \alpha \right] \otimes \left( W \left[ i_1, z_1 \right] \otimes W \left[ z_1, o_1 \right] \right), \qquad \text{ or } \qquad X \left[ x_1, 1, 1, \alpha \right] \otimes \left( W \left[ i_1, x_1 \right] \otimes W \left[ x_1, o_1 \right] \right),
\end{equation}
can be interpreted as unitary maps (i.e. rotations about the Z- or X-axes of the Bloch sphere by angle ${\alpha}$; for ${\alpha = \pi}$, these therefore correspond to the Z and X Pauli matrices, respectively), as shown in Figure \ref{fig:Figure16}. More specifically, one has the unitary maps:

\begin{equation}
\ket{0} \bra{0} + e^{i \alpha} \ket{1} \bra{1}, \qquad \text{ and } \qquad \ket{+} \bra{+} + e^{i \alpha} \ket{-} \bra{-},
\end{equation}
for the Z- and X-spiders, respectively.

\begin{figure}[ht]
\centering
\includegraphics[width=0.255\textwidth]{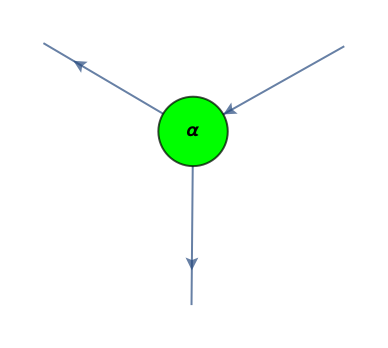}\hspace{0.25\textwidth}
\includegraphics[width=0.255\textwidth]{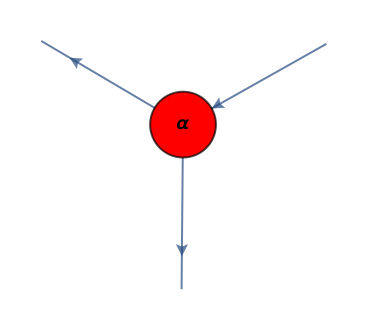}
\caption{Z- and X-spiders with input arities of 1 and output arities of 2, corresponding to the expressions ${Z \left[ z_1, 1, 2, \alpha \right] \otimes \left( W \left[ i_1, z_1 \right] \otimes \left( W \left[ z_1, o_1 \right] \otimes W \left[ z_1, o_2 \right] \right) \right)}$ and ${X \left[ x_1, 1, 2, \alpha \right] \otimes \left( W \left[ i_1, x_1 \right] \otimes \left( W \left[ x_1, o_1 \right] \otimes W \left[ x_1, o_2 \right] \right) \right)}$, respectively, are interpreted as linear isometries.}
\label{fig:Figure17}
\end{figure}

Spiders with input arities of 1 and output arities of 2, i.e. expressions of the general form:

\begin{equation}
Z \left[ z_1, 1, 2, \alpha \right] \otimes \left( W \left[ i_1, z_1 \right] \otimes \left( W \left[ z_1, o_1 \right] \otimes W \left[ z_1, o_2 \right] \right) \right),
\end{equation}
or:

\begin{equation}
X \left[ x_1, 1, 2, \alpha \right] \otimes \left( W \left[ i_1, x_1 \right] \otimes \left( W \left[ x_1, o_1 \right] \otimes W \left[ x_1, o_2 \right] \right) \right),
\end{equation}
can be interpreted as linear isometries (i.e. for ${\alpha = 0}$, these correspond to simple copy operations in either the computational or the Hadamard-transformed bases), as shown in Figure \ref{fig:Figure17}. More specifically, one has the linear isometries:

\begin{equation}
\ket{0 0} \bra{0} + e^{i \alpha} \ket{1 1} \bra{1}, \qquad \text{ and } \qquad \ket{+ +} \bra{+} + e^{i \alpha} \ket{- -} \bra{-},
\end{equation}
for the Z- and X-spiders, respectively.

\begin{figure}[ht]
\centering
\includegraphics[width=0.255\textwidth]{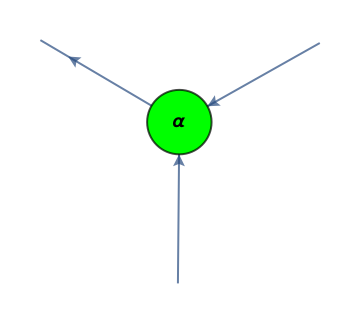}\hspace{0.25\textwidth}
\includegraphics[width=0.255\textwidth]{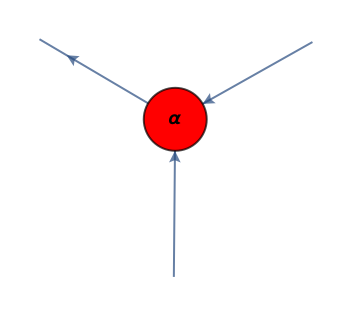}
\caption{Z- and X-spiders with input arities of 2 and output arities of 1, corresponding to the expressions ${Z \left[ z_1, 2, 1, \alpha \right] \otimes \left( W \left[ i_1, z_1 \right] \otimes \left( W \left[ i_2, z_1 \right] \otimes W \left[ z_1, o_1 \right] \right) \right)}$ and ${X \left[ x_1, 2, 1, \alpha \right] \otimes \left( W \left[ i_1, x_1 \right] \otimes \left( W \left[ i_2, x_1 \right] \otimes W \left[ x_1, o_1 \right] \right) \right)}$, respectively, are interpreted as partial linear isometries.}
\label{fig:Figure18}
\end{figure}

Spiders with input arities of 2 and output arities of 1, i.e expressions of the general form:

\begin{equation}
Z \left[ z_1, 2, 1, \alpha \right] \otimes \left( W \left[ i_1, z_1 \right] \otimes \left( W \left[ i_2, z_1 \right] \otimes W \left[ z_1, o_1 \right] \right) \right),
\end{equation}
or:

\begin{equation}
X \left[ x_1, 2, 1, \alpha \right] \otimes \left( W \left[ i_1, x_1 \right] \otimes \left( W \left[ i_2, x_1 \right] \otimes W \left[ x_1, o_1 \right] \right) \right),
\end{equation}
can be interpreted as partial linear isometries (i.e. for ${\alpha = 0}$, these correspond to CNOT operations followed by destructive Z- or X-measurements applied to a single qubit, post-selected to either state ${\ket{0}}$ or ${\ket{+}}$), as shown in Figure \ref{fig:Figure18}. More specifically, one has the partial linear isometries:

\begin{equation}
\ket{0} \bra{0 0} + e^{i \alpha} \ket{1} \bra{1 1}, \qquad \text{ and } \qquad \ket{+} \bra{+ +} + e^{i \alpha} \ket{-} \bra{- -},
\end{equation}
for the Z- and X-spiders, respectively.

\begin{figure}[ht]
\centering
\includegraphics[width=0.205\textwidth]{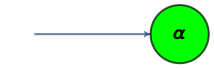}\hspace{0.25\textwidth}
\includegraphics[width=0.205\textwidth]{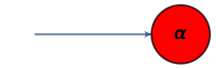}
\caption{Z- and X-spiders with input arities of 1 and output arities of 0, corresponding to the expressions ${Z \left[ z_1, 1, 0, \alpha \right] \otimes W \left[ i_1, z_1 \right]}$ and ${X \left[ x_1, 1, 0, \alpha \right] \otimes W \left[ i_1, x_1 \right]}$, respectively, are interpreted as projections.}
\label{fig:Figure19}
\end{figure}

Spiders with input arities of 1 and output arities of 0, i.e. expressions of the general form:

\begin{equation}
Z \left[ z_1, 1, 0, \alpha \right] \otimes W \left[ i_1, z_1 \right], \qquad \text{ or } \qquad X \left[ x_1, 1, 0, \alpha \right] \otimes W \left[ i_1, x_1 \right],
\end{equation}
can be interpreted as projections (i.e. for ${\alpha = 0}$ or ${\alpha = \pi}$, these correspond to destructive Z- or X-measurements applied to a single qubit, post-selected to either state ${\ket{+}}$, ${\ket{-}}$, ${\ket{0}}$ or ${\ket{1}}$), as shown in Figure \ref{fig:Figure19}. More specifically, one has the projections:

\begin{equation}
\bra{0} + e^{i \alpha} \bra{1}, \qquad \text{ and } \qquad \bra{+} + e^{i \alpha} \bra{-},
\end{equation}
for the Z- and X-spiders, respectively.

\begin{figure}[ht]
\centering
\includegraphics[width=0.255\textwidth]{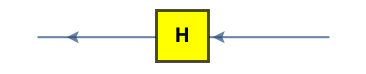}\hspace{0.25\textwidth}
\includegraphics[width=0.145\textwidth]{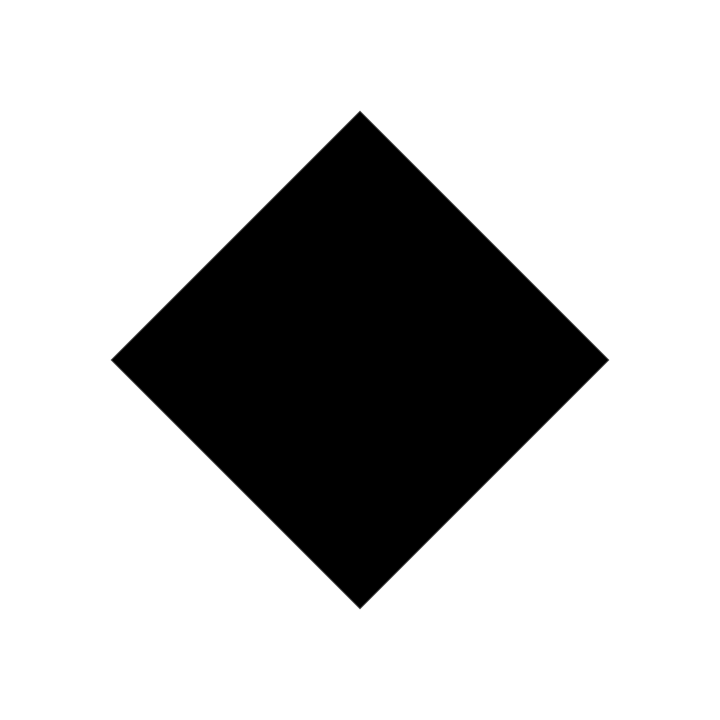}
\caption{The Hadamard and black diamond spiders, corresponding to the expressions ${H \left[ h_1, \right] \otimes \left( W \left[ i_1, h_1 \right] \otimes W \left[ h_1, o_1 \right] \right)}$ and ${B \left[ d_1 \right]}$, are interpreted as Hadamard transformations and overall (multiplicative) scalar factors, respectively.}
\label{fig:Figure20}
\end{figure}

More generally, we can define the Z- and X-spider interpretations inductively as follows: a Z-spider with an input arity of $n$ and an output arity of $m$ corresponds to a linear map of the form:

\begin{equation}
\ket{0}^{\otimes m} \bra{0}^{\otimes n} + e^{i \alpha} \ket{1}^{\otimes m} \bra{1}^{\otimes n},
\end{equation}
whilst an X-spider with an input arity of $n$ and an output arity of $m$ corresponds to a linear map of the form:

\begin{equation}
\ket{+}^{\otimes m} \bra{+}^{\otimes n} + e^{i \alpha} \ket{-}^{\otimes m} \bra{-}^{\otimes n}.
\end{equation}
A wire with an input arity of 1 and an output arity of 1 acts like an identity map:

\begin{equation}
\ket{0} \bra{0} + \ket{1} \bra{1};
\end{equation}
a wire with an input arity of 0 and an output arity of 2 acts like a Bell state:

\begin{equation}
\ket{0 0} + \ket{1 1};
\end{equation}
a wire with an input arity of 2 and an output arity of 0 acts like a Bell effect:

\begin{equation}
\bra{0 0} + \bra{1 1};
\end{equation}
and a pair of twisted wires acts like a SWAP gate (with input arity 2 and output arity 2):

\begin{equation}
\ket{0 0} \bra{0 0} + \ket{0 1} \bra{1 0} + \ket{1 0} \bra{0 1} + \ket{1 1} \bra{1 1}.
\end{equation}
There also exist two additional ``special'' kinds of generators that we have not yet considered: the yellow Hadamard gates with input arities of 1 and output arities of 1, given by expressions of the form:

\begin{equation}
H \left[ h_1 \right] \otimes \left( W \left[ i_1, h_1 \right] \otimes W \left[ h_1, o_1 \right] \right),
\end{equation}
and which act as unitary Hadamard transformations:

\begin{equation}
\ket{+} \bra{0} + \ket{-} \bra{1},
\end{equation}
and also the black diamonds with input and output arities always equal to zero (i.e. they always appear disconnected from the rest of the graph), given by expressions of the form ${B \left[ d_1 \right]}$, and which act as overall (multiplicative) scalar factors of ${\sqrt{D}}$ applied to the whole linear map (where $D$ is the dimensionality of the corresponding Hilbert space). In the above, ${h_1}$ and ${d_1}$ designate the ``names'' of the Hadamard and black diamond spiders, respectively (again, to be used as unique identifiers). The two additional generators are shown in Figure \ref{fig:Figure20}.

The standard equational rewriting rules for the ZX-calculus, as presented in the tutorial of Coecke and Duncan\cite{coecke3}\cite{vandewetering}, are thus implemented directly as equational (i.e. bidirectional) rewriting rules for our multiway operator system; since many of the standard rewriting rules of the ZX-calculus (such as the Z- and X-spider fusion rules) apply to spiders with arbitrary arities and arbitrary wire configurations, they are actually treated by our code not as single rules but rather as infinite rule \textit{schemas} within the multiway operator formalism, analogous to the use of the \textit{bang-box} notation in Quantomatic\cite{quick}. For instance, each instance of a Z- or X-spider fusion rule is computed using a schema function of 5 arguments - namely, the input and output arities of the first Z/X-spider, the input and output arities of the second Z/X-spider, and finally the number of wires connecting the two spiders.

\begin{figure}[ht]
\centering
\includegraphics[width=0.395\textwidth]{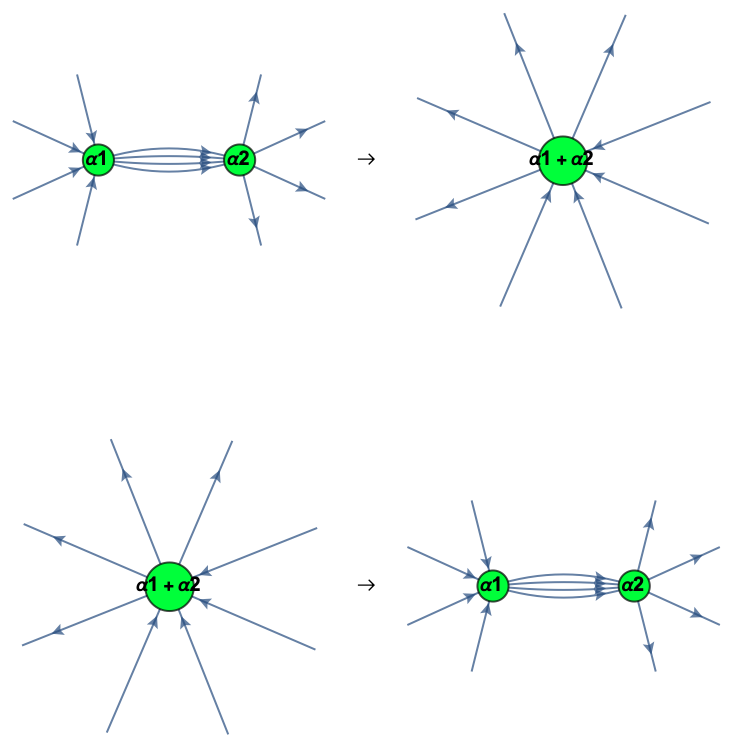}\hspace{0.1\textwidth}
\includegraphics[width=0.395\textwidth]{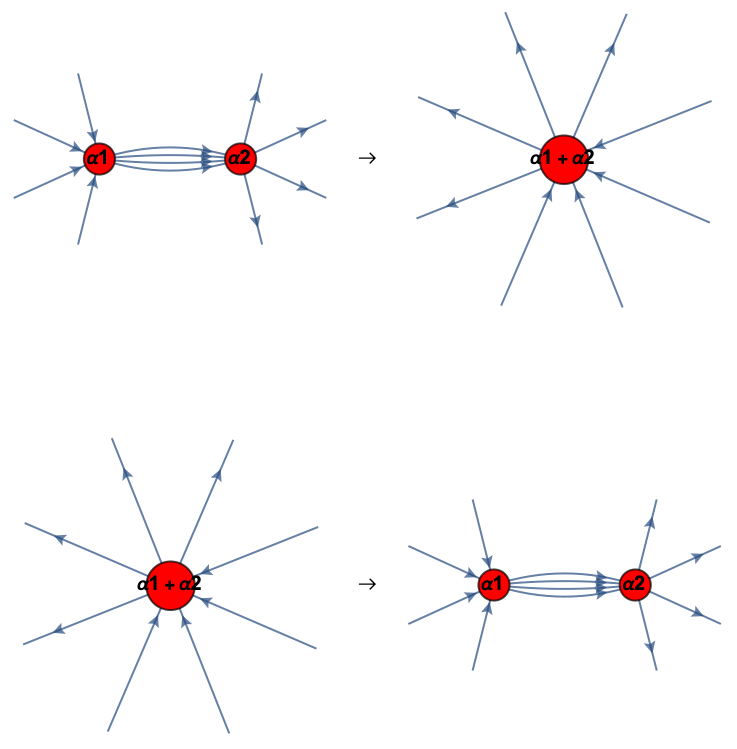}
\includegraphics[width=0.495\textwidth]{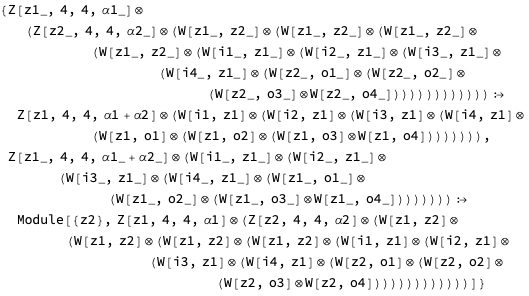}
\includegraphics[width=0.495\textwidth]{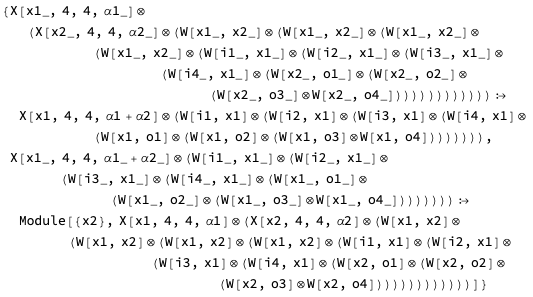}
\caption{The S1 rules, or the Z- and X-spider fusion rules (for the case in which both spiders have input and output arities equal to 4, and in which the two spiders are connected by exactly 4 wires), along with their associated operator forms. These rules correspond to the statement that the Z- and X-spiders represent orthonormal bases (computational and Hadamard-transformed, respectively).}
\label{fig:Figure21}
\end{figure}

The S1 rules, or the Z- and X-spider fusion rules (for the case in which both spiders have input and output arities equal to 4, and in which the two spiders are connected by exactly 4 wires), along with their associated operator forms, are shown in Figure \ref{fig:Figure21}. The algebraic interpretation of these rules is that both the Z- and X-spiders represent orthonormal bases (namely the computational and Hadamard-transformed bases, respectively), and therefore whenever two spiders of the same type touch, they can merge, with their respective phases combining additively.  

\begin{figure}[ht]
\centering
\includegraphics[width=0.495\textwidth]{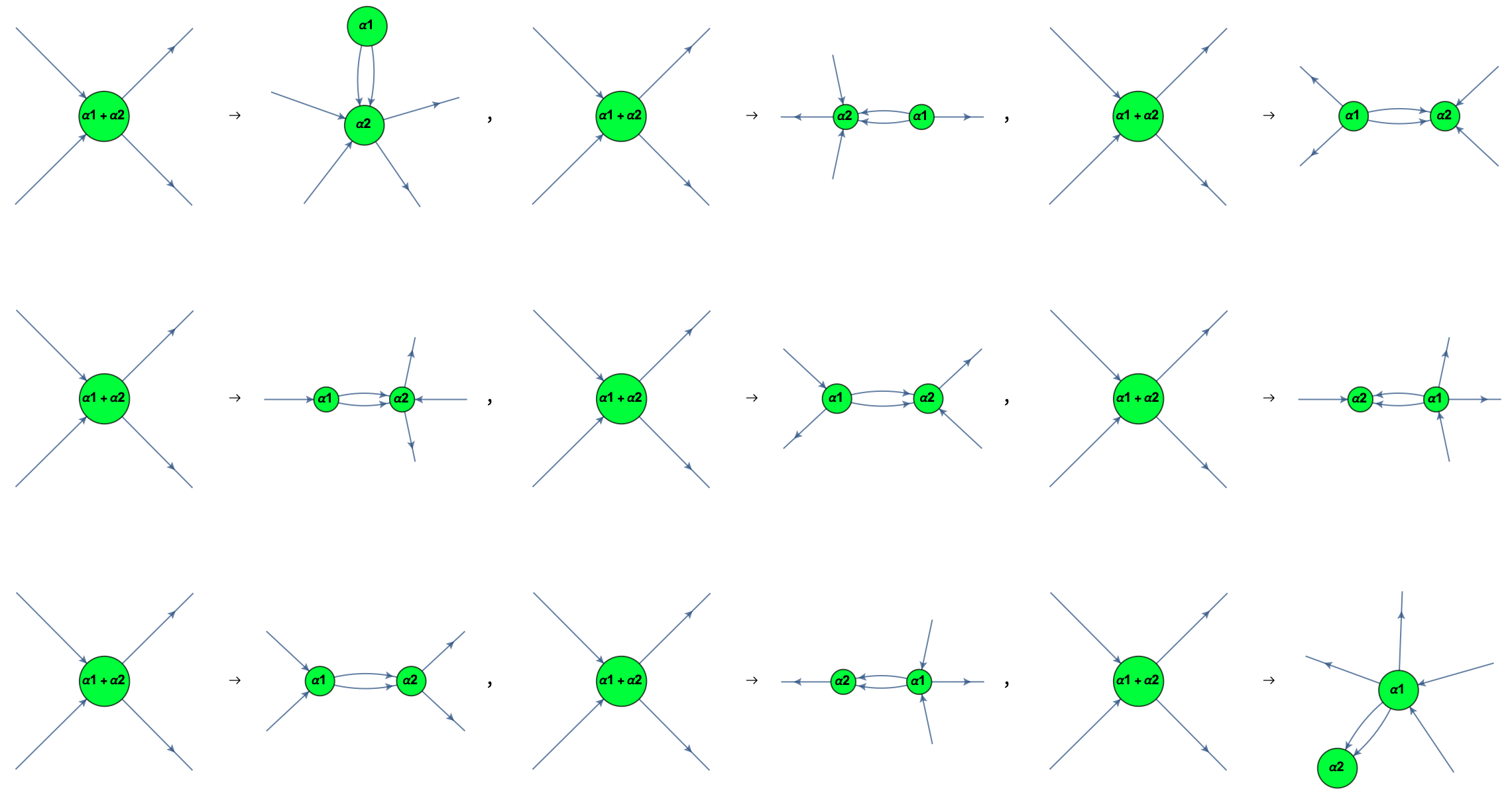}
\includegraphics[width=0.495\textwidth]{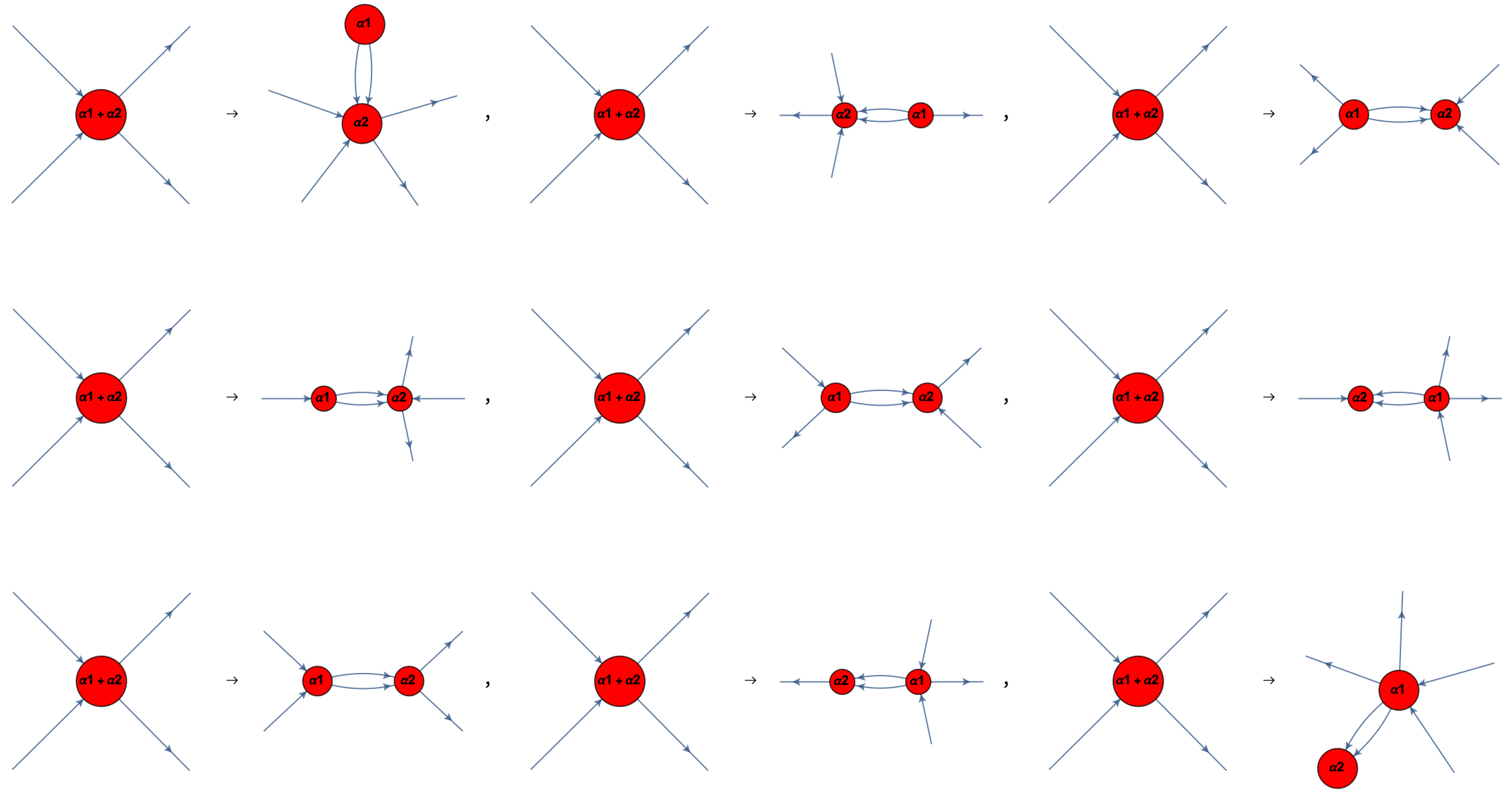}
\includegraphics[width=0.495\textwidth]{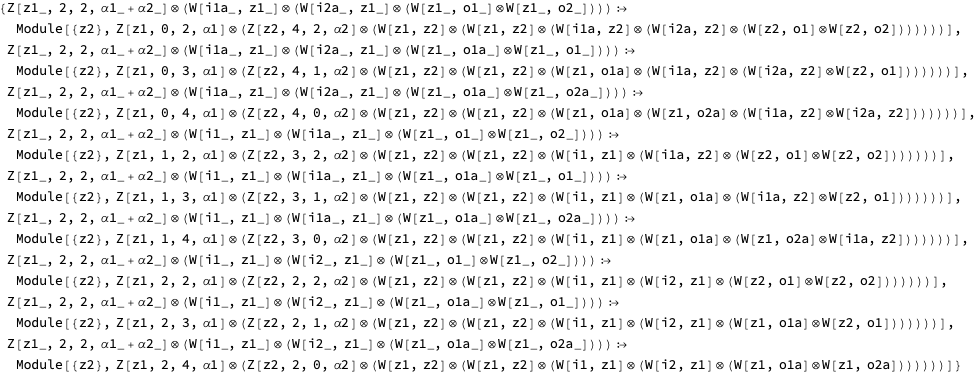}
\includegraphics[width=0.495\textwidth]{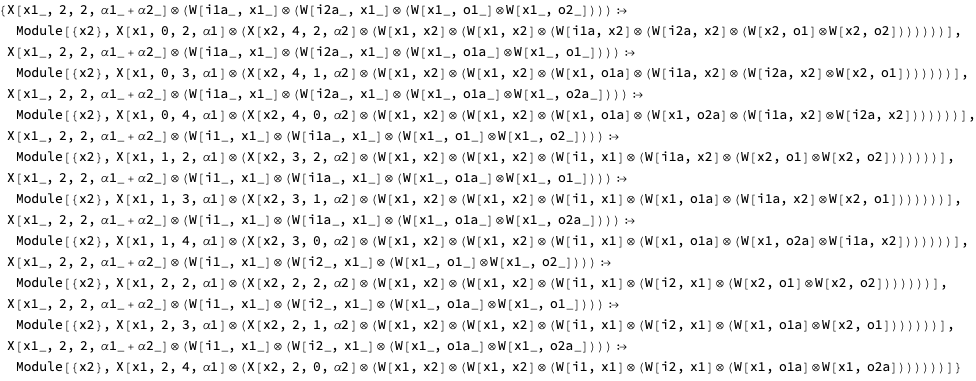}
\caption{Samples of the enumeration of all possible Z- and X-spider ``fission'' rules (shown here for the case in which both spiders have input and output arities equal to 2, and in which the two spiders are connected by exactly 2 wires), along with their associated operator forms. These rules are the time-reversed versions of the S1, or Z- and X-spider fusion rules, shown above.}
\label{fig:Figure59}
\end{figure}

Note that, for a spider of input arity $n$ and output arity $m$, there will in general be ${\left( n + 1 \right) \times \left( m + 1 \right)}$ possible ways for it to ``fission'' into two different spiders, so the time-reversed versions of the S1 rules are actually a little more subtle than the ones presented above. Although we will not worry too much about this subtlety here (as it is implicitly handled by our construction of the canonical rule enumeration later), it is worth noting that, for a true time-reversal of the S1 rules above, one must in fact enact a systematic enumeration of all possible ``fission'' rules, as shown in Figure \ref{fig:Figure59}.

\begin{figure}[ht]
\centering
\includegraphics[width=0.395\textwidth]{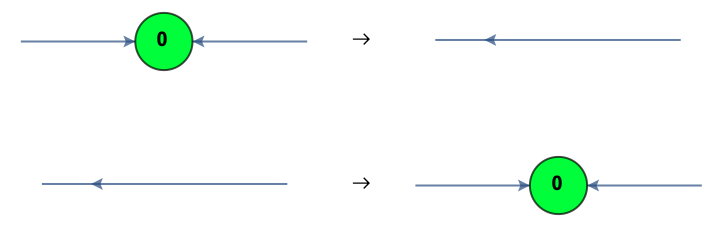}\hspace{0.1\textwidth}
\includegraphics[width=0.395\textwidth]{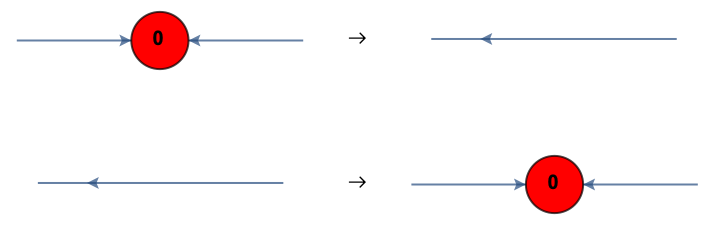}
\includegraphics[width=0.495\textwidth]{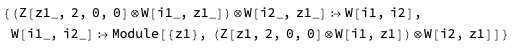}
\includegraphics[width=0.495\textwidth]{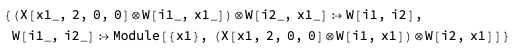}
\includegraphics[width=0.395\textwidth]{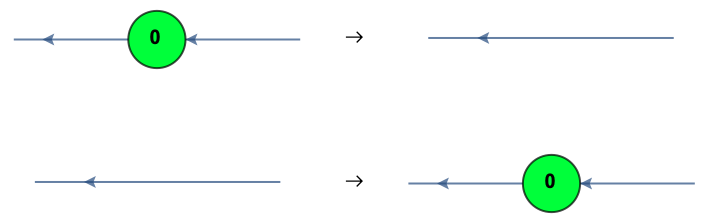}\hspace{0.1\textwidth}
\includegraphics[width=0.395\textwidth]{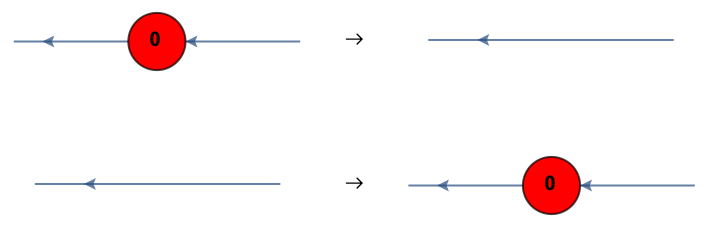}
\includegraphics[width=0.495\textwidth]{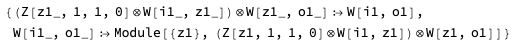}
\includegraphics[width=0.495\textwidth]{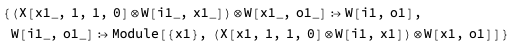}
\includegraphics[width=0.395\textwidth]{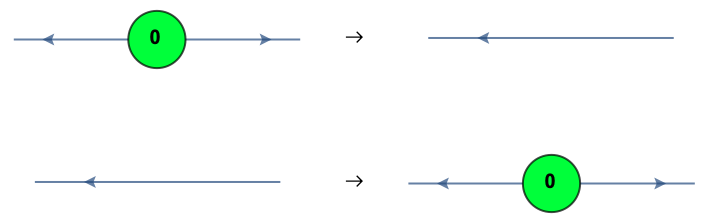}\hspace{0.1\textwidth}
\includegraphics[width=0.395\textwidth]{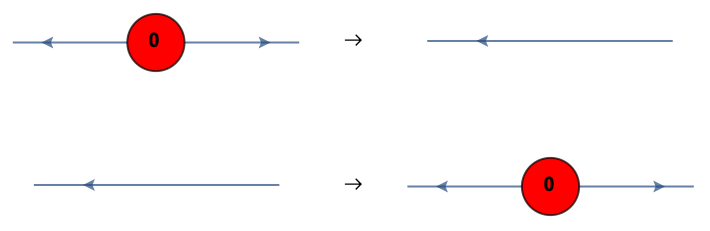}
\includegraphics[width=0.495\textwidth]{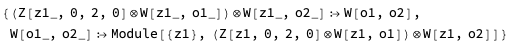}
\includegraphics[width=0.495\textwidth]{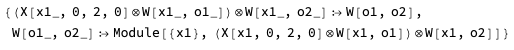}
\caption{The S2 rules, or the Z- and X-spider identity rules (the most general case), along with their respective operator forms. These rules correspond to the statement that the Bell state is always identical, irrespective of whether it is represented in the computational or Hadamard-transformed basis.}
\label{fig:Figure22}
\end{figure}

The S2 rules, or the Z- and X-spider identity rules (in the most general case), along with their associated operator forms, are shown in Figure \ref{fig:Figure22}. The algebraic interpretation of these rules is that the Bell state is always identical, irrespective of whether it is represented in the computational basis or the Hadamard-transformed basis, and therefore any phaseless Z- or X-spider can be replaced by the identity map. Category-theoretically, this rule is asserting that the Z- and X-spiders induce the same compact structure.

\begin{figure}[ht]
\centering
\includegraphics[width=0.395\textwidth]{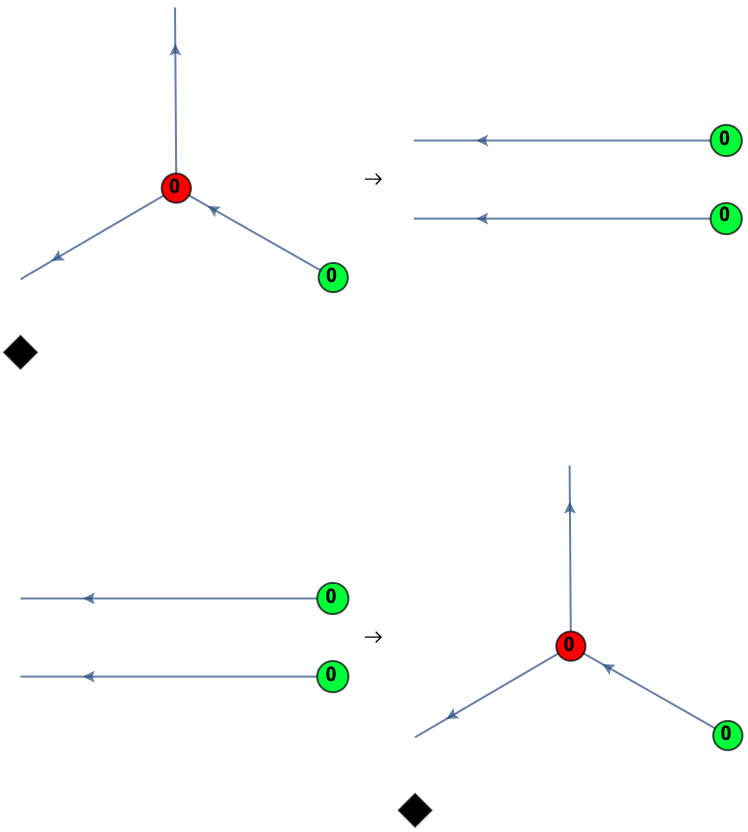}\hspace{0.1\textwidth}
\includegraphics[width=0.395\textwidth]{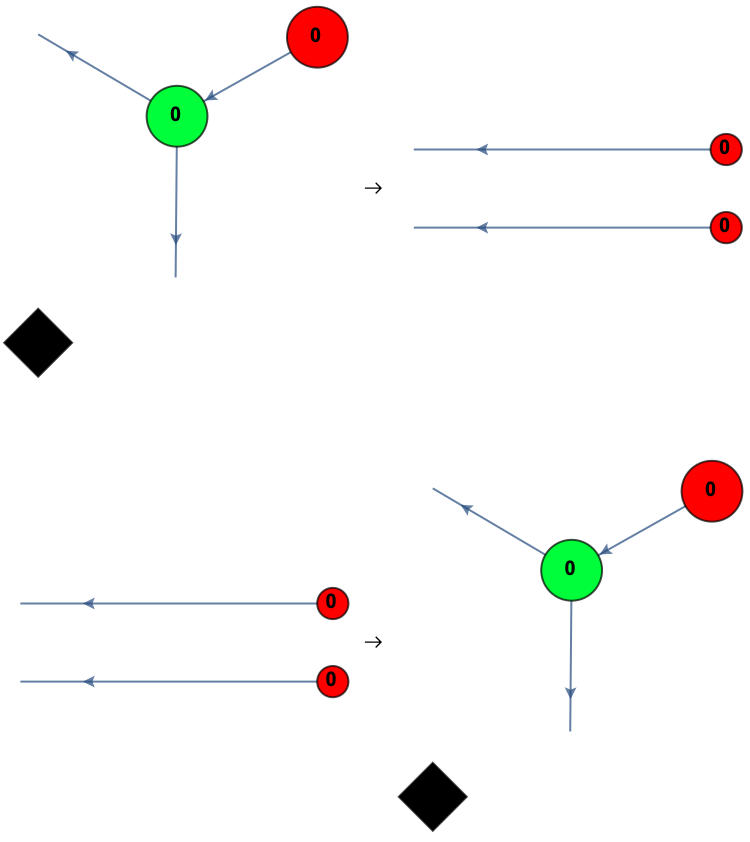}
\includegraphics[width=0.495\textwidth]{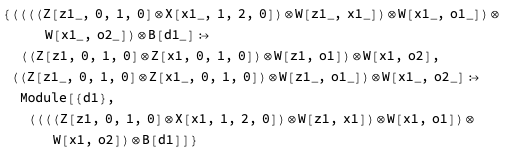}
\includegraphics[width=0.495\textwidth]{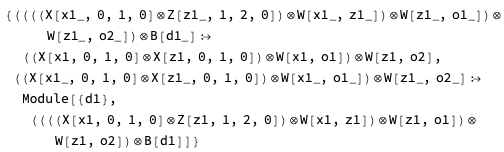}
\caption{The B1 rules, or the Z- and X-spider copy rules (the most general case), along with their respective operator forms. These rules correspond to the statement that Z- and X-spiders of arity 1 are proportional to Hadamard-transformed and computational basis states (namely ${\ket{+}}$ and ${\ket{0}}$), respectively, up to a multiplicative constant (a black diamond).}
\label{fig:Figure23}
\end{figure}

The B1 rules, or the Z- and X-spider copy rules (in the most general case), along with their associated operator forms, are shown in Figure \ref{fig:Figure23}. The algebraic interpretation of these rules is that a Z-spider of arity 1 is proportional to a Hadamard-transformed basis state (namely ${\ket{+}}$), and an X-spider of arity 1 is proportional to a computational basis state (namely ${\ket{0}}$), up to a multiplicative constant (i.e. up to a black diamond), and therefore arity 1 Z-spiders get ``copied through'' X-spiders, and vice versa.

\begin{figure}[ht]
\centering
\includegraphics[width=0.395\textwidth]{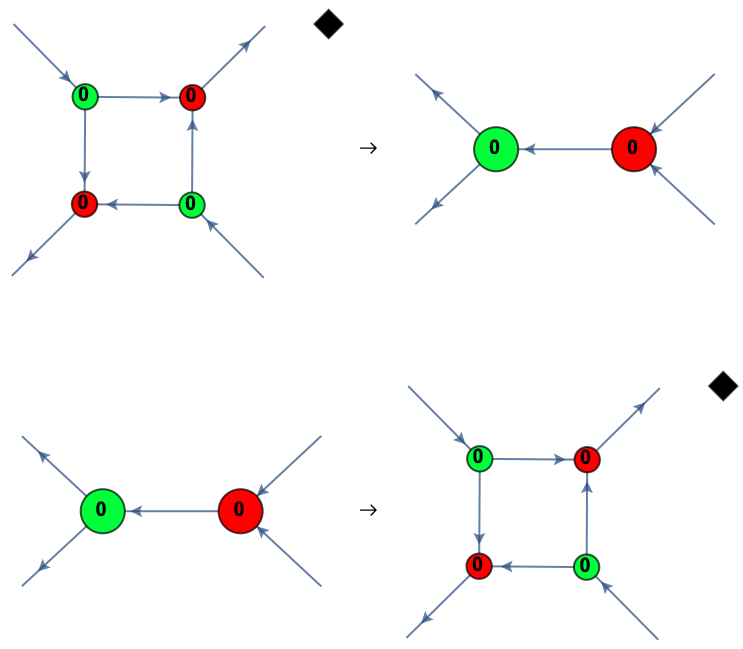}\hspace{0.1\textwidth}
\includegraphics[width=0.395\textwidth]{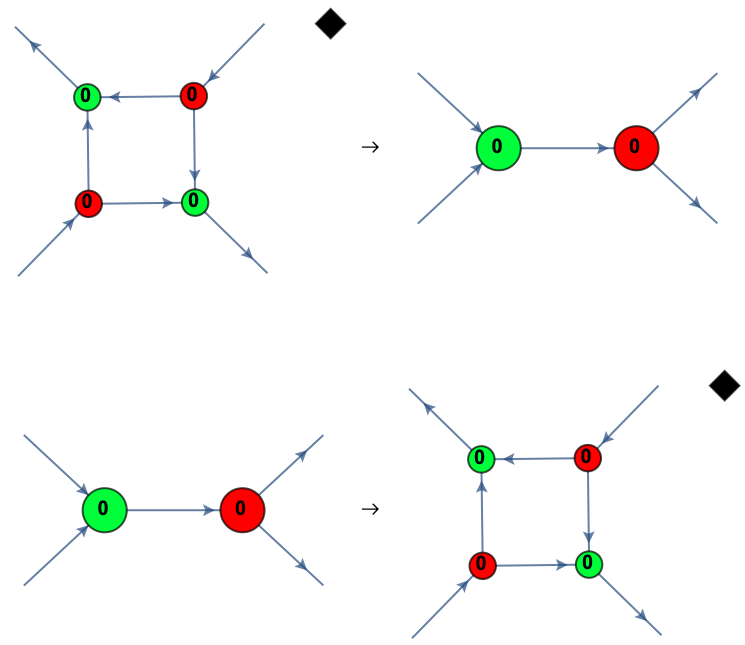}
\includegraphics[width=0.495\textwidth]{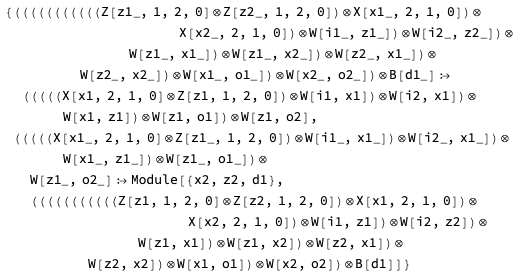}
\includegraphics[width=0.495\textwidth]{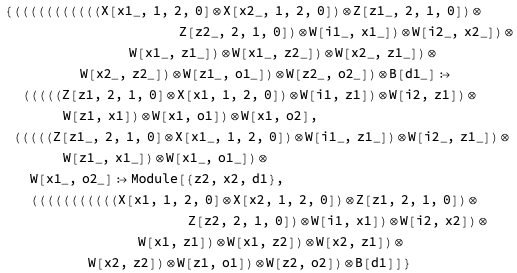}
\caption{The B2 rules, or the bialgebra simplification rules (the most general case), along with their respective operator forms. These rules correspond to the statement that the computational and Hadamard-transformed bases are strongly complementary.}
\label{fig:Figure24}
\end{figure}

The B2 rules, or the bialgebra simplification rules (in the most general case), along with their associated operator forms, are shown in Figure \ref{fig:Figure24}. The algebraic/categorical interpretation of these rules is that the computational and Hadamard-transformed bases are \textit{strongly complementary}, and therefore any 2-cycle of Z- and X-spiders must simplify. The concept of strong complementarity of observables in categorical quantum mechanics, as developed by Coecke, Duncan, Kissinger and Wang\cite{coecke4}, and its relationship to dagger special (commutative) Frobenius algebras\cite{coecke5}, is briefly outlined below.

\begin{definition}
A ``monoid object'', denoted ${\left( M, \mu, \eta \right)}$, in a monoidal category ${\left( \mathbf{C}, \otimes, I \right)}$, is an object $M$ in ${\mathrm{ob} \left( \mathbf{C} \right)}$, equipped with a pair of morphisms:

\begin{equation}
\mu : M \otimes M \to M, \qquad \text{ and } \eta : I \to M,
\end{equation}
in ${\hom \left( \mathbf{C} \right)}$, known as ``multiplication'' and ``unit'', respectively, such that the following pair of diagrams both commute:

\begin{equation}
\begin{tikzcd}
\left( M \otimes M \right) \otimes M \arrow[r, "\alpha"] \arrow[d, "\mu \otimes id"] & M \otimes \left( M \otimes M \right) \arrow[r, "id \otimes \mu"] & M \otimes M \arrow[d, "\mu"]\\
M \otimes M \arrow[rr, "\mu"] & & M
\end{tikzcd},
\end{equation}
and:

\begin{equation}
\begin{tikzcd}
I \otimes M \arrow[r, "\eta \otimes id"]  \arrow[dr, "\lambda"] & M \otimes M \arrow[d, "\mu"] & M \otimes I \arrow[l, "id \otimes \eta"] \arrow[dl, "\rho"]\\
& M &
\end{tikzcd},
\end{equation}
where, in the above, ${\alpha}$, ${\lambda}$ and ${\rho}$ denote the usual associator isomorphism and the left and right unitor isomorphisms of the monoidal category, respectively, and $I$ is the monoidal identity object.
\end{definition}

\begin{definition}
A ``comonoid object'', denoted ${\left( M, \delta, \epsilon \right)}$, in a monoidal category ${\left( \mathbf{C}, \otimes, I \right)}$, is the dual of a monoid object, i.e. it is a monoid in the opposite/dual category ${\mathbf{C}^{op}}$.
\end{definition}

\begin{definition}
A monoid object (or, dually, a comonoid object) ${\left( M, \mu, \eta \right)}$ in a symmetric monoidal category ${\left( \mathbf{C}, \otimes, I \right)}$ is ``commutative'' if:

\begin{equation}
\mu \circ \sigma = \mu,
\end{equation}
where ${\sigma}$ denotes the usual monoidal symmetry isomorphism.
\end{definition}

\begin{definition}
A ``Frobenius algebra'', denoted ${\left( A, \mu, \eta, \delta, \epsilon \right)}$, in a monoidal category ${\left( \mathbf{C}, \otimes, I \right)}$, is an object $A$ in ${\mathrm{ob} \left( \mathbf{C} \right)}$, equipped with two pairs of morphisms, namely:

\begin{equation}
\mu : A \otimes A \to A, \qquad \text{ and } \qquad \eta : I \to A,
\end{equation}
and:

\begin{equation}
\delta : A \to A \otimes A, \qquad \text{ and } \qquad \epsilon : A \to I,
\end{equation}
in ${\hom \left( \mathbf{C} \right)}$, such that ${\left( A, \mu, \eta \right)}$ forms a monoid object in ${\mathbf{C}}$ and ${\left( A, \delta, \epsilon \right)}$ dually forms a comonoid object in ${\mathbf{C}}$, in such a way that the following pair of diagrams both commute:

\begin{equation}
\begin{tikzcd}
A \otimes A \arrow[r, "\delta \otimes A"] \arrow[d, "\mu"] & A \otimes A \otimes A \arrow[d, "A \otimes \mu"]\\
A \arrow[r, "\delta"] & A \otimes A
\end{tikzcd},
\end{equation}
and:

\begin{equation}
\begin{tikzcd}
A \otimes A \arrow[r, "A \otimes \delta"] \arrow[d, "\mu"] & A \otimes A \otimes A \arrow[d, "\mu \otimes A"]\\
A \arrow[r, "\delta"] & A \otimes A
\end{tikzcd},
\end{equation}
which naturally generalizes the notion of a finite-dimensional unital associative algebra equipped with a bilinear form.
\end{definition}
Note that, in the above definition, we have assumed (without loss of generality) that the associated monoidal category is \textit{strict}:

\begin{definition}
A ``strict monoidal category'', denoted ${\left( \mathbf{C}, \otimes, I \right)}$, is a monoidal category in which the associator, left unitor and right unitor isomorphisms ${\alpha}$, ${\lambda}$ and ${\rho}$ are all identity isomorphisms.
\end{definition}

\begin{definition}
Within a dagger symmetric monoidal category ${\left( \mathbf{C}, \otimes, I, \dagger \right)}$, a ``dagger special (commutative) Frobenius algebra'', denoted ${\mathcal{O}_{\color{green}{\circ}}}$, is a (commutative) Frobenius algebra $A$ defined in the usual way by the pairs of morphisms:

\begin{equation}
\mu_{\color{green}{\circ}} : A \otimes A \to A, \qquad \text{ and } \qquad \eta_{\color{green}{\circ}} : I \to A,
\end{equation}
and:

\begin{equation}
\delta_{\color{green}{\circ}} : A \to A \otimes A, \qquad \text{ and } \qquad \epsilon_{\color{green}{\circ}} : A \to I,
\end{equation}
in ${\hom \left( \mathbf{C} \right)}$, such that one has the following compatibility conditions with the dagger structure:

\begin{equation}
\delta_{\color{green}{\circ}} = \left( \mu_{\color{green}{\circ}} \right)^{\dagger}, \qquad \text{ and } \qquad \epsilon_{\color{green}{\circ}} = \left( \eta_{\color{green}{\circ}} \right)^{\dagger},
\end{equation}
and moreover, the diagrammatic equality shown in Figure \ref{fig:Figure53} holds.

\begin{figure}[ht]
\centering
\includegraphics[width=0.695\textwidth]{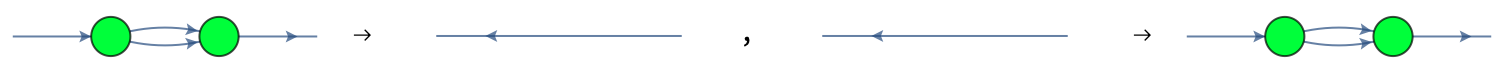}
\includegraphics[width=0.695\textwidth]{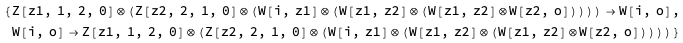}
\caption{The defining diagrammatic equality for a dagger special (commutative) Frobenius algebra.}
\label{fig:Figure53}
\end{figure}
\end{definition}
The connection between dagger special commutative Frobenius algebras and quantum observables lies in the fact that, within the standard mathematical formalism of quantum mechanics, every non-degenerate observable forms an orthonormal basis of eigenstates, and in the category ${\mathbf{FdHilb}}$ of finite-dimensional Hilbert spaces, the orthonormal bases are in bijective correspondence with the dagger special commutative Frobenius algebras\cite{coecke6}\cite{coecke7}. For this reason, we shall henceforth adopt the standard convention of referring to dagger special commutative Frobenius algebras as ``observable structures''.

\begin{definition}
A pair of observable structures ${\left( \mathcal{O}_{\color{green}{\circ}}, \mathcal{O}_{\color{red}{\circ}} \right)}$, acting on a common object $A$, are ``complementary'' if and only if the diagrammatic equality shown in Figure \ref{fig:Figure54} holds.

\begin{figure}[ht]
\centering
\includegraphics[width=0.695\textwidth]{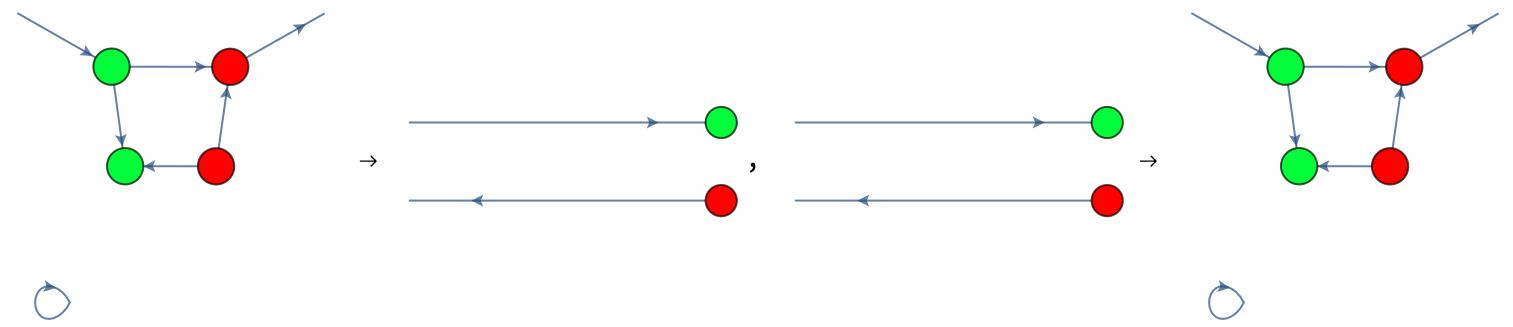}
\includegraphics[width=0.695\textwidth]{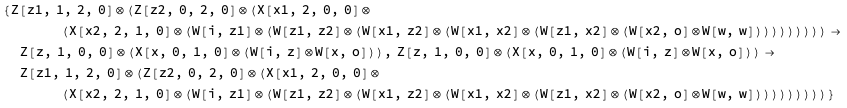}
\caption{The defining diagrammatic equality for the complementarity of observable structures.}
\label{fig:Figure54}
\end{figure}
\end{definition}
In the category ${\mathbf{FdHilb}}$, this reduces to the standard quantum mechanical notion of complementarity between observables.

\begin{definition}
A pair of observable structures ${\left( \mathcal{O}_{\color{green}{\circ}}, \mathcal{O}_{\color{red}{\circ}} \right)}$, acting on a common object $A$, are ``coherent'' if and only if the diagrammatic equalities shown in Figures \ref{fig:Figure55}, \ref{fig:Figure56} and \ref{fig:Figure57} all hold.

\begin{figure}[ht]
\centering
\includegraphics[width=0.695\textwidth]{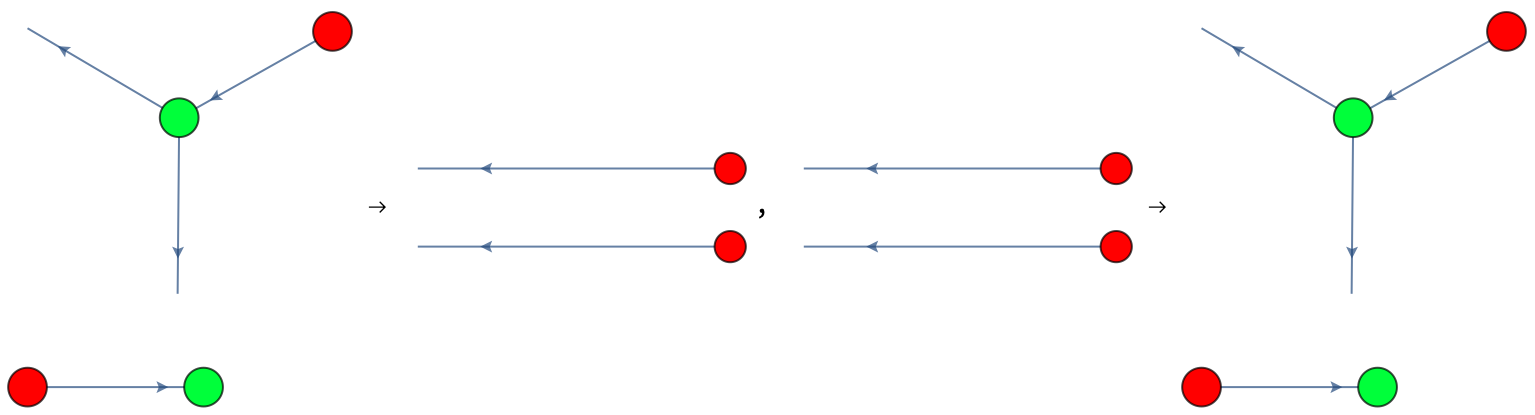}
\includegraphics[width=0.695\textwidth]{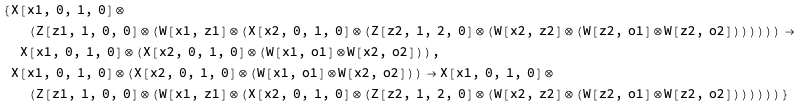}
\caption{The first defining diagrammatic equality for the coherence of observable structures.}
\label{fig:Figure55}
\end{figure}

\begin{figure}[ht]
\centering
\includegraphics[width=0.695\textwidth]{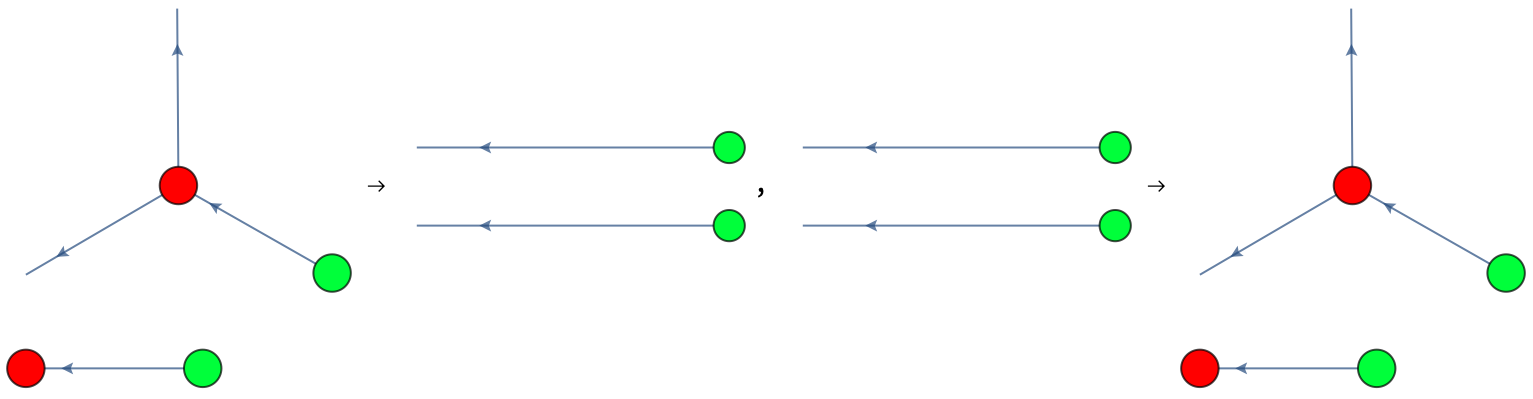}
\includegraphics[width=0.695\textwidth]{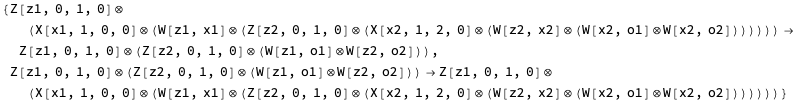}
\caption{The second defining diagrammatic equality for the coherence of observable structures.}
\label{fig:Figure56}
\end{figure}

\begin{figure}[ht]
\centering
\includegraphics[width=0.695\textwidth]{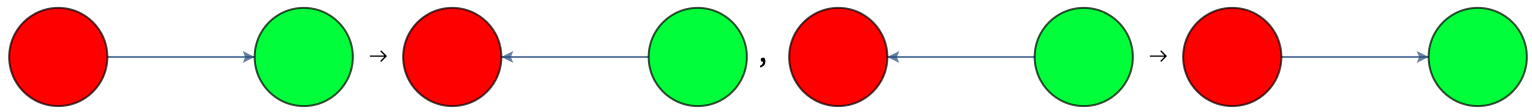}
\includegraphics[width=0.695\textwidth]{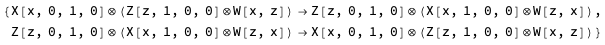}
\caption{The third defining diagrammatic equality for the coherence of observable structures.}
\label{fig:Figure57}
\end{figure}
\end{definition}

\begin{definition}
A pair of observable structures ${\left( \mathcal{O}_{\color{green}{\circ}}, \mathcal{O}_{\color{red}{\circ}} \right)}$, acting on a common object $A$, are ``strongly complementary'' if and only if they are coherent and the diagrammatic equality shown in Figure \ref{fig:Figure58} holds.

\begin{figure}[ht]
\centering
\includegraphics[width=0.695\textwidth]{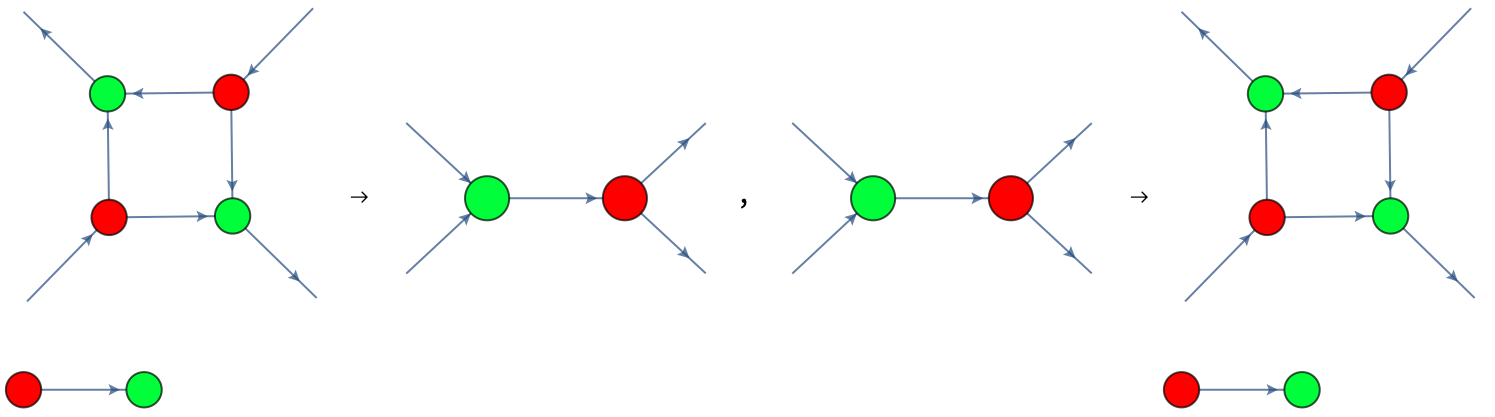}
\includegraphics[width=0.695\textwidth]{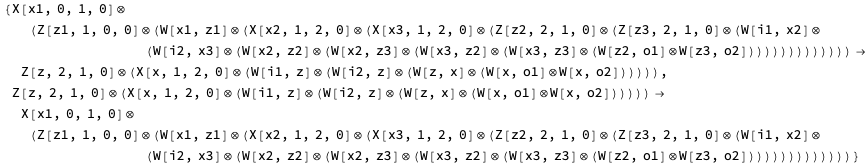}
\caption{The defining diagrammatic equality for the strong complementarity of observable structures.}
\label{fig:Figure58}
\end{figure}
\end{definition}
From this last equality, we see that the bialgebra simplification rules follow immediately from the fact that the computational and Hadamard-transformed orthonormal bases (when interpreted as observable structures/dagger special commutative Frobenius algebras) are strongly complementary.

\begin{figure}[ht]
\centering
\includegraphics[width=0.395\textwidth]{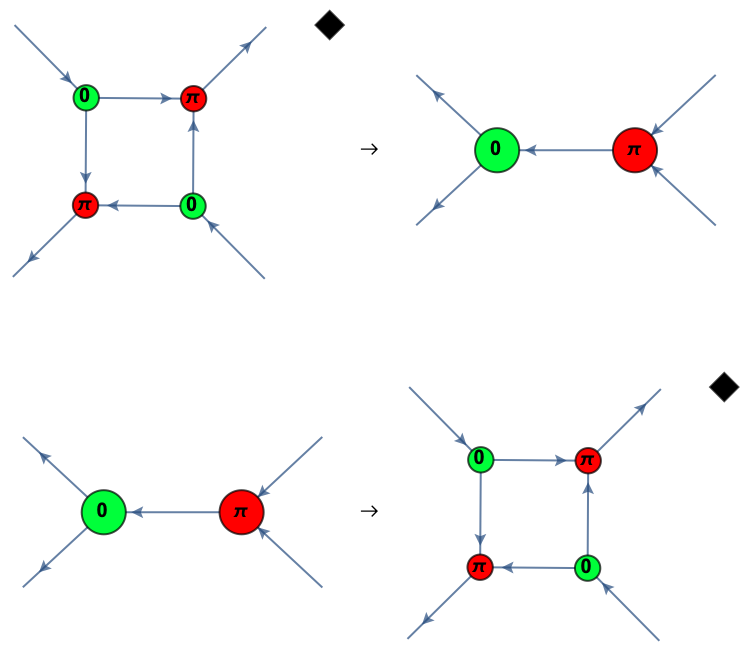}\hspace{0.1\textwidth}
\includegraphics[width=0.395\textwidth]{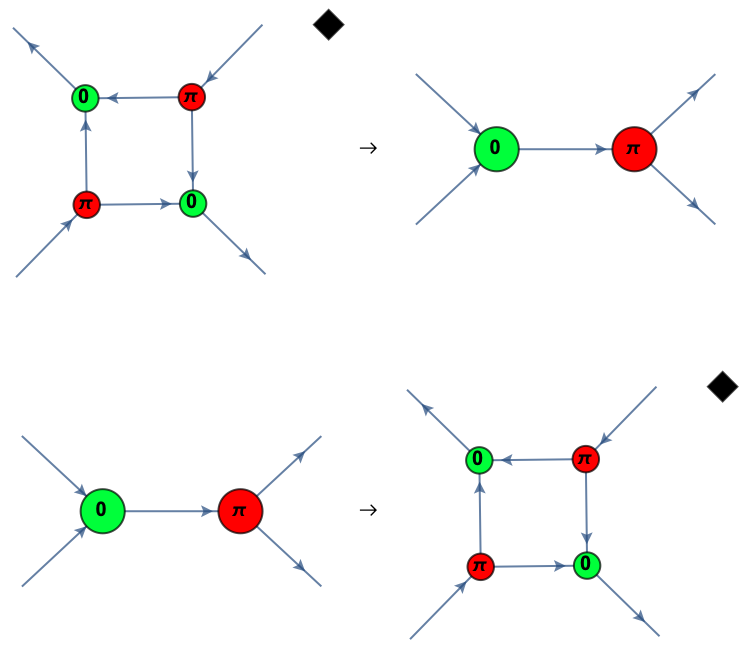}
\includegraphics[width=0.495\textwidth]{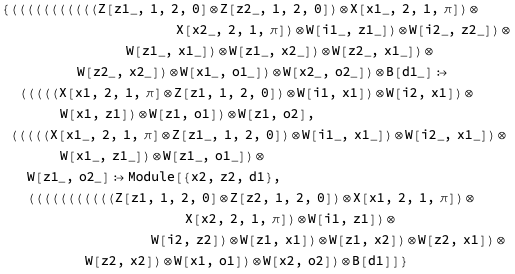}
\includegraphics[width=0.495\textwidth]{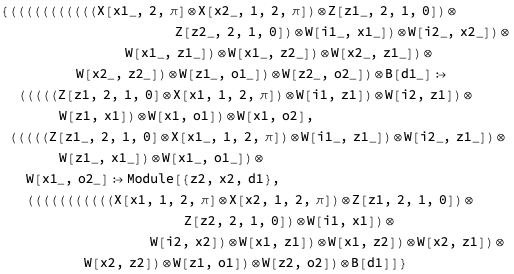}
\caption{A variant of the B2 rules, or the bialgebra simplification rules (the most general case), along with their respective operator forms. These rules modify the standard B2 rules for the case in which the X-spider phases are equal to ${\pi}$, with the Z-spider phases being equal to zero.}
\label{fig:Figure60}
\end{figure}

\begin{figure}[ht]
\centering
\includegraphics[width=0.395\textwidth]{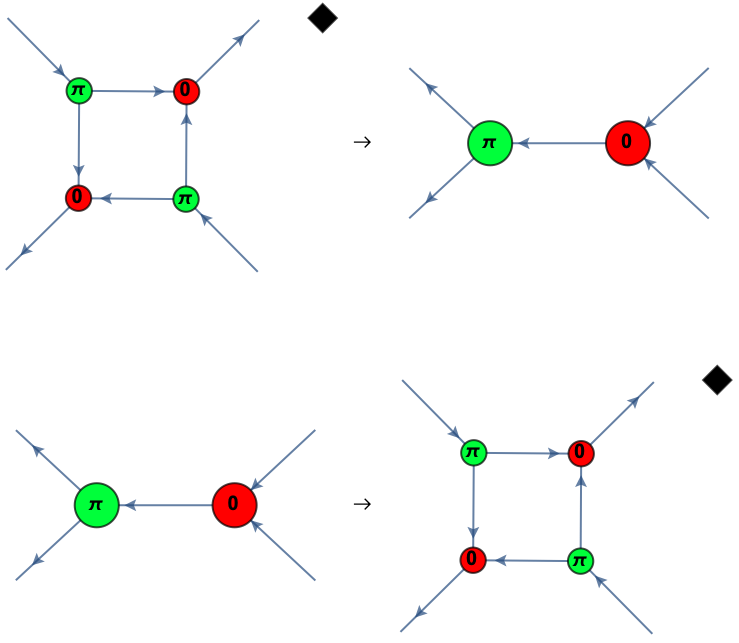}
\includegraphics[width=0.395\textwidth]{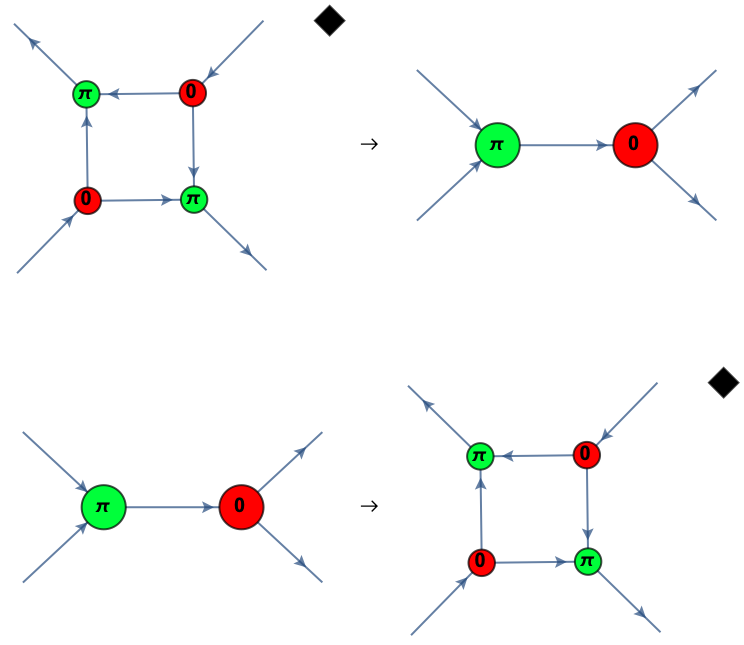}
\includegraphics[width=0.495\textwidth]{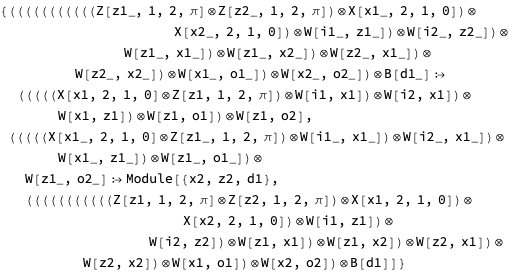}
\includegraphics[width=0.495\textwidth]{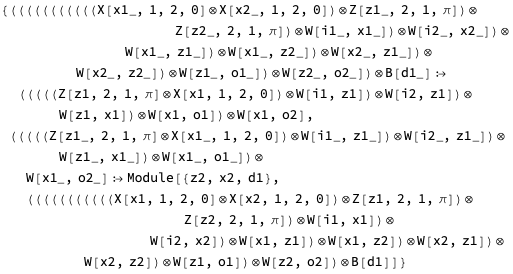}
\caption{A variant of the B2 rules, or the bialgebra simplification rules (the most general case), along with their respective operator forms. These rules modify the standard B2 rules for the case in which the Z-spider phases are equal to ${\pi}$, with the X-spider phases being equal to zero.}
\label{fig:Figure61}
\end{figure}

It is worth noting that, in addition to the case presented above in which both the Z- and X-spiders are strictly phaseless, there exist two other cases in which the bialgebra rule holds, namely whenever either of the Z-spider or X-spider phases are equal to ${\pi}$, with the other phases being equal to zero. This can be easily verified algebraically using the following explicit representations of the spiders:

\begin{equation}
\mathbf{H} = \frac{1}{\sqrt{2}} \begin{bmatrix}
1 & 1\\
1 & -1
\end{bmatrix}, \qquad \mathbf{Z} \left( \alpha \right) = \begin{bmatrix}
1 & 0\\
0 & 0\\
0 & 0\\
0 & e^{i \alpha}
\end{bmatrix}, \qquad \mathbf{X} \left( \beta \right) = \mathbf{H}   \begin{bmatrix}
1 & 0 & 0 & 0\\
0 & 0 & 0 & e^{i \beta} 
\end{bmatrix} \left( \mathbf{H} \otimes \mathbf{H} \right),
\end{equation}
where spiders $\mathbf{Z} \left( \alpha \right)$ and $\mathbf{X} \left( \beta \right)$ are both expressed in the computational basis. \\ 
Furthermore, swapping wires are represented using:

\begin{equation}
\mathbf{SWAP} = \begin{bmatrix}
1 & 0 & 0 & 0\\
0 & 0 & 1 & 0\\
0 & 1 & 0 & 0\\
0 & 0 & 0 & 1
\end{bmatrix},
\end{equation}
which, when combined together, indeed verifies:

\begin{equation}
\sqrt{2} \left( \mathbf{Z} \left( \alpha \right) \otimes \mathbf{Z} \left( \alpha \right) \right) \left( \mathbf{I}_2 \otimes \mathbf{SWAP} \otimes \mathbf{I}_2 \right) \left( \mathbf{X} \left( \beta \right) \otimes \mathbf{X} \left( \beta \right) \right) = \mathbf{X} \left( \beta \right) \mathbf{Z} \left( \alpha \right),
\end{equation}
for ${\left( \alpha = \pi, \beta = 0 \right)}$ and ${\left( \alpha = 0, \beta = \pi \right)}$, where ${\mathbf{I}_2}$ denotes the ${2 \times 2}$ identity matrix. Consequently, the variants of the B2 rules in which the X-spider and Z-spider phases are equal to ${\pi}$, with all other phases equal to zero, are shown in Figures \ref{fig:Figure60} and \ref{fig:Figure61}, respectively.

\begin{figure}[ht]
\centering
\includegraphics[width=0.395\textwidth]{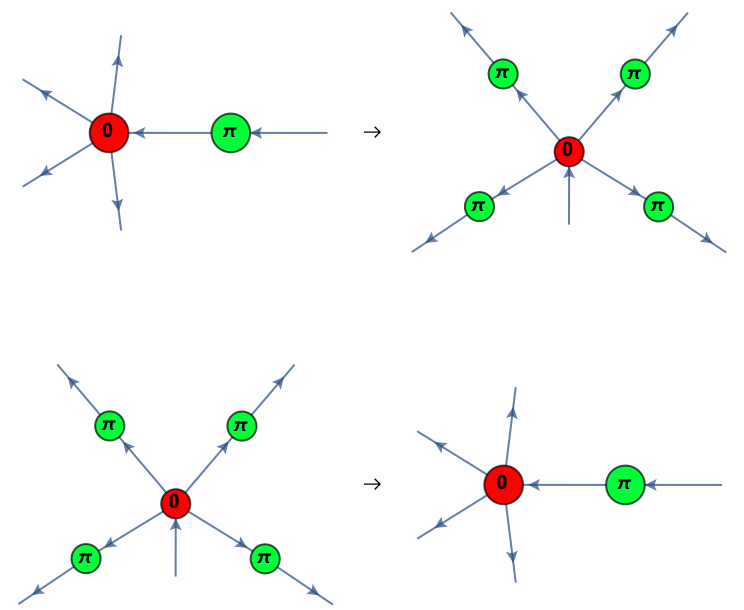}\hspace{0.1\textwidth}
\includegraphics[width=0.395\textwidth]{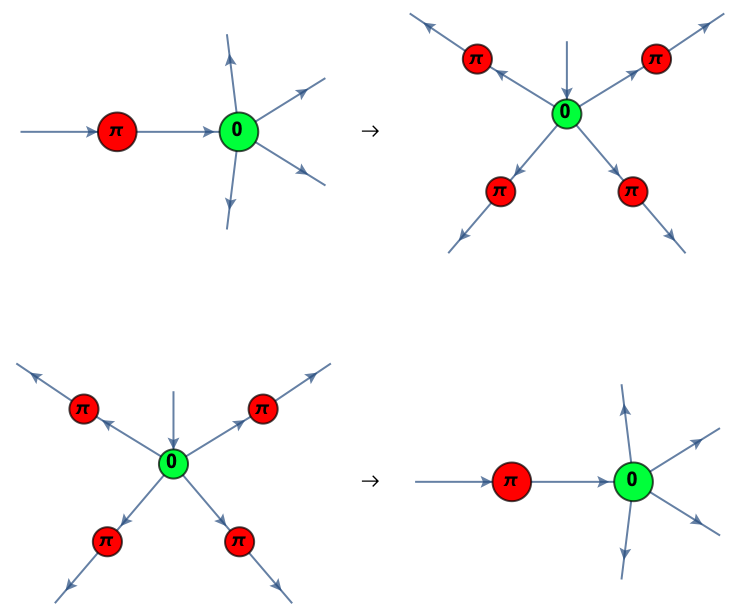}
\includegraphics[width=0.495\textwidth]{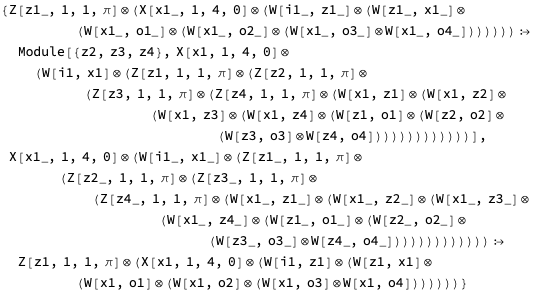}
\includegraphics[width=0.495\textwidth]{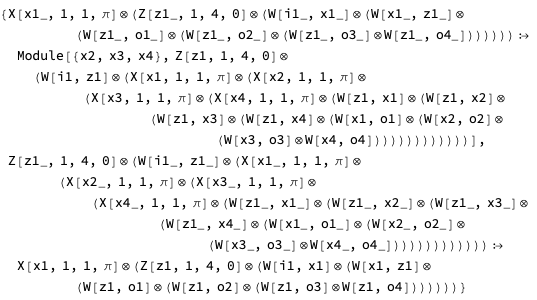}
\caption{The K1 rules, or the Z- and X-spider ${\pi}$-copy rules (for the case in which the X- and Z-spiders have output arities equal to 4, respectively), along with their associated operator forms. These rules correspond to the statement that a Hadamard NOT gate (i.e. an X-spider with an input arity of 1, an output arity of 1 and a phase of ${\pi}$) is a function map of the Hadamard-transformed basis, and that a computational NOT gate (i.e. a Z-spider with an input arity of 1, an output arity of 1 and a phase of ${\pi}$) is a function map of the computational basis, respectively.}
\label{fig:Figure25}
\end{figure}

The K1 rules, or the Z- and X-spider ${\pi}$-copy rules (for the case in which the X- and Z-spiders have output arities equal to 4, respectively), along with their associated operator forms, are shown in Figure \ref{fig:Figure25}. The algebraic interpretation of these rules is that a Hadamard NOT gate (i.e. an X-spider with an input arity of 1, an output arity of 1 and a phase of ${\pi}$) is a function map of the Hadamard-transformed basis (i.e. it maps Hadamard-transformed basis states to Hadamard-transformed basis states), and therefore it copies through a Z-spider, whereas a computational NOT gate (i.e. a Z-spider with an input arity of 1, an output arity of 1 and a phase of ${\pi}$) is a function map of the computational basis (i.e. it maps computational basis states to computational basis states), and therefore it copies through an X-spider.

\begin{figure}[ht]
\centering
\includegraphics[width=0.395\textwidth]{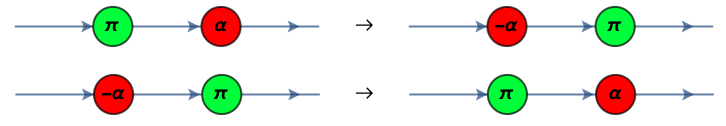}\hspace{0.1\textwidth}
\includegraphics[width=0.395\textwidth]{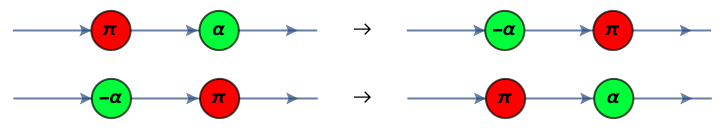}
\includegraphics[width=0.495\textwidth]{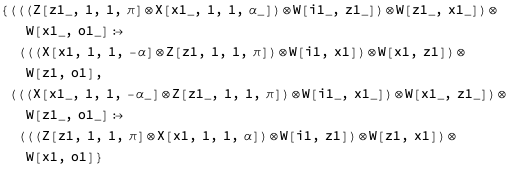}
\includegraphics[width=0.495\textwidth]{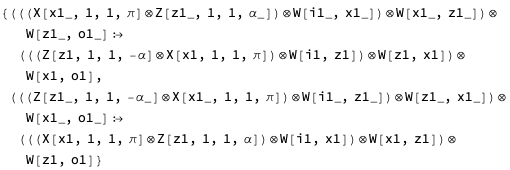}
\caption{The K2 rules, or the Z- and X-spider phase flip rules (the most general case), along with their associated operator forms. These rules correspond to the statement that when a Hadamard NOT gate is commuted through a Z-rotation gate, or when a computational NOT gate is commuted through an X-rotation gate, the rotation of the gate flips.}
\label{fig:Figure26}
\end{figure}

The K2 rules, or the Z- and X-spider phase flip rules (in the most general case), along with their associated operator forms, are shown in Figure \ref{fig:Figure26}. The algebraic interpretation of these rules is that, when a Hadamard NOT gate (i.e. an X-spider with an input arity of 1, an output arity of 1 and a phase of ${\pi}$) is commuted through a Z-rotation gate, or when a computational NOT gate (i.e. a Z-spider with an input arity of 1, an output arity of 1 and a phase of ${\pi}$) is commuted through an X-rotation gate, the rotation of the latter gate flips, and therefore the phase of the latter spider is negated.

\begin{figure}[ht]
\centering
\includegraphics[width=0.395\textwidth]{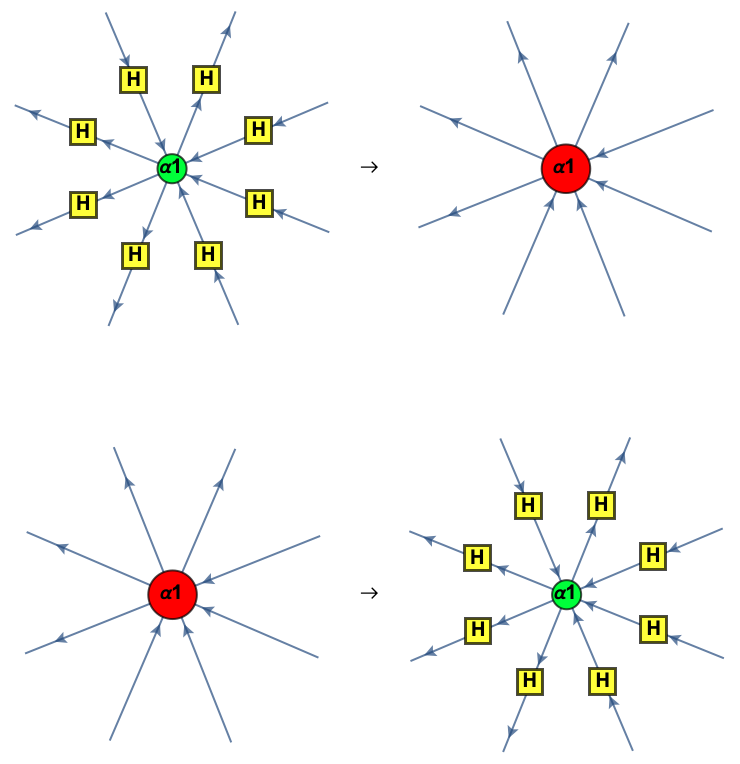}\hspace{0.1\textwidth}
\includegraphics[width=0.395\textwidth]{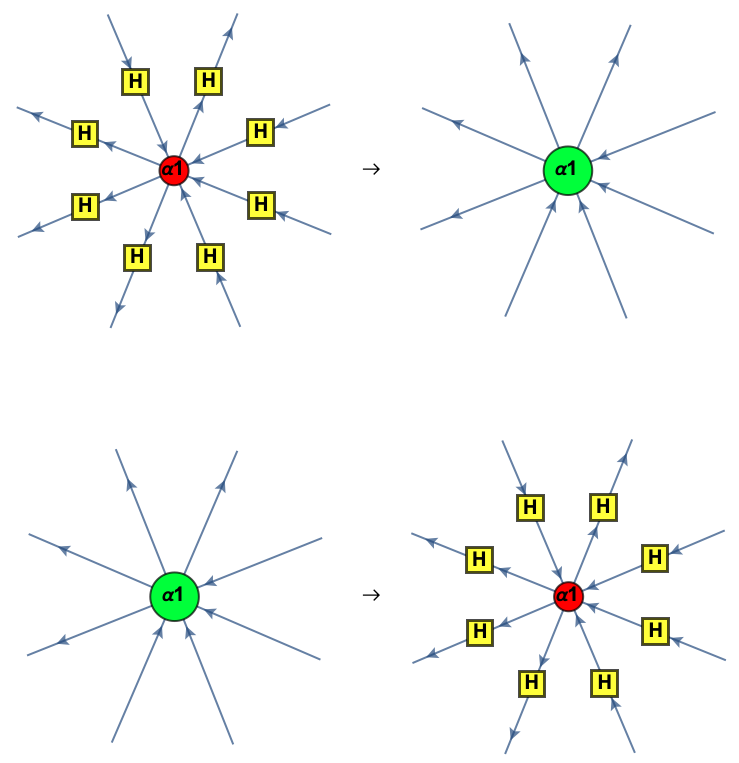}
\includegraphics[width=0.495\textwidth]{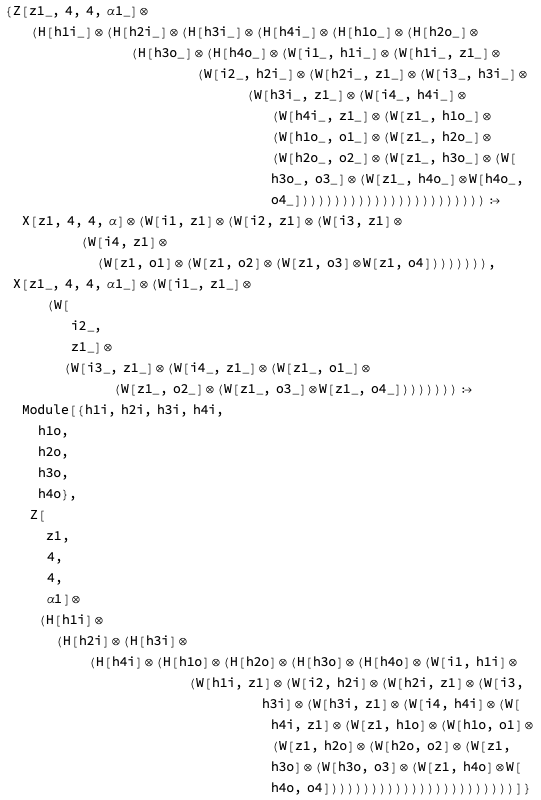}
\includegraphics[width=0.495\textwidth]{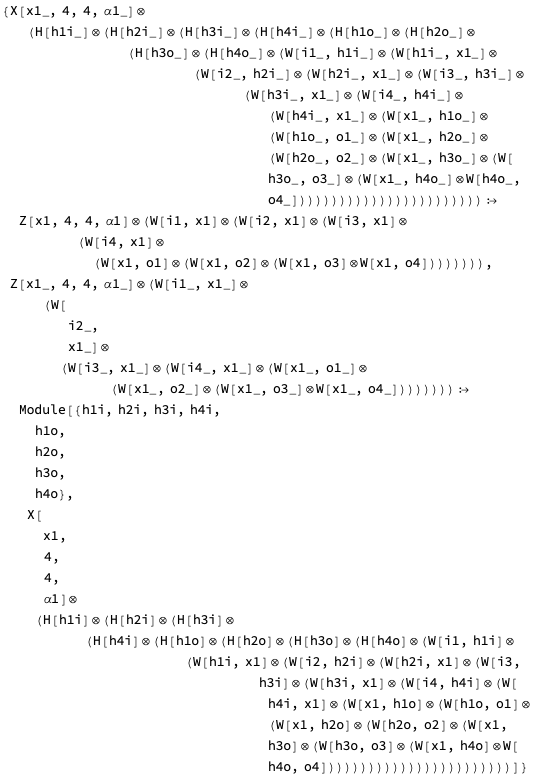}
\caption{The C rules, or the Z- and X-spider color change rules (for the case in which the Z- and X-spiders have input and output arities equal to 4, respectively), along with their associated operator forms. These rules correspond to the statement that Hadamard gates map from the computational basis to the Hadamard-transformed basis, and back again.}
\label{fig:Figure27}
\end{figure}

The C rules, or the Z- and X-spider color change rules (for the case in which the Z- and X-spiders have input and output arities equal to 4, respectively), along with their associated operator forms, are shown in Figure \ref{fig:Figure27}. The algebraic interpretation of these rules is that a Hadamard gate maps from the computational basis to the Hadamard-transformed basis, and back again, and therefore the color of the spider is inverted.

\begin{figure}[ht]
\centering
\includegraphics[width=0.395\textwidth]{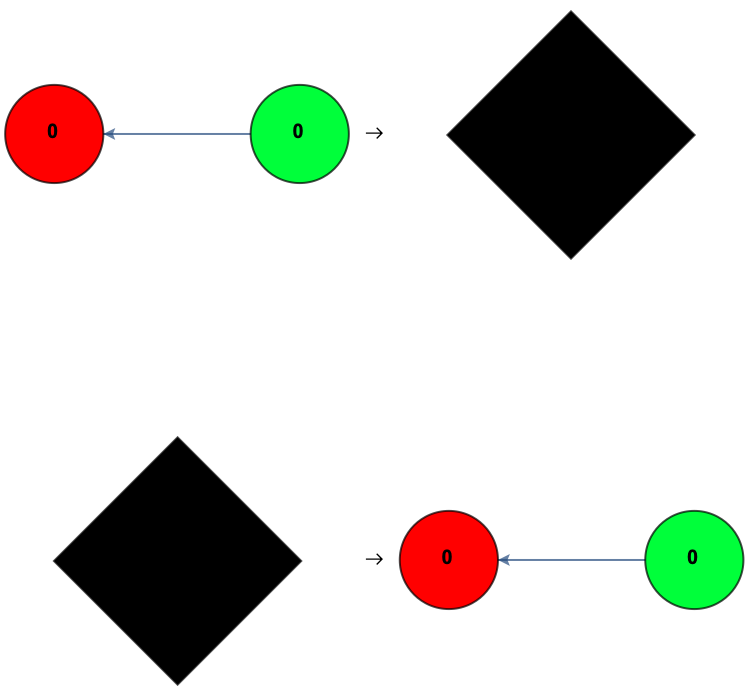}\hspace{0.1\textwidth}
\includegraphics[width=0.395\textwidth]{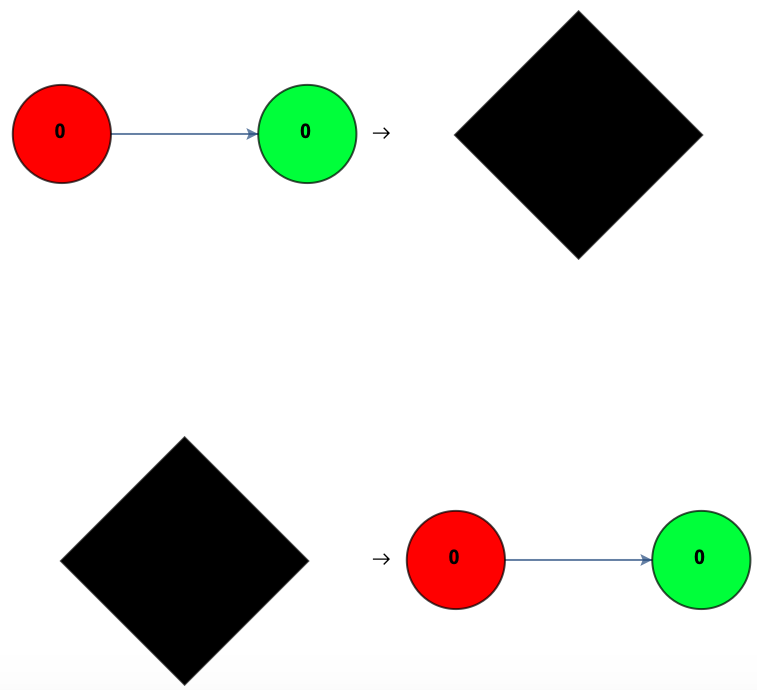}
\includegraphics[width=0.495\textwidth]{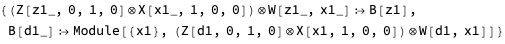}
\includegraphics[width=0.495\textwidth]{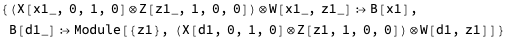}
\caption{The D1 rules, or the spider cancellation rules (the most general case), along with their associated operator forms. These rules correspond to the statement that Z- and X-spiders can cancel to yield a single scalar factor.}
\label{fig:Figure28}
\end{figure}

\begin{figure}[ht]
\centering
\includegraphics[width=0.595\textwidth]{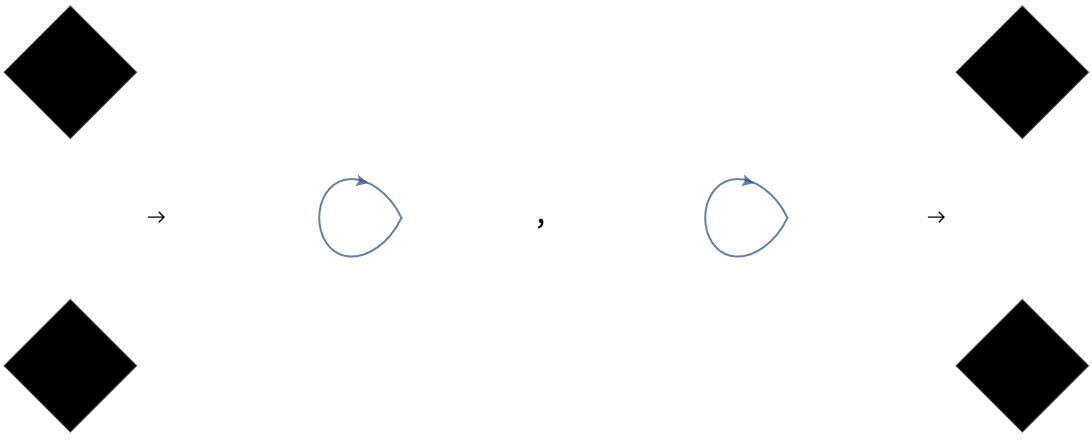}
\includegraphics[width=0.495\textwidth]{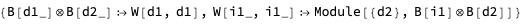}
\caption{The D2 rules, or the scalar multiplication rules (the most general case), along with their associated operator forms. These rules correspond to the statement that two ${\sqrt{D}}$ scalar factors multiply together to yield $D$.}
\label{fig:Figure29}
\end{figure}

The D1 and D2 rules, or the spider cancellation and scalar multiplication rules (in the most general case), along with their associated operator forms, are shown in Figures \ref{fig:Figure28} and \ref{fig:Figure29}. The algebraic interpretation of the spider cancellation rules is that Z- and X-spiders can mutually cancel, so as to yield a single scalar factor, therefore reducing down to single black diamond, whilst the interpretation of the scalar multiplication rules is that two scalar factors of ${\sqrt{D}}$ multiply together to yield $D$, and therefore two black diamonds can combine to yield a single loop of wire (representing an overall multiplicative scalar factor of $D$).

\begin{figure}[ht]
\centering
\includegraphics[width=0.495\textwidth]{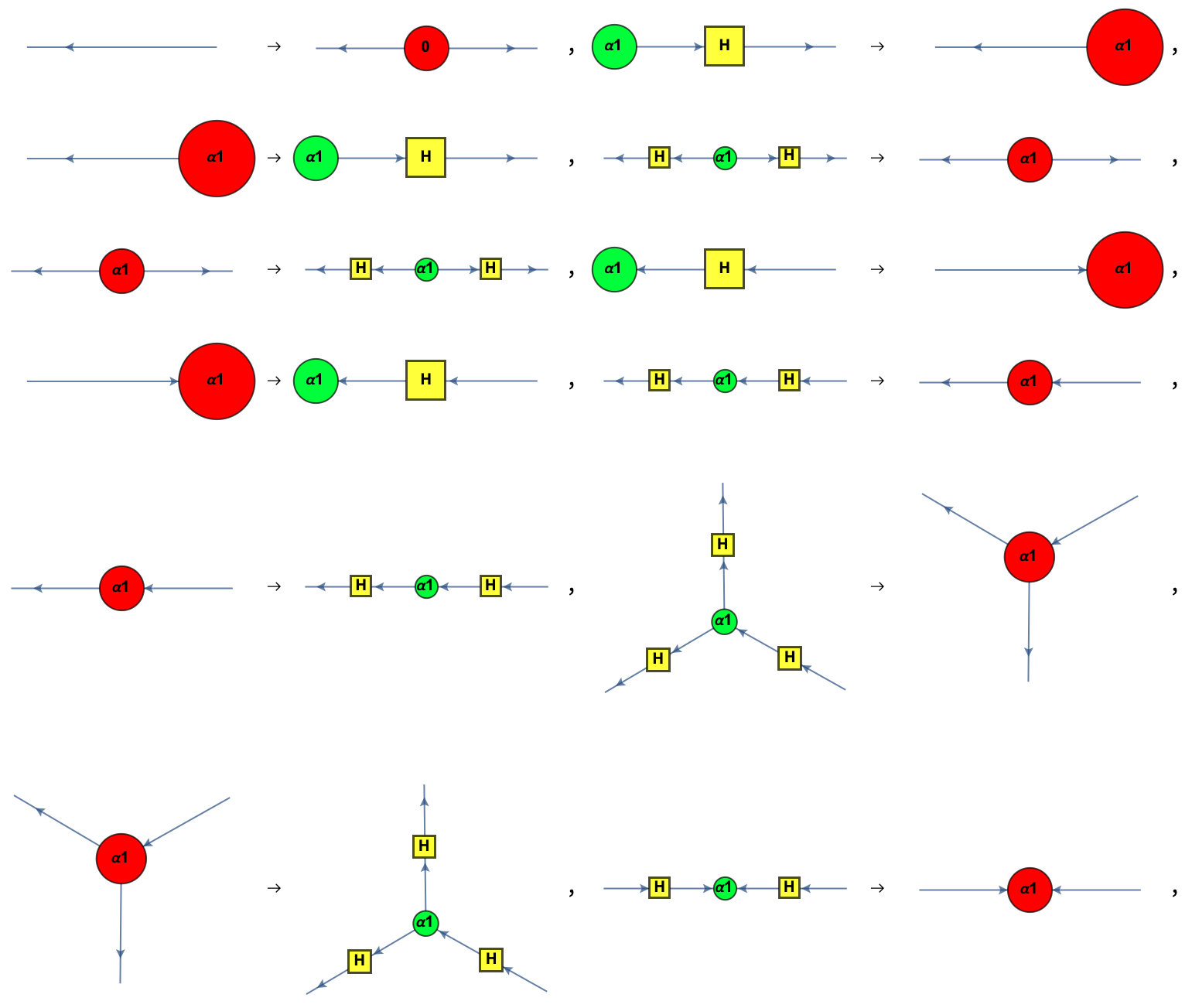}
\includegraphics[width=0.495\textwidth]{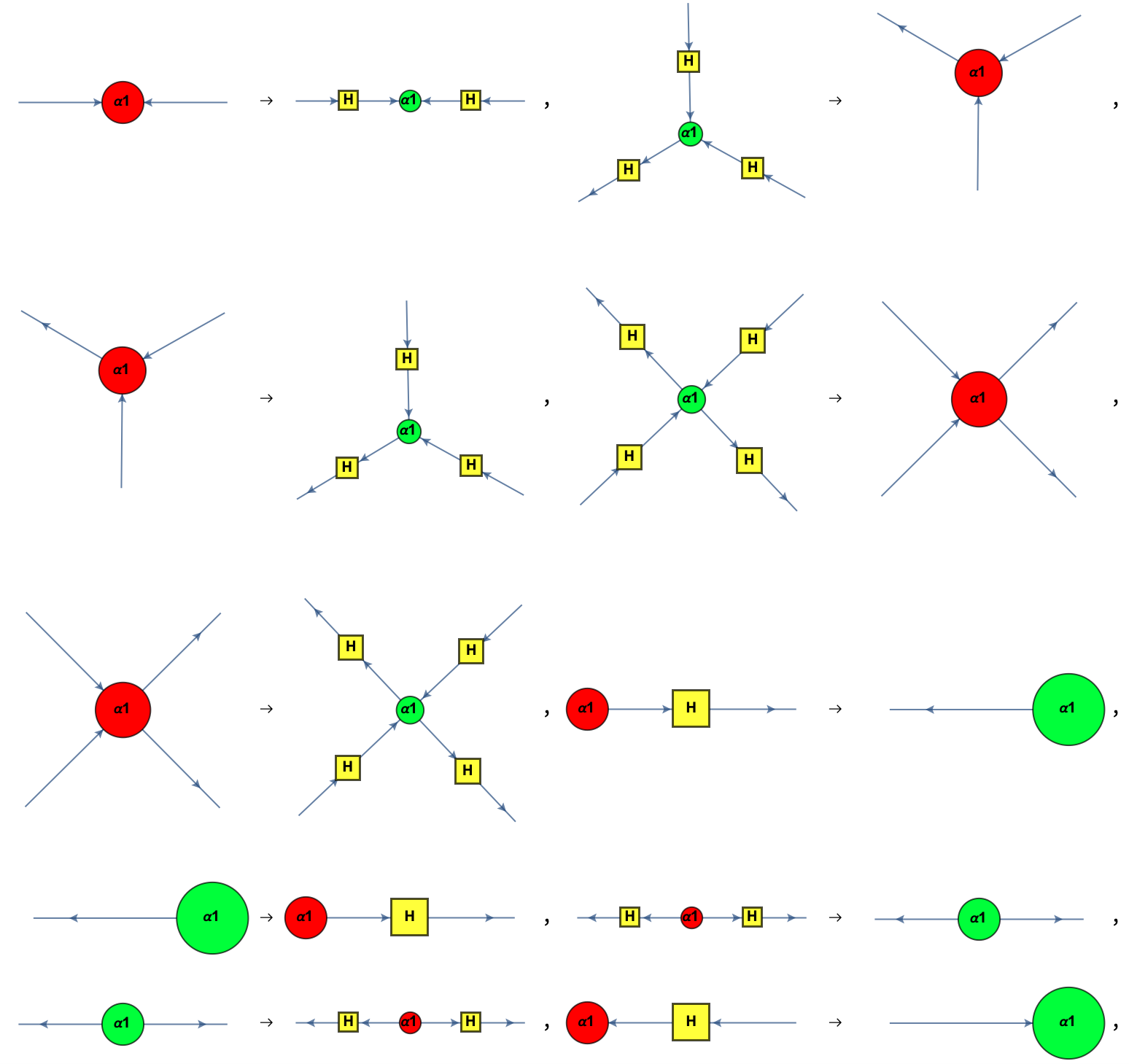}
\caption{Samples of the complete rule enumeration for the ZX-calculus, up to input arity 2 and output arity 2 across all generators.}
\label{fig:Figure30}
\end{figure}

Finally, one has a set of non-diagrammatic axioms for the composition operator ${\otimes}$, ensuring that the monoidal structure induced by ${\otimes}$ satisfies the requisite associativity and symmetry properties (and therefore that the corresponding monoidal category is equipped with the requisite associator and symmetry isomorphisms):

\begin{equation}
\left\lbrace a\_ \otimes \left( b\_ \otimes c\_ \right) :> \left( a \otimes b \right) \otimes c, \left( a\_ \otimes b\_ \right) \otimes c\_ :> a \otimes \left( b \otimes c \right), a\_ \otimes b\_ :> b \otimes a \right\rbrace.
\end{equation}
With the multiway operator rules (and rule schemas) thus defined, one can consequently enact a systematic enumeration of all possible rules in the ZX-calculus up to a given arity, as shown in Figure \ref{fig:Figure30} for the case of enumeration up to input arity 2 and output arity 2 across all generators.

\clearpage

\subsection{Multiway Evolution Graphs of ZX-Diagrams}

We now proceed to demonstrate how multiway systems provide a natural embedding space for the collection of all possible deduction chains in the ZX-calculus, by constructing explicit examples of multiway evolution graphs for the rewriting of certain ZX-diagrams. Much like the paths discussed previously for the case of more abstract rewriting systems, a single path in a multiway evolution graph here represents a proof of equivalence between two ZX-diagrams. The multiway evolution graph as a whole can therefore be interpreted as representing the space of all possible proofs in the ZX-calculus, starting from a given initial diagram. 

\begin{figure}[ht]
\centering
\includegraphics[width=0.595\textwidth]{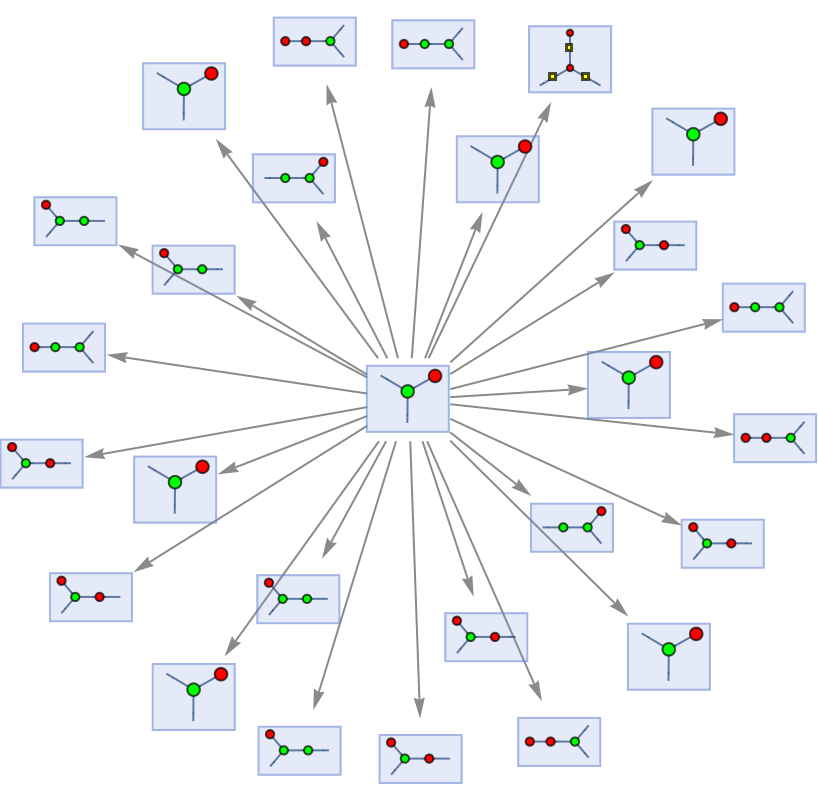}
\caption{The multiway evolution graph corresponding to the first step in the non-deterministic evolution of the ZX-calculus multiway operator system, starting from a simple two-spider initial diagram.}
\label{fig:Figure31}
\end{figure}

\begin{figure}[ht]
\centering
\includegraphics[width=0.895\textwidth]{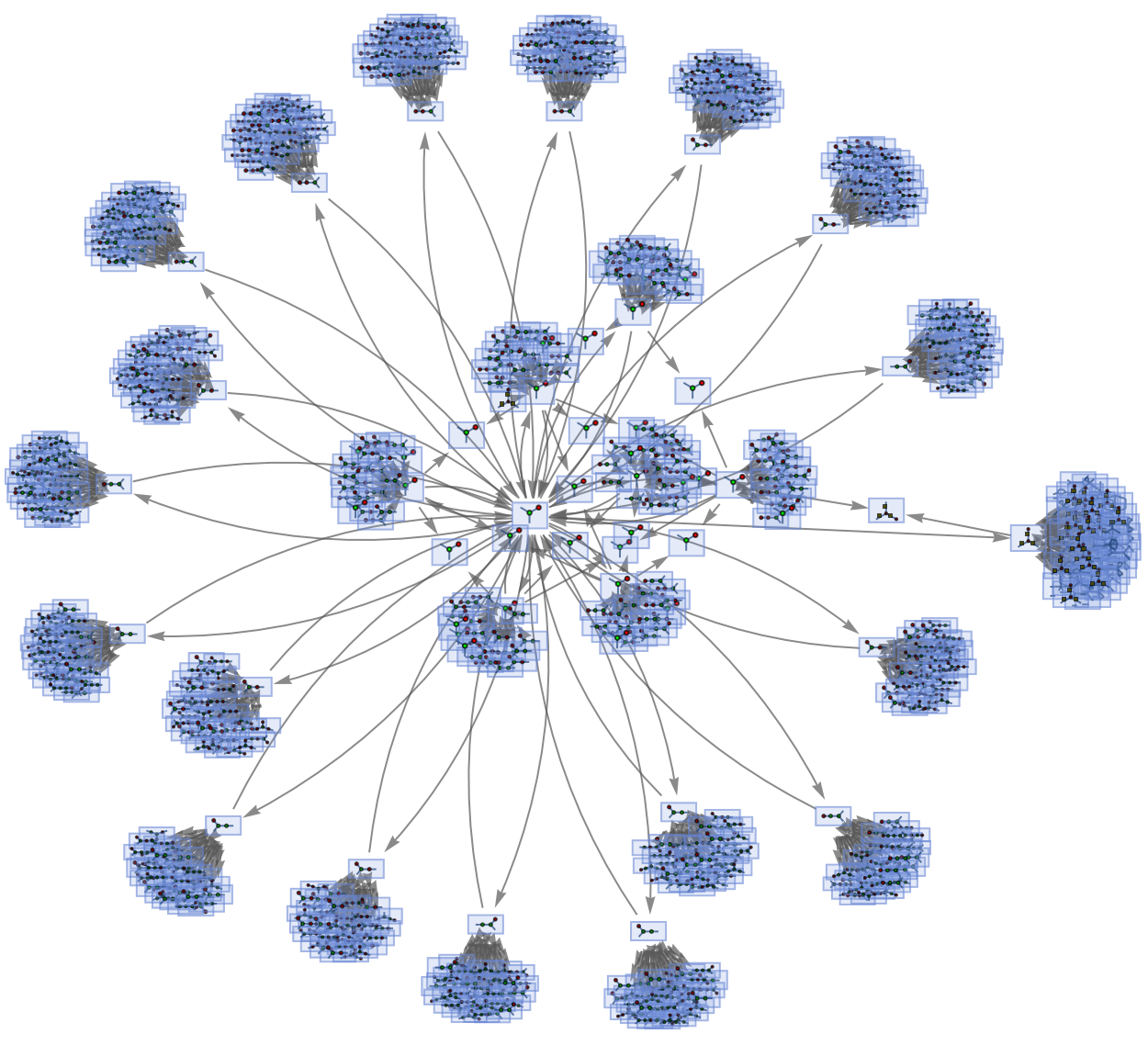}
\caption{The multiway states graph (i.e. a variant of a multiway evolution graph in which cycles are permitted) corresponding to the first two steps in the non-deterministic evolution of the ZX-calculus multiway operator system, starting from a simple two-spider initial diagram.}
\label{fig:Figure32}
\end{figure}

\begin{figure}[ht]
\centering
\includegraphics[width=0.595\textwidth]{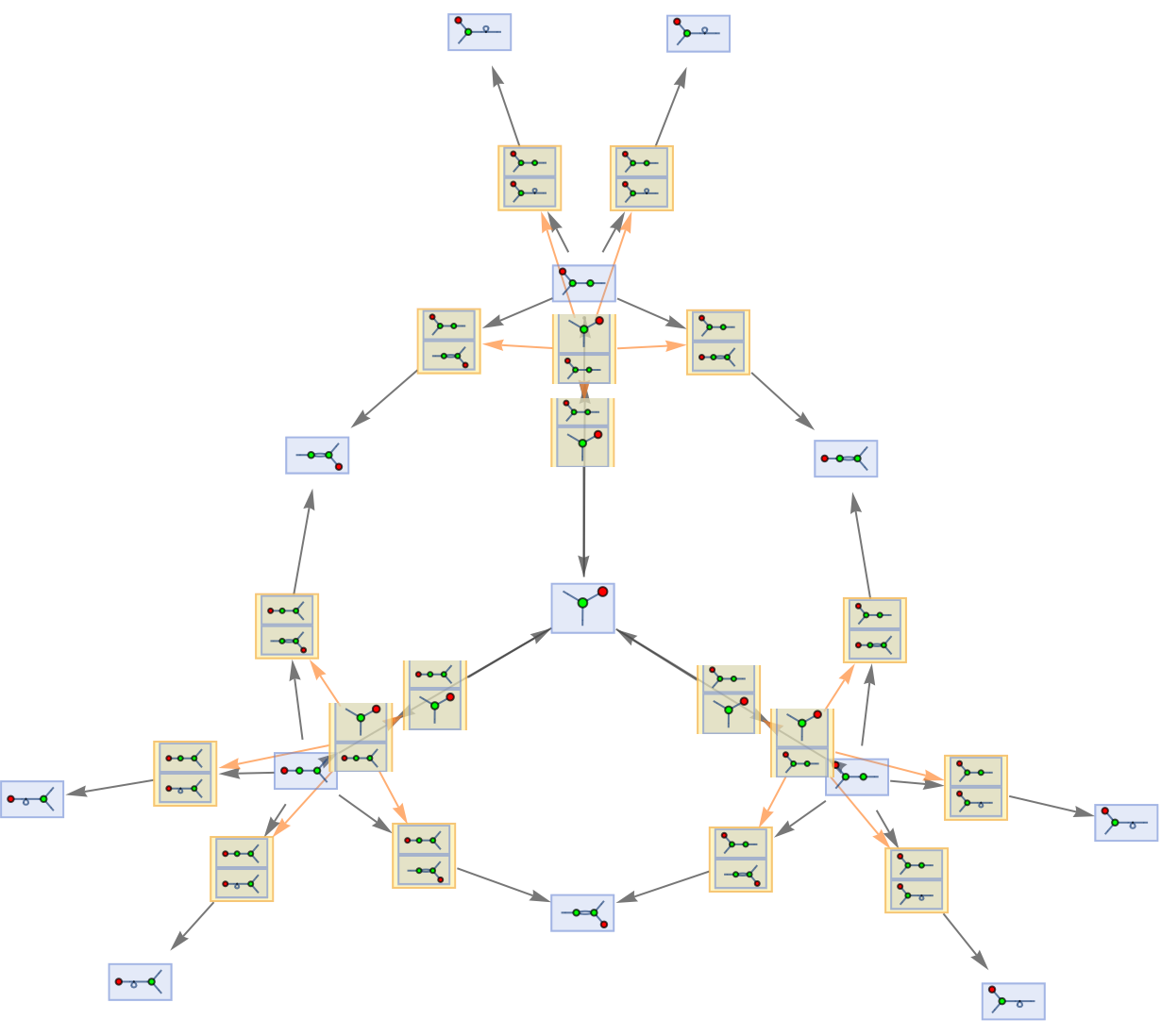}
\caption{A subgraph of the multiway evolution causal graph (with evolution edges shown in gray, and causal edges shown in orange) corresponding to the first two steps in the non-determinsitic evolution of the ZX-calculus multiway operator system, starting from a simple two-spider initial diagram (and here restricted to use only Z-spider identity rules of a particular arity).}
\label{fig:Figure62}
\end{figure}

\begin{figure}[ht]
\centering
\includegraphics[width=0.595\textwidth]{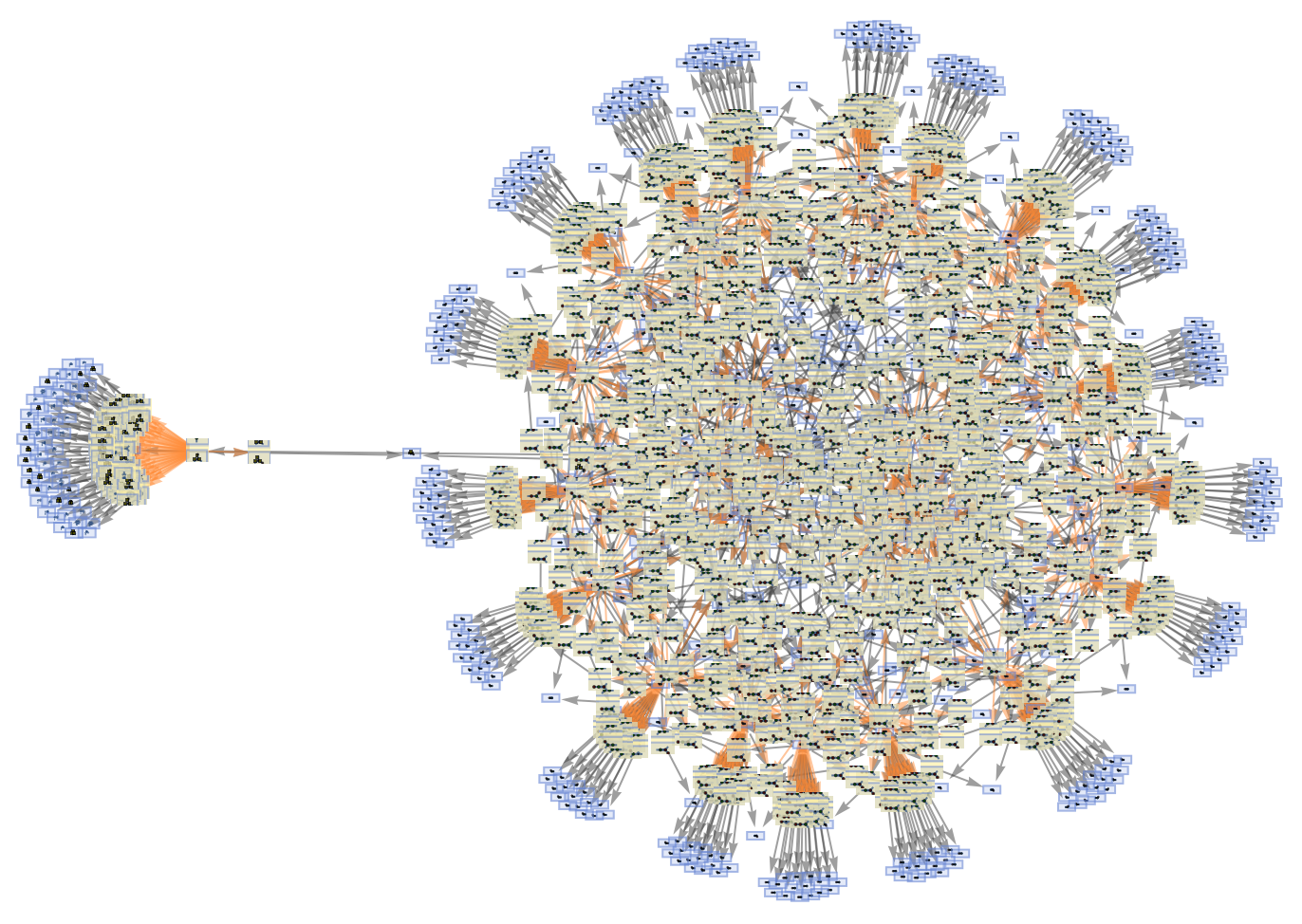}
\caption{The multiway evolution causal graph (with evolution edges shown in gray, and causal edges shown in orange) corresponding to the first two steps in the non-deterministic evolution of the ZX-calculus multiway operator system, starting from a simple two-spider initial diagram (with no restrictions).}
\label{fig:Figure33}
\end{figure}

Let us illustrate this more concretely with the help of an initial expression of the form:

\begin{equation}
X \left[ x_1, 0, 1, 0 \right] \otimes \left( Z \left[ z_1, 1, 2, 0 \right] \otimes \left( W \left[ x_1, z_1 \right] \otimes \left( W \left[ z_1, o_1 \right] \otimes W \left[ z_1, o_2 \right] \right) \right) \right),
\end{equation}
i.e. a simple two-spider initial diagram. Applying all possible diagrammatic rewritings allowable by the ZX-calculus, one obtains, after one and two steps of evolution respectively, the multiway evolution graphs shown in Figures \ref{fig:Figure31} and \ref{fig:Figure32} (although strictly speaking, the second figure shows a multiway \textit{states} graph, i.e. a variant of a multiway evolution graph in which cycles are permitted). Moreover, a subgraph of the multiway evolution causal graph (with evolution edges shown in gray, and causal edges shown in orange) for the multiway operator system after two evolution steps is shown in Figure \ref{fig:Figure62}, with the subgraph induced by applying only Z-spider identity rules of a particular arity. The full evolution causal graph, obtained by applying all possible rules without restrictions, is shown in Figure \ref{fig:Figure33}, exhibiting a highly non-trivial pattern of causal relationships between the diagrammatic rewriting events.

Recasting the diagrammatic rewritings of the ZX-calculus into the framework of multiway operator systems confers a multitude of both conceptual and practical advantages over the more traditional formulation in terms of abstract rewriting systems. One immediate conceptual benefit of the multiway formalism is that, by being a very concrete instantiation of an otherwise rather abstract metamathematical construct (namely the ``space of all possible proofs'' that one can construct using the ZX rules), the multiway evolution graph allows one to reformulate questions regarding the soundness, consistency, completeness, etc., of the ZX rules in terms of simple combinatorial properties of the associated evolution graph. For instance, to take a highly idealized case, one could imagine a simple proof calculus in which each proposition is simply a string of $0$s and $1$s, with the proposition ${010}$ constituting the negation of proposition ${101}$, etc., and with the inference rules of the calculus being given by elementary string substitution rules, such as ${\left\lbrace 1 \to 01, 0 \to 10 \right\rbrace}$\cite{wolfram}.

\begin{figure}[ht]
\centering
\includegraphics[width=0.395\textwidth]{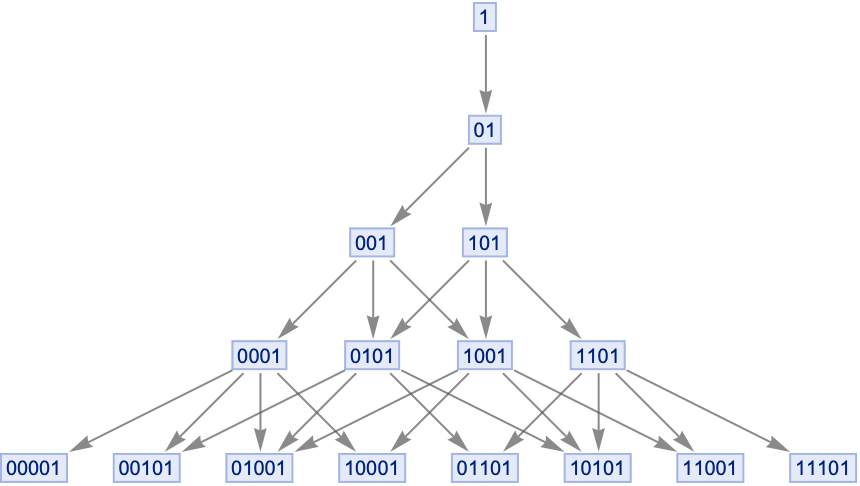}\hspace{0.05\textwidth}
\includegraphics[width=0.545\textwidth]{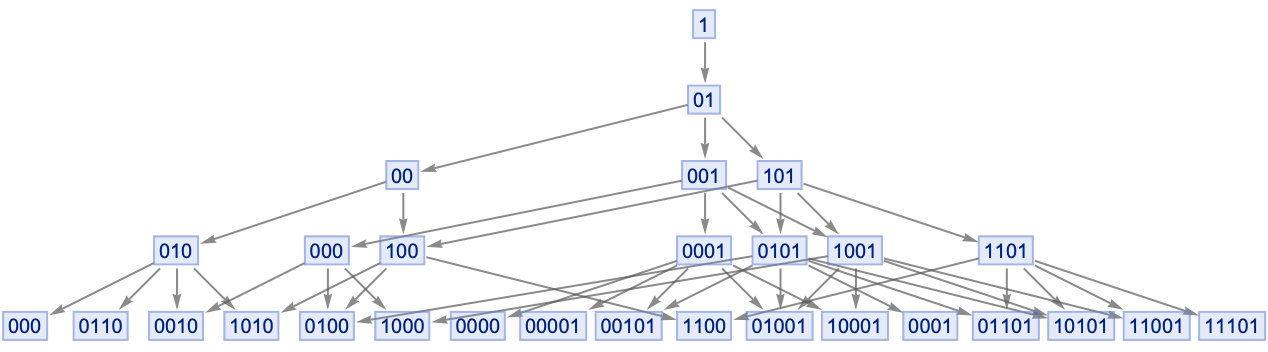}
\includegraphics[width=0.695\textwidth]{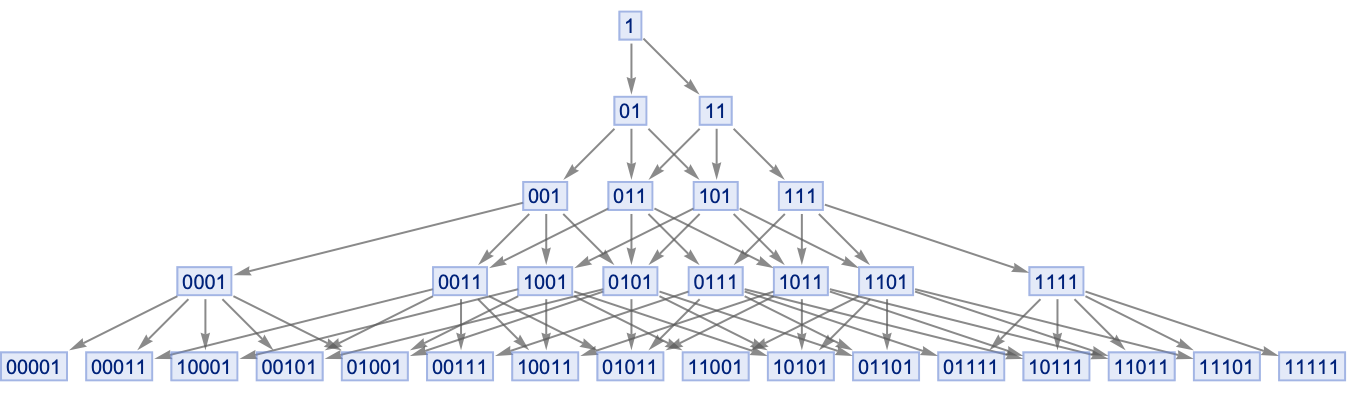}
\caption{Multiway evolution graphs for three elementary string substitution systems, namely ${\left\lbrace 1 \to 01, 0 \to 10 \right\rbrace}$, ${\left\lbrace 1 \to 01, 0 \to 10, 01 \to 00 \right\rbrace}$ and ${\left\lbrace 1 \to 01, 0 \to 10, 1 \to 11 \right\rbrace}$, respectively. The first multiway system is incomplete, in the sense that there exist certain strings, such as ${111}$, for which the system generates neither the string nor its negation ${000}$. The second multiway system is inconsistent, in the sense that there exist certain strings, such as ${010}$, for which the system generates both the string and its negation ${101}$. Finally, the third multiway system is both complete and consistent - for every possible string, either it or its negation is generated, but never both.}
\label{fig:Figure63}
\end{figure}

Then, as shown in Figure \ref{fig:Figure63} one can very straightforwardly infer the completeness and consistency properties of the calculus by simply inspecting the state vertices that can be reached in the multiway evolution graph starting from the single vertex $1$. The first multiway system, corresponding to the string substitution system ${\left\lbrace 1 \to 0,1 0 \to 10 \right\rbrace}$, represents an incomplete inference system, since there exist certain strings, such as ${111}$, for which the multiway system neither generates the string nor its negation ${000}$. The second multiway system, corresponding to the string substitution system ${\left\lbrace 1 \to 01, 0 \to 10, 01 \to 00 \right\rbrace}$, represents an inconsistent inference system, since there exist certain strings such as ${010}$, for which the multiway system generates both the string and its negation ${101}$. Finally, the third multiway system, corresponding to the string substitution system ${\left\lbrace 1 \to 01, 0 \to 10, 1 \to 11 \right\rbrace}$, represents an inference system that is both complete and consistent, since for every possible string, the multiway system either generates it or its negation, but never both. Thus, the multiway formalism potentially provides one with a general procedure for proving certain metamathematical properties of the ZX-calculus in terms of combinatorial properties of the space of possible proofs.

On a much more pragmatic level, the fact that multiway operator systems come naturally equipped with a causal partial order on updating events also potentially paves the way for new parallelization methods in the automated rewriting of ZX-diagrams. If one has a particularly large ZX-diagram and wishes to parallelize its rewriting by first splitting it into subdiagrams, each of which gets rewritten by a parallel computational thread, and then reassembling it at the end, then in order to ensure the consistency of the reassembly process, one must first ensure that the parallel rewritings applied to the individual subdiagrams do not conflict with each other (i.e. two updating events occurring in different subdiagrams must not have any non-trivial causal relationship). The natural structure of the multiway evolution causal graphs reduces the algorithmic and computational complexity of making such determinations quite considerably, as compared to the ordinary abstract rewriting approach. Moreover, the ability to infer information regarding the causal partial order on updating events can also be used to assist with the process of ``lemma selection'' in automated theorem-proving over ZX-diagrams, since when selecting which particular ZX-diagram equivalences to select as lemmas, one is attempting to select the lemmas which will exhibit the greatest effect on shortening subsequent proofs (which are, in turn, the lemmas which exert the greatest causal influence on future rewritings). We intend to explore many of these topics in greater detail in a forthcoming article regarding the application of multiway systems to automated reasoning over ZX-diagrams.

\clearpage

\section{Connection to Quantum Foundations: The Monoidal Structure of Multiway Systems}
\label{sec:section3}

\subsection{Representation of Quantum Processes Using Multiway and Branchial Graphs}

\begin{figure}[ht]
\centering
\includegraphics[width=0.895\textwidth]{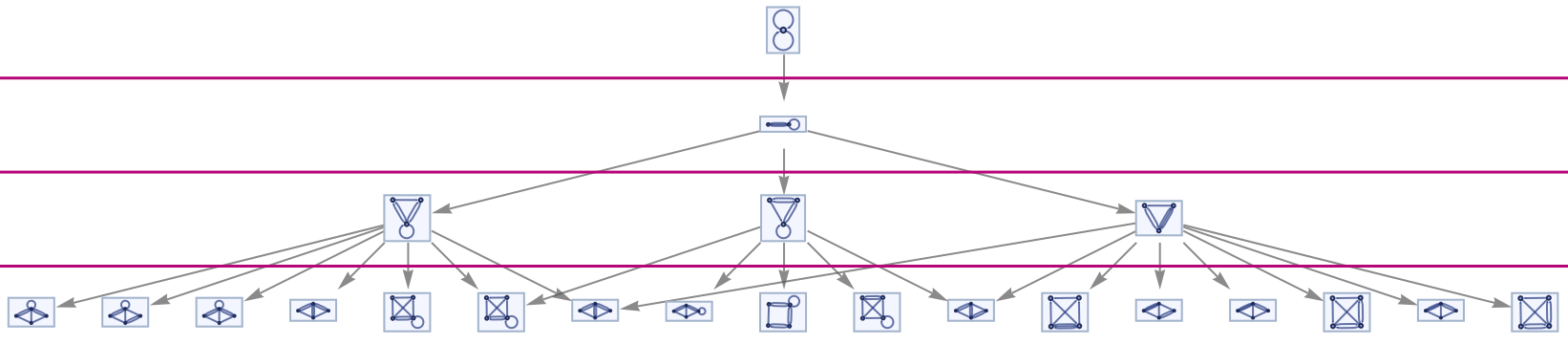}
\caption{The default foliation of the multiway evolution graph for the set substitution system ${\left\lbrace \left\lbrace x, y \right\rbrace, \left\lbrace y, z \right\rbrace \right\rbrace \to \left\lbrace \left\lbrace w, y \right\rbrace, \left\lbrace y, w \right\rbrace, \left\lbrace x, w \right\rbrace \right\rbrace}$.}
\label{fig:Figure36}
\end{figure}

\begin{figure}[ht]
\centering
\includegraphics[width=0.395\textwidth]{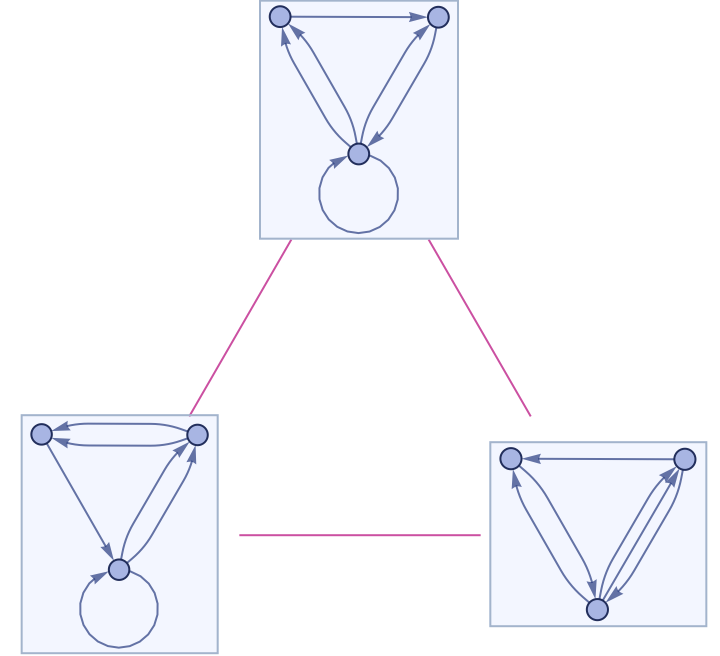}\hspace{0.1\textwidth}
\includegraphics[width=0.495\textwidth]{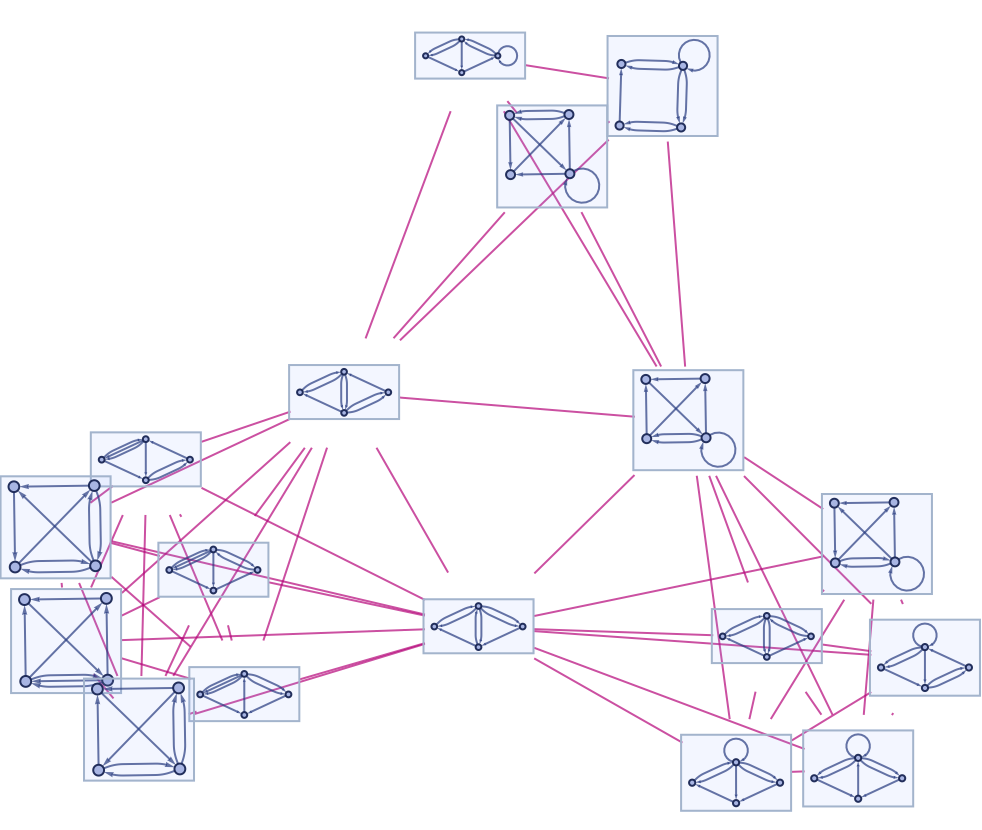}
\caption{The corresponding branchial graphs (i.e. branchlike hypersurfaces), as witnessed by an ``observer'' embedded within the default foliation of the multiway evolution graph for the set substitution system ${\left\lbrace \left\lbrace x, y \right\rbrace, \left\lbrace y, z \right\rbrace \right\rbrace \to \left\lbrace \left\lbrace w, y \right\rbrace, \left\lbrace y, w \right\rbrace, \left\lbrace x, w \right\rbrace \right\rbrace}$.}
\label{fig:Figure37}
\end{figure}

Within the current hypothesis for how quantum mechanics can be formulated within the Wolfram model \cite{gorard2}, quantum \textit{observers} (as distinguished from relativistic observers) are assumed to coordinatize multiway evolution graphs by assigning to each state vertex an integer time value, in such a way as to \textit{foliate} the multiway evolution graph into the level sets of this universal time function, known as ``branchial graphs'' or ``branchlike hypersurfaces'' (which are assumed to be analogous to the discrete Cauchy surfaces that one obtains by foliating a causal network into spacelike hypersurfaces). Indeed, the rationale behind referring to such choices of universal time function as ``observers'' stems from the formal analogy to choices of reference frame in relativity. More precisely, one has: 

\begin{definition}
An ``observer'' in a multiway system is any ordered sequence of non-intersecting ``branchlike hypersurfaces'' ${\Sigma_t}$ that covers the entire multiway evolution graph, with the ordering defined by a universal time function:

\begin{equation}
t : \mathcal{M} \to \mathbb{Z}, \qquad \text{ such that } \Delta t \neq 0 \text{ everywhere},
\end{equation}
such that the branchlike hypersurfaces are exactly the level sets of this function, satisfying:

\begin{equation}
\forall t_1, t_2 \in \mathbb{Z}, \qquad \Sigma_{t_1} = \left\lbrace p \in \mathcal{M} : t \left( p \right) = t_1 \right\rbrace, \text{ and } \Sigma_{t_1} \cap \Sigma_{t_2} = \emptyset \iff t_1 \neq t_2,
\end{equation}
where ${\mathcal{M}}$ denotes the vertex set of the multiway evolution graph.
\end{definition}
The branchial graphs themselves effectively show common the ancestry distance between multiway states for a given value of the universal time function; in other words, vertices $A$ and $B$ are connected by an undirected edge in the branchial graph if and only if they share a common ancestor $C$ in the multiway evolution graph. An example of a default choice of foliation for the multiway evolution graph of a Wolfram model system is shown in Figure \ref{fig:Figure36}, along with the corresponding sequence of branchial graphs (i.e. branchlike hypersurfaces) as witnessed by an ``observer'' embedded within that foliation in Figure \ref{fig:Figure37}.

The conjectural significance of these branchial graphs is that, under the assumptions of this formalism, each state vertex within the multiway evolution graph corresponds to a different pure state for the universe (under the assumption that the overall state of the universe is described by some generalized Hartle-Hawking wave function\cite{hartle}), with each branchial graph thus indicating the instantaneous superposition of certain eigenstates at a given moment of time, and with the evolution of the multiway system from one branchlike hypersurface to the next thus corresponding to the unitary evolution of this wave function. The magnitude of the amplitude associated with each eigenstate is taken to be related to that state's path weighting (i.e. the number of distinct evolution paths which lead to that state) within the multiway evolution graph, as illustrated via a toy example in the next subsection. 

Geometrically, much like the points on a discrete spacelike hypersurface for a given causal network are presumed to correspond to points in some associated continuous Riemann manifold (with the discrete spatial distance metric converging to a Riemannian metric in the continuum limit), points on a discrete branchlike hypersurface for a given multiway evolution graph are presumed to correspond to points in some associated continuous projective Hilbert space, with the discrete branchial distance metric defined above converging to a \textit{Fubini-Study metric} on ${\mathbb{CP}^n}$ in the continuum limit:

\begin{definition}
The ``Fubini-Study metric'' on the complex projective Hilbert space ${\mathbb{CP}^n}$, written in terms of the homogeneous coordinates:

\begin{equation}
\mathbf{Z} = \left[ Z_0, \dots, Z_n \right],
\end{equation}
i.e. the standard coordinate notation for projective varieties in algebraic geometry, is defined by the line element:

\begin{equation}
ds^2 = \frac{\left\lvert \mathbf{Z} \right\rvert^2 \left\lvert d \mathbf{Z} \right\rvert^2 - \left( \mathbf{\bar{Z}} \cdot d \mathbf{Z} \right) \left( \mathbf{Z} \cdot d \mathbf{\bar{Z}} \right)}{\left\lvert \mathbf{Z} \right\rvert^4},
\end{equation}
or, in a more explicit form:

\begin{equation}
ds^2 = \frac{Z_{\alpha} \bar{Z}^{\alpha} d Z_{\beta} d \bar{Z}^{\beta} - \bar{Z}^{\alpha} Z_{\beta} d Z_{\alpha} \bar{Z}^{\beta}}{\left( Z_{\alpha} \bar{Z}^{\alpha} \right)^2}.
\end{equation}
\end{definition}
In other words, assuming that the points on each branchlike hypersurface represent pure quantum states of the form:

\begin{equation}
\ket{\psi} = \sum_{k = 0}^{n} Z_k \ket{e_k} = \left[ Z_0 : Z_1 : \cdots : Z_n \right],
\end{equation}
for some orthonormal basis set ${\left\lbrace \ket{e_k} \right\rbrace}$ for the Hilbert space, the discrete branchial distance metric converges to the following infinitesimal line element:

\begin{equation}
ds^2 = \frac{\braket{\delta \psi}{\delta \psi}}{\braket{\psi}{\psi}} - \frac{\braket{\delta \psi}{\psi} \braket{\psi}{\delta \psi}}{\braket{\psi}{\psi}^2}.
\end{equation}
For further details regarding the formal correspondence between branchlike hypersurfaces and projective Hilbert spaces, see \cite{gorard2}.

\subsubsection{Illustration of Quantum State Transitions Using Multiway Graphs}

\begin{figure}[ht]
\centering
\includegraphics[width=0.495\textwidth]{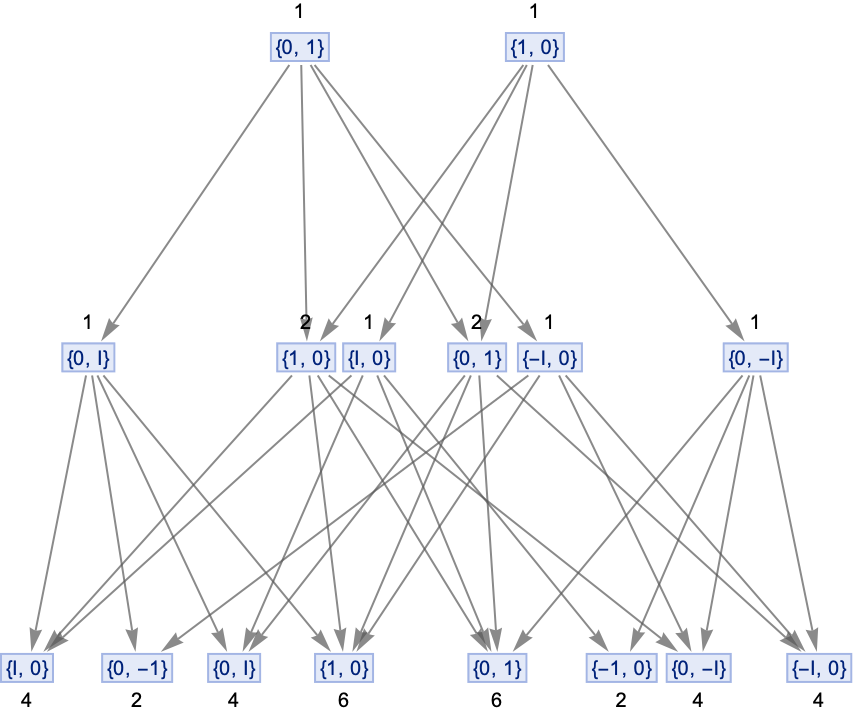}
\caption{The multiway evolution graph for a toy quantum system based upon a root-NOT gate being applied to an initial ${\frac{1}{\sqrt{2}} \left( \ket{0} + \ket{1} \right)}$ superposition state, with vertex weights given by the number of distinct evolution paths that lead to a given state.}
\label{fig:Figure38}
\end{figure}

\begin{figure}[ht]
\centering
\includegraphics[width=0.395\textwidth]{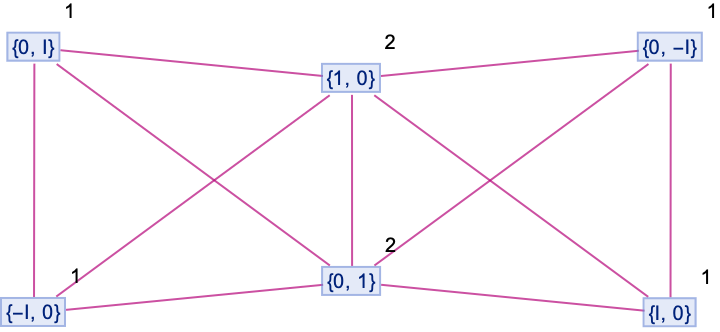}\hspace{0.1\textwidth}
\includegraphics[width=0.395\textwidth]{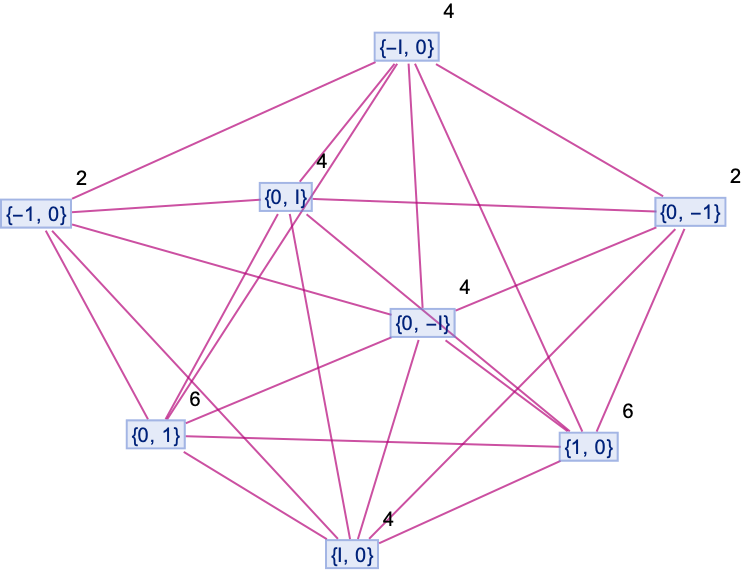}
\caption{The sequence of branchial graphs (i.e. branchlike hypersurfaces), as witnessed within the default foliation of the multiway evolution graph for the toy root-NOT quantum system starting from an initial ${\frac{1}{\sqrt{2}} \left( \ket{0} + \ket{1} \right)}$ superposition state, with vertex weights given by the number of distinct evolution paths that lead to a given state.}
\label{fig:Figure39}
\end{figure}

As an initial toy example, let us first illustrate how a typical multiway evolution graph can be used to represent state transitions between qubits, as obtained by applying a standard root-NOT quantum gate:

\begin{equation}
\sqrt{NOT} = \frac{1}{2} \begin{bmatrix}
1 + i & 1 - i\\
1 - i & 1 + i
\end{bmatrix},
\end{equation}
to an initial superposition state of the form ${\frac{1}{\sqrt{2}} \left( \ket{0} + \ket{1} \right)}$. This evolution is shown in Figures \ref{fig:Figure38} and \ref{fig:Figure39}, produced using a simple Wolfram Language software framework that we have developed known as \textit{QuantumToMultiwaySystem}\cite{wfr4}.

By simply summing over the collection of states on each branchlike hypersurface (with each state vector being multiplied by path weight of its associated vertex), one immediately verifies that this representation of the quantum evolution is faithful (in the sense that it agrees with the ordinary representation of the evolution in terms of basic matrix multiplication), and that, under the interpretation that each branchial graph designates a superposition of eigenstates, the path weights of the vertices naturally play the role of the quantum amplitudes, as required.

\clearpage

\subsection{Rule Compositions and Rulial Multiway Systems}

\begin{figure}[ht]
\centering
\includegraphics[width=0.495\textwidth]{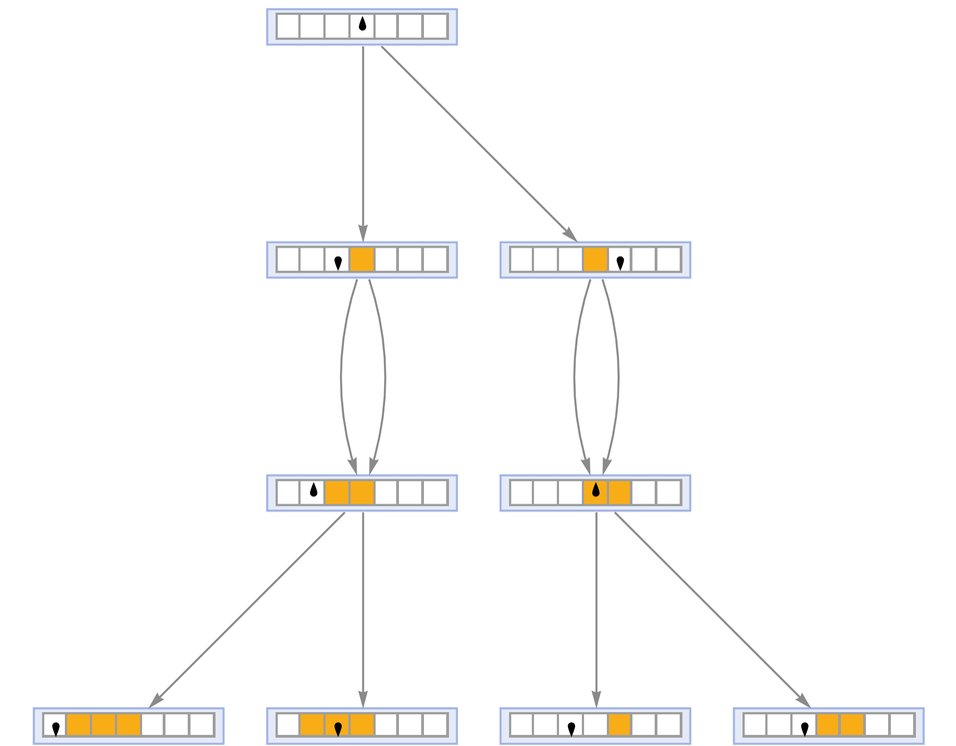}\\
\includegraphics[width=0.495\textwidth]{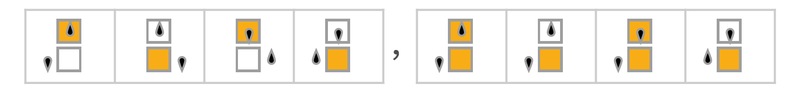}
\caption{The multiway evolution graph for a 2-state, 2-color non-deterministic Turing machine constructed by parallel composition of the two deterministic Turing machine rules shown below it.}
\label{fig:Figure40}
\end{figure}

\begin{figure}[ht]
\centering
\includegraphics[width=0.495\textwidth]{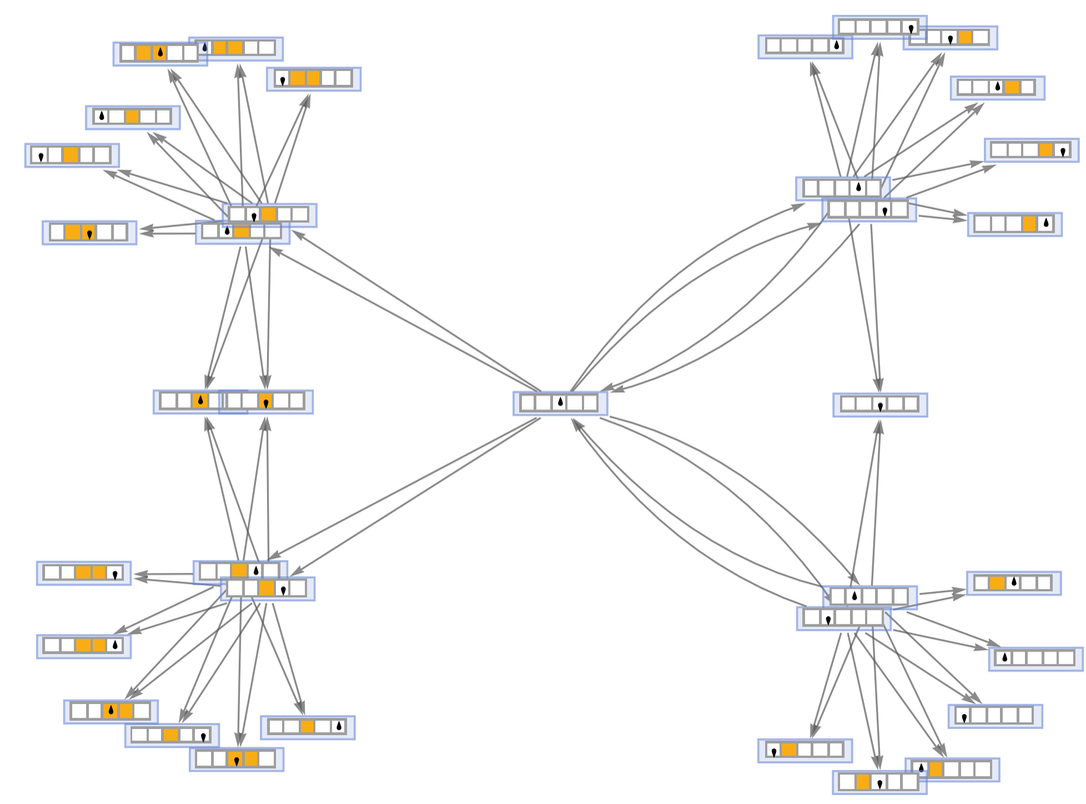}
\caption{The rulial multiway states graph (i.e. a variant of a multiway evolution graph in which cycles are permitted), as obtained by the parallel composition of all possible 2-state, 2-color Turing machine rules.}
\label{fig:Figure41}
\end{figure}

One important consistency condition required by our formulation is that the category of branchial graphs, which we tentatively denote ${\mathbf{BrGraph}}$ (exploiting the fact that the branchial graphs, and thus, by extension, the multiway evolution graphs, tentatively denoted ${\mathbf{MuGraph}}$, trivially form subcategories of the category ${\mathbf{Graph}}$ of graphs, to be defined subsequently) be equipped with a monoidal structure; in other words, it is required that one be able to take monoidal products of branchial graphs and multiway evolution graphs, just as one can take tensor products of states and operators in standard quantum mechanics. In the following subsection, we will show how this monoidal structure arises naturally from the parallel composition of rewriting rules. This can be made mathematically precise by considering the combinatorial properties of so-called ``rulial'' multiway systems - that is, multiway systems obtained by the parallel composition of all possible rules of a given size. For instance, a multiway evolution graph for a particular parallel composition of 2-state, 2-color Turing machine rules (thus yielding a non-deterministic Turing machine evolution) is shown in Figure \ref{fig:Figure40}, whilst the \textit{rulial} multiway states graph (i.e. a variant of a multiway evolution graph in which cycles are permitted), obtained by the parallel composition of all possible 2-state, 2-color Turing machine rules, is shown in Figure \ref{fig:Figure41}. In the above, a \textit{non-deterministic Turing machine} is taken to refer to a generalization of an ordinary (deterministic) Turing machine in which two or more deterministic Turing machine rules can be applied at any given step, thus weakening the partial transition function that defines an ordinary Turing machine:

\begin{equation}
\delta : \left( Q \setminus F \right) \times F \to Q \times \Gamma \times \left\lbrace L, R \right\rbrace,
\end{equation}
to a partial transition \textit{relation} of the same basic form:

\begin{equation}
\delta \subseteq \left( \left( Q \setminus F \right) \times F \right) \times \left( Q \times \Gamma \times \left\lbrace L, S, R \right\rbrace \right),
\end{equation}
where, as usual, $Q$ denotes a finite, non-empty set of states, ${\Gamma}$ denotes a finite, non-empty alphabet, $F$ denotes a set of final states, and $L$, $R$ and $S$ denote the possible shift directions for the tape head, namely left shift, right shift and no shift, respectively. Note that this description of a non-deterministic Turing machine differs fundamentally from that of a \textit{quantum} Turing machine, in which the transition function becomes a transition monoid:

\begin{equation}
\delta : \Sigma \times Q \otimes \Gamma \to \Sigma \times Q \otimes \Gamma \times \left\lbrace L, R \right\rbrace,
\end{equation}
and where the set of states $Q$ is interpreted as a finite-dimensional Hilbert space, such that ${\delta}$ reduces to a collection of unitary matrices corresponding to automorphisms of $Q$. In the context of our multiway formulation, deterministic Turing machines correspond to particular paths through the multiway evolution graph that are chosen by \textit{deterministic} evolution rules; non-deterministic Turing machines also correspond to particular paths, albeit ones that are chosen by \textit{non-deterministic} evolution rules; and quantum Turing machines correspond to the entire multiway evolution graph itself (i.e. the superposition of all possible paths), as represented in terms of an evolution between branchlike hypersurfaces (where each branchlike hypersurface corresponds to a pure superposition of classical Turing machine states).

As we shall outline formally in the next subsection, within this particular formulation, a non-deterministic Turing machine can therefore be interpreted geometrically as corresponding to a fiber bundle which has been constructed by taking a monoidal product of ordinary (deterministic) Turing machines. The original motivation for considering this particular model for the monoidal structure was that it would entail that, upon appropriate foliation of the rulial multiway system, each rulial branchial graph would consequently inherit, via the category ${\mathbf{FdHilb}}$, the structure of a tensor product of finite-dimensional Hilbert spaces\cite{arsiwalla}\cite{gorard3}.

\clearpage

\subsection{Monoidal Products of Multiway Systems}

\begin{figure}[ht]
\centering
\includegraphics[width=0.395\textwidth]{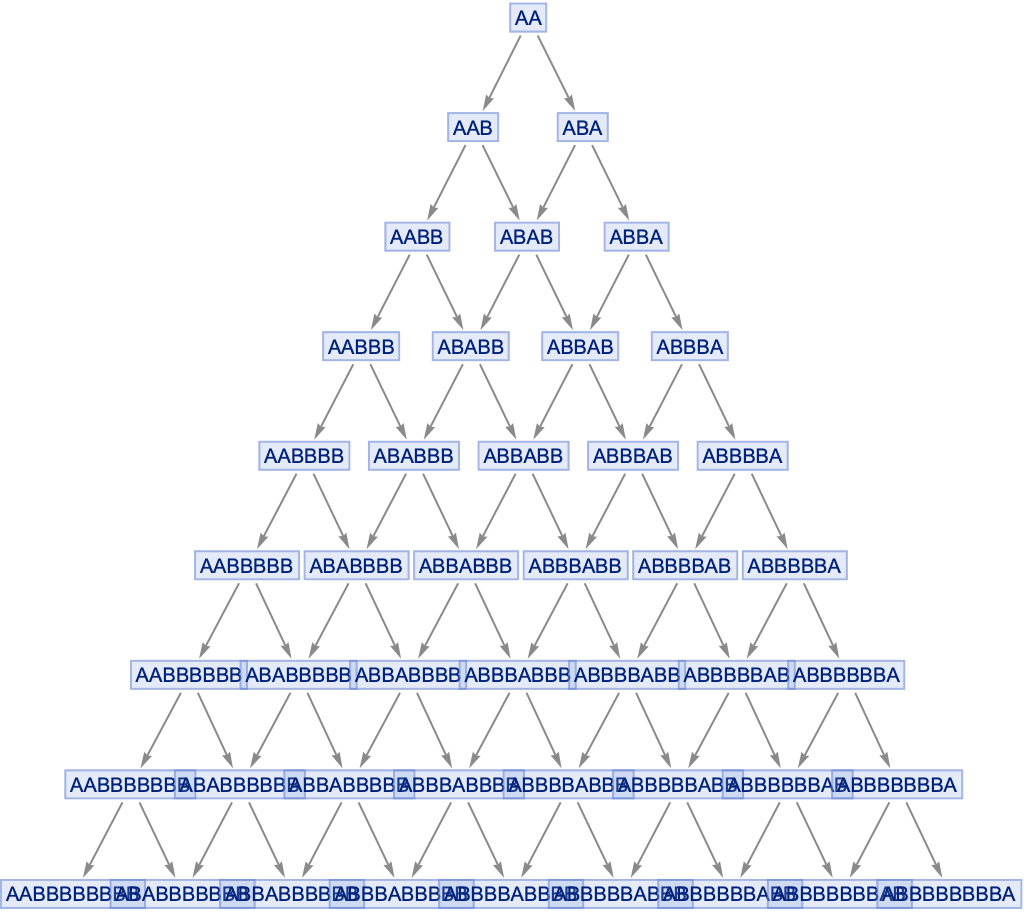}\hspace{0.1\textwidth}
\includegraphics[width=0.395\textwidth]{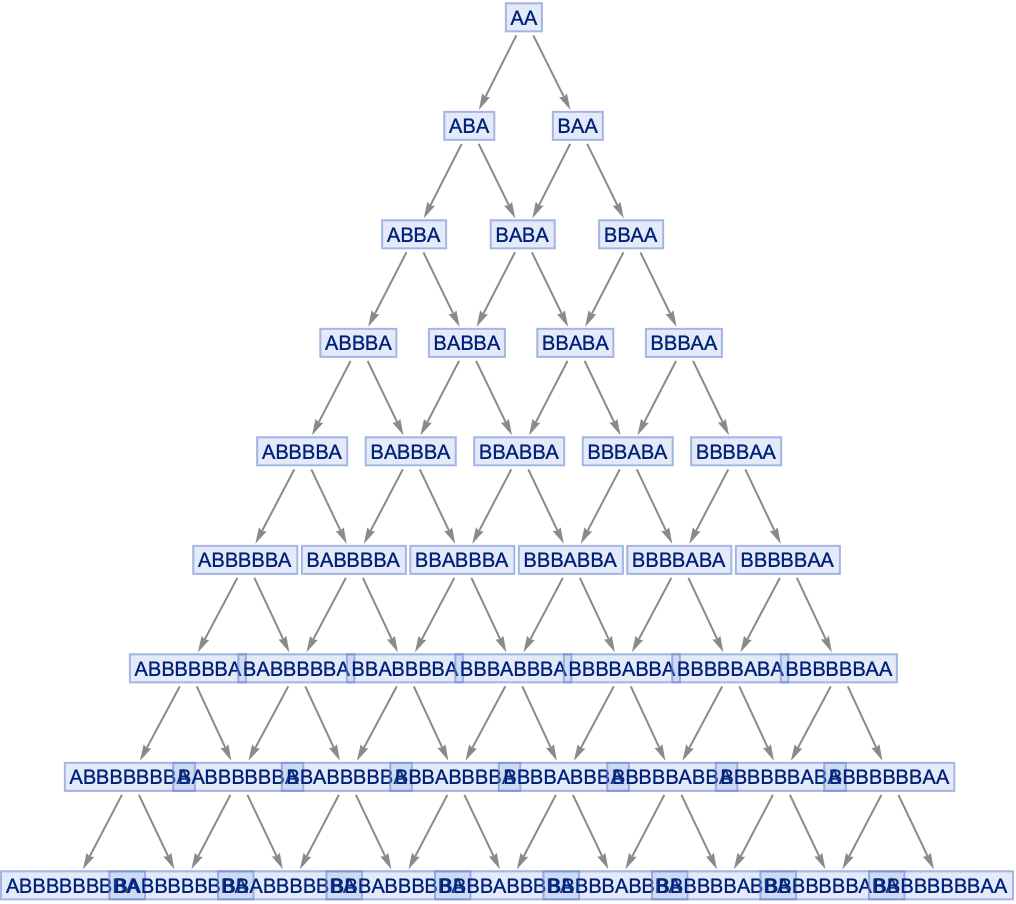}
\caption{The two-dimensional multiway evolution graphs for two elementary string substitution systems, namely ${\left\lbrace A \to AB \right\rbrace}$ and ${\left\lbrace A \to BA \right\rbrace}$, respectively.}
\label{fig:Figure42}
\end{figure}

\begin{figure}[ht]
\centering
\includegraphics[width=0.595\textwidth]{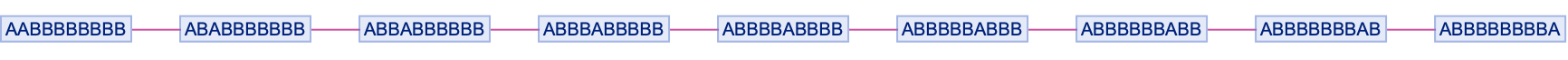}
\includegraphics[width=0.595\textwidth]{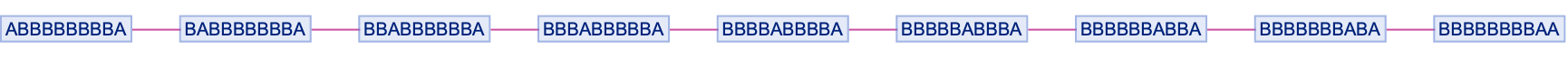}
\caption{The one-dimensional branchial graphs (i.e. branchlike hypersurfaces), as witnessed after 8 evolution steps, in the default foliation of the multiway evolution graph for two elementary string substitution systems, namely ${\left\lbrace A \to AB \right\rbrace}$ and ${\left\lbrace A \to BA \right\rbrace}$, respectively.}
\label{fig:Figure43}
\end{figure}

\begin{figure}[ht]
\centering
\includegraphics[width=0.395\textwidth]{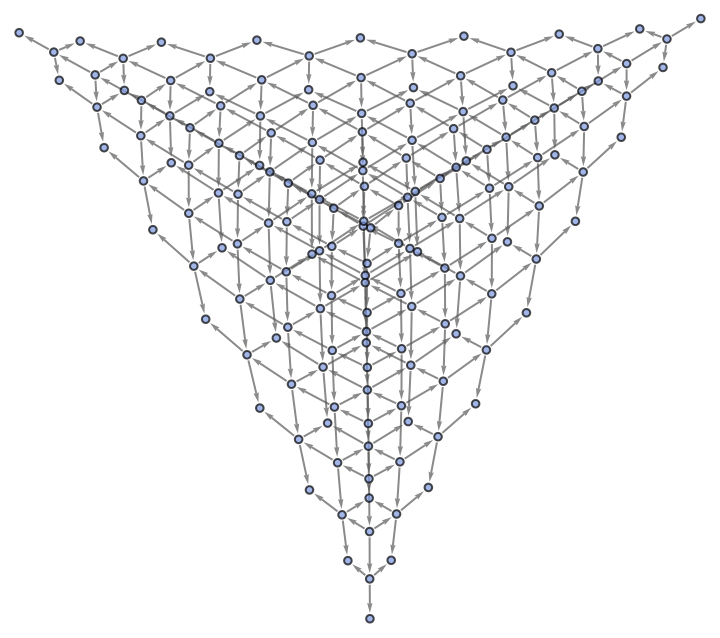}\hspace{0.1\textwidth}
\includegraphics[width=0.395\textwidth]{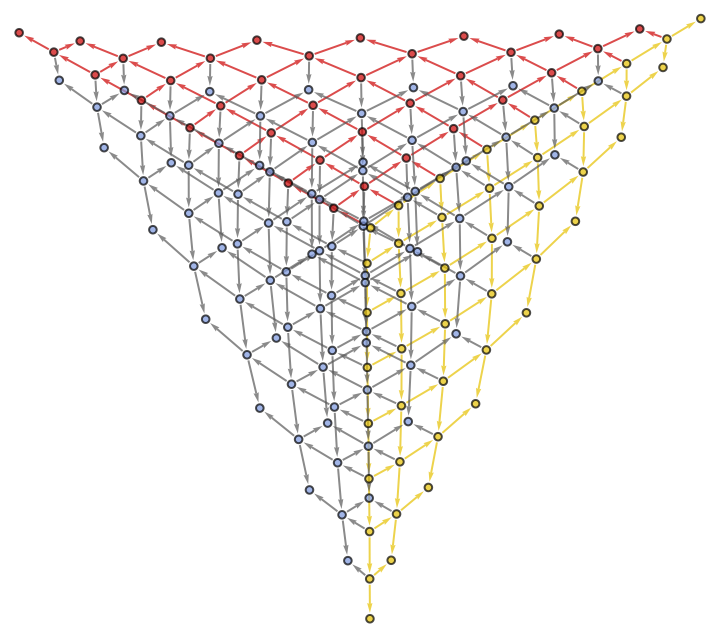}
\caption{On the left, the higher-dimensional multiway evolution graph for the composite string substitution system ${\left\lbrace A \to AB, A \to BA \right\rbrace}$, obtained by parallel composition of the two elementary string substitution systems shown above. On the right, the two constituent elementary multiway systems are highlighted.}
\label{fig:Figure44}
\end{figure}

\begin{figure}[ht]
\centering
\includegraphics[width=0.295\textwidth]{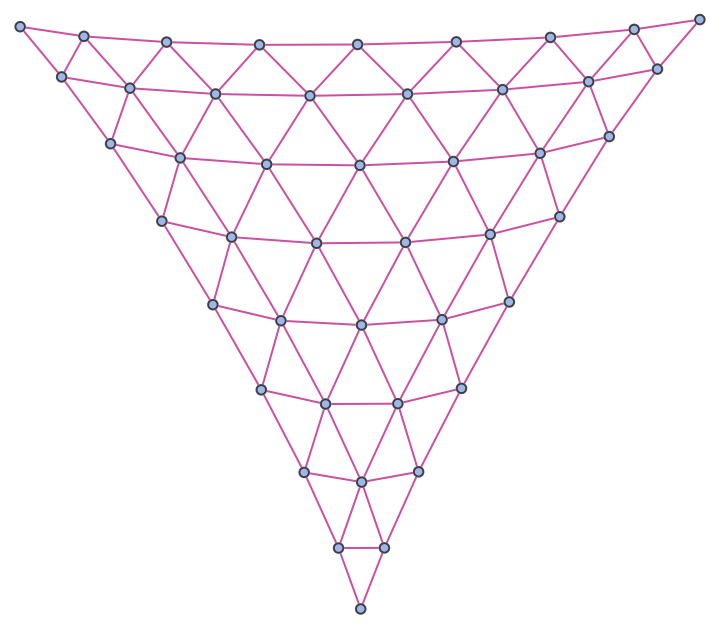}\hspace{0.25\textwidth}
\includegraphics[width=0.295\textwidth]{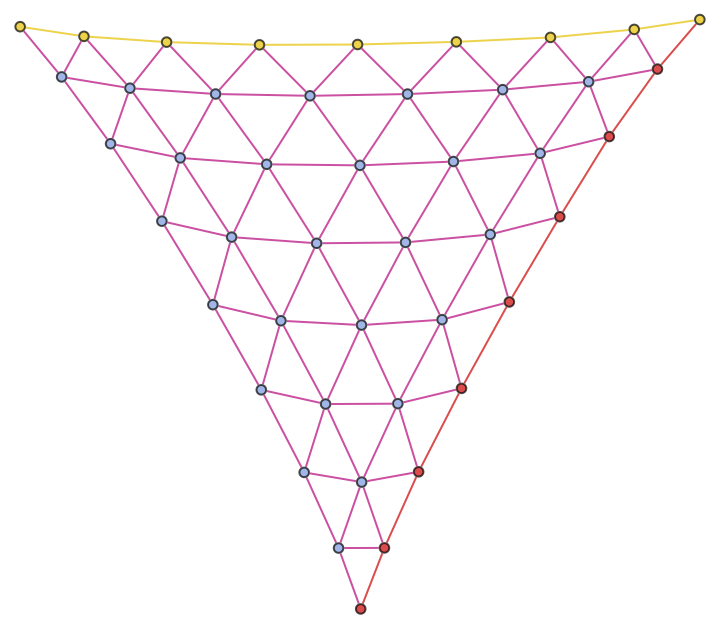}
\caption{On the left, the two-dimensional branchial graph (i.e. branchlike hypersurface), as witnessed after 8 evolution steps, in the default foliation of the multiway evolution graph for the composite string substitution system ${\left\lbrace A \to AB, A \to BA \right\rbrace}$, obtained by parallel composition of the two elementary string substitution systems shown above. On the right, the two constituent elementary branchial graphs are highlighted.}
\label{fig:Figure45}
\end{figure}

In this subsection we will demonstrate explicitly how the disjoint union of rules endows the rulial multiway system with a natural monoidal structure.  As an illustrative example, consider the two-dimensional multiway evolution graphs produced by two elementary string substitution systems, namely ${\left\lbrace A \to AB \right\rbrace}$ and ${\left\lbrace A \to BA \right\rbrace}$, as shown in Figure \ref{fig:Figure42}, with the associated one-dimensional branchial graphs (as witnessed after 8 evolution steps in the default foliation of the multiway evolution graph) shown in Figure \ref{fig:Figure43}. By composing the two rules together in parallel, we thus obtain a composite string multiway system ${\left\lbrace A \to AB, A \to BA \right\rbrace}$, with the two elementary two-dimensional string multiway systems forming a spanning set for this new higher-dimensional space, as shown in Figure \ref{fig:Figure44}; similarly, the two elementary one-dimensional string branchial graphs form a spanning set for the two-dimensional composite branchial graph, as shown in Figure \ref{fig:Figure45}. We now wish to illustrate how such composite multiway and  branchial graphs may be considered to be the result of a monoidal product of the two original elementary multiway and branchial graphs respectively.

Trivially, the parallel composition of rules in the context of a rulial multiway system is both symmetric and associative (since it is merely a disjoint union of rule sets), so all that remains is to prove is that this putative monoidal structure on the branchial and multiway evolution graphs is, in fact, compatible with the monoidal structure of the category of ZX-diagrams, and then we will have successfully established a rigorous connection between the formalism of multiway systems and the formalism of categorical quantum mechanics. 

First, let us consider how the monoidal structure makes itself manifest in the two two-dimensional multiway evolution graphs shown in Figure \ref{fig:Figure42}, resulting in the three-dimensional multiway evolution graph shown in Figure \ref{fig:Figure44}. We can see immediately that this composite multiway evolution graph does not correspond to either the Cartesian product nor the Kronecker product of the two constituent evolution graphs, since the number of nodes that one sees in the composite graph is far fewer than one would expect from a typical graph product. However, noting instead that all graph products can be interpreted as fiber bundles, with the Cartesian product yielding a trivial bundle and the Kronecker product yielding a non-trivial bundle, we can see that the Kronecker product is only one possible such construction, since we can choose to merge sections of graphs in a variety of non-trivial ways. The precise operation playing out in our example here is that, when two rules act on each of the multiway states shown in Figure \ref{fig:Figure42}, each of the multiway evolution graphs acquires additional edges and vertices extending in a third direction, corresponding to the places where the second rule is applicable. Then, these two extended evolution graphs are merged, with identifications made between vertices that correspond to the same string state, yielding the resulting multiway system shown in Figure \ref{fig:Figure44}. The branchial graph shown in Figure \ref{fig:Figure45} is simply a slice of the multiway product shown in Figure \ref{fig:Figure44}, and its structure is therefore directly induced from this monoidal construction.

\begin{figure}[ht]
\centering
\includegraphics[width=0.395\textwidth]{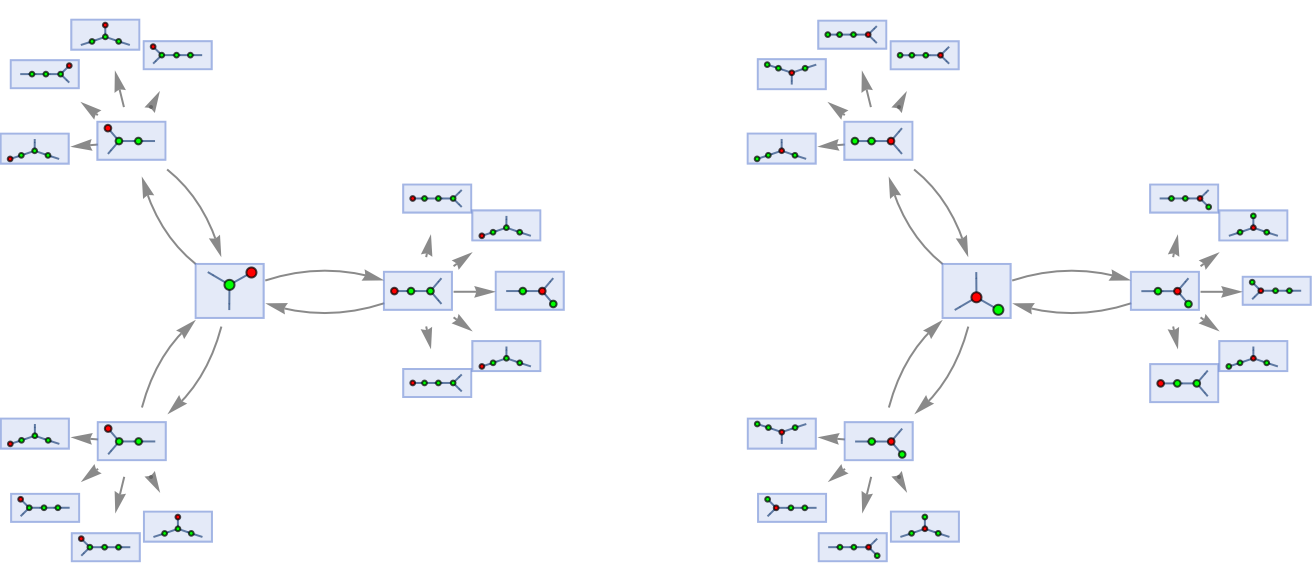}\hspace{0.1\textwidth}
\includegraphics[width=0.395\textwidth]{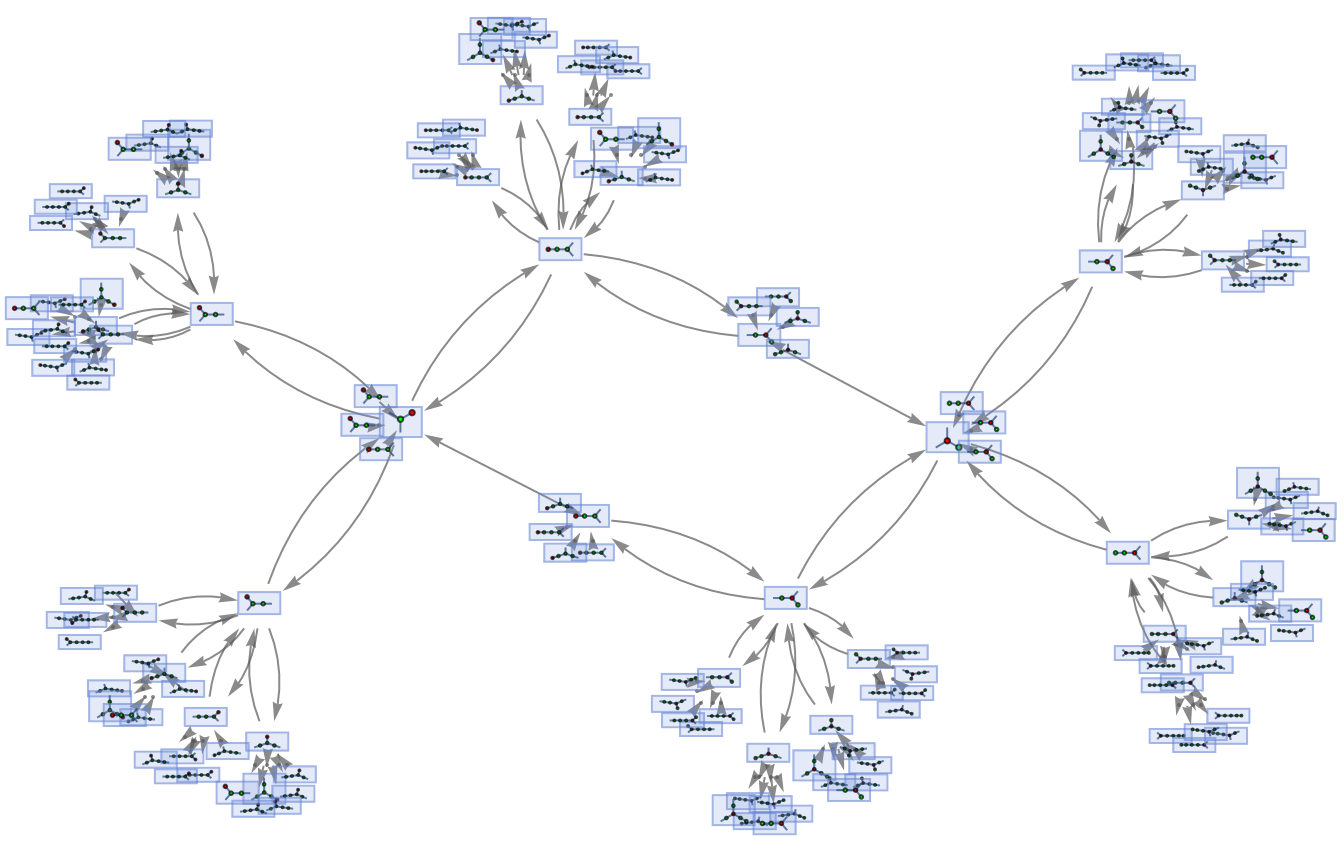}
\caption{On the left, the unmerged multiway states graph (i.e. a variant of the multiway evolution graph in which cycles are permitted) for the first two steps in the non-deterministic evolution of a ZX-calculus multiway operator system, starting from two independent two-spider initial conditions. On the right, the merged multiway states graph for the same ZX-calculus multiway operator system, showing the effects of completion.}
\label{fig:Figure65}
\end{figure}

Indeed, as we shall prove formally in the following subsection, these mergings of multiway evolution graphs based on identifications between specific vertices correspond precisely to the monoidal products described by the category of directed cospans on selective adhesive rules. In fact, as we shall also subsequently describe, the monoidal structure arising from the category of directed cospans is applicable more generally to any abstract rewriting system that is equipped with a notion of \textit{completion} (as is the case with the Knuth-Bendix completions employed within the context of the Wolfram model). An example of such a merging of multiway states graphs (as a consequence of completion) for a simple ZX-calculus multiway operator system starting from two independent two-spider initial conditions is shown in Figure \ref{fig:Figure65}. In this particular case, the completion (i.e. the minimal set of additional rewriting rules that must be added to the multiway system in order to guarantee causal invariance) corresponds to the set:

\begin{multline}
\left\lbrace X \left[ x_1, 0, 1, 0 \right] \otimes \left( Z \left[ z_1, 1, 2, 0 \right] \otimes \left( W \left[ x_1, z_1 \right] \otimes \left( W \left[ z_1, o_1 \right] \otimes W \left[ z_1, o_2 \right] \right) \right) \right) \to \right.\\
Z \left[ z_1, 0, 1, 0 \right] \otimes \left( X \left[ x_1, 1, 2, 0 \right] \otimes \left( W \left[ z_1, x_1 \right] \otimes \left( W \left[ x_1, o_1 \right] \otimes W \left[ x_1, o_2 \right] \right) \right) \right),\\
Z \left[ z_1, 0, 1, 0 \right] \otimes \left( X \left[ x_1, 1, 2, 0 \right] \otimes \left( W \left[ z_1, x_1 \right] \otimes \left( W \left[ x_1, o_1 \right] \otimes W \left[ x_1, o_2 \right] \right) \right) \right) \to\\
\left. X \left[ x_1, 0, 1, 0 \right] \otimes \left( Z \left[ z_1, 1, 2, 0 \right] \otimes \left( W \left[ x_1, z_1 \right] \otimes \left( W \left[ z_1, o_1 \right] \otimes W \left[ z_1, o_2 \right] \right) \right) \right) \right\rbrace.
\end{multline}

\begin{figure}[ht]
\centering
\includegraphics[width=0.495\textwidth]{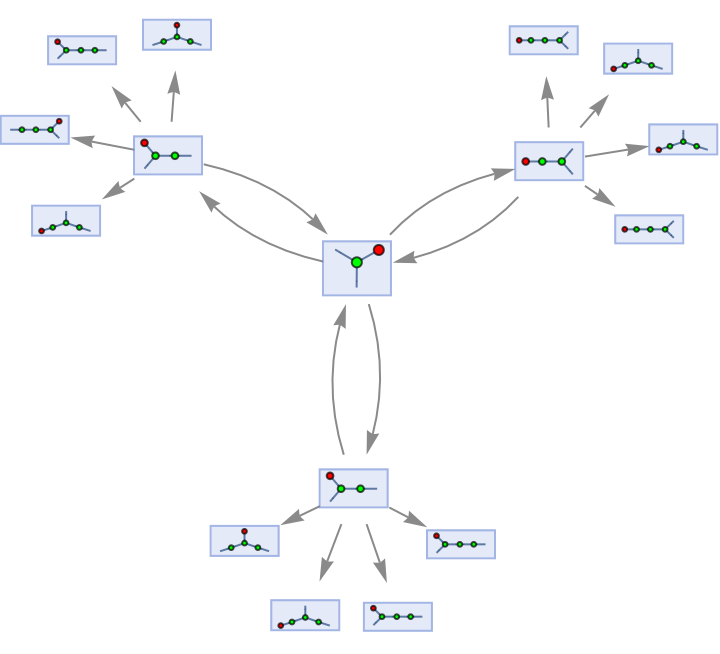}
\includegraphics[width=0.495\textwidth]{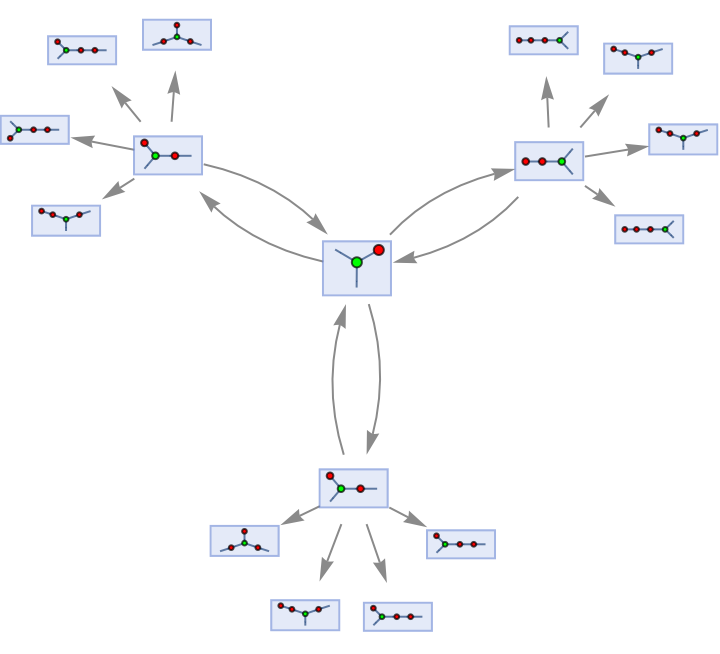}
\caption{The multiway states graphs (i.e. variants of multiway evolution graphs in which cycles are permitted) corresponding to the first two steps in the non-deterministic evolution of a ZX-calculus multiway operator system, starting from a simple two-spider initial diagram, using only the (input arity 2 variants of the) Z- and X-spider identity rules, respectively.}
\label{fig:Figure46}
\end{figure}

\begin{figure}[ht]
\centering
\includegraphics[width=0.395\textwidth]{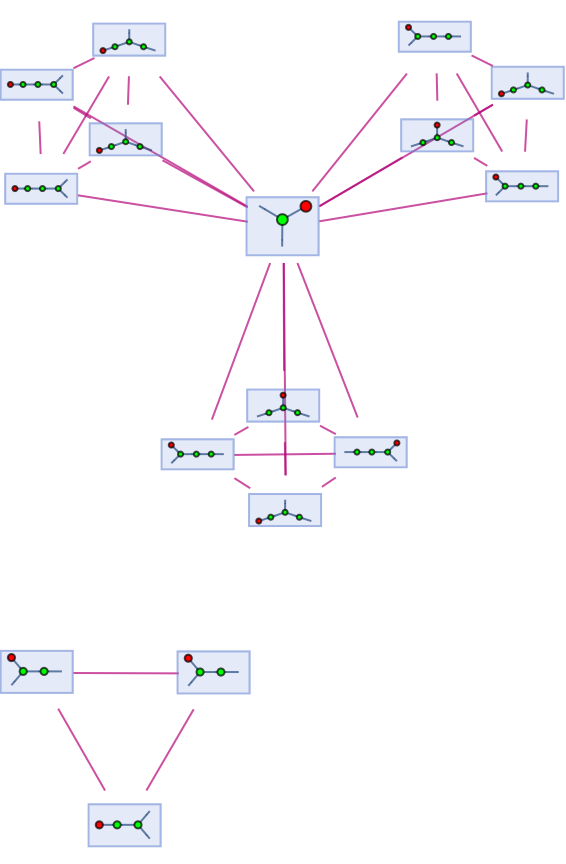}\hspace{0.1\textwidth}
\includegraphics[width=0.395\textwidth]{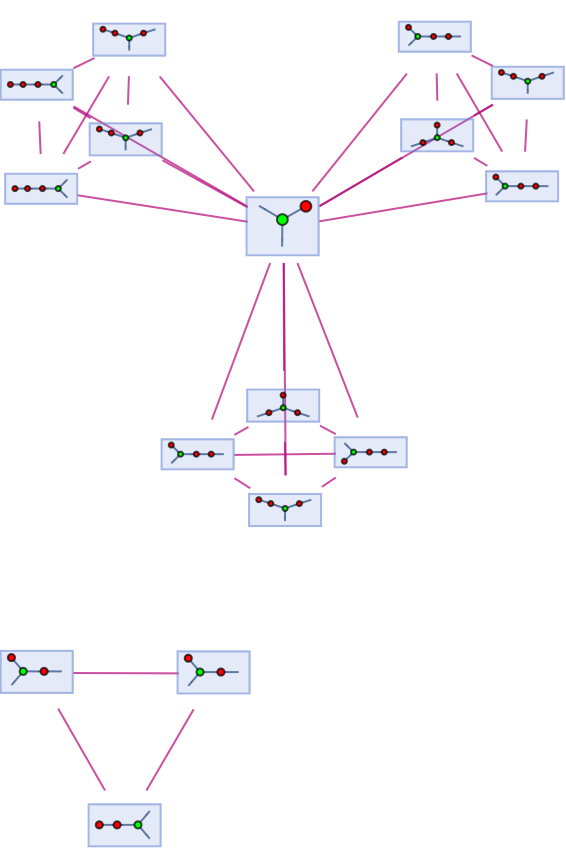}
\caption{The corresponding branchial graphs (i.e. branchlike hypersurfaces), as witnessed within the default foliation of the multiway states graph for the first two steps in the non-deterministic evolution of a ZX-calculus multiway operator system, starting from a simple two-spider initial diagram, using only the (input arity 2 variants of the) Z- and X-spider identity rules, respectively.}
\label{fig:Figure47}
\end{figure}

Let us now proceed to illustrate computationally how such a monoidal structure, operating within ${\mathbf{MuGraph}}$, the putative category of multiway systems, would act on the multiway evolution graphs produced by standard rewritings of ZX-diagrams. Starting from an initial expression of the form:

\begin{equation}
X[x_1, 0, 1, 0] \otimes \left( Z \left[ z_1, 1, 2, 0 \right] \otimes \left( W \left[ x_1, z_1 \right] \otimes \left( W \left[ z_1, o_1 \right] \otimes W \left[ z_1, o_2 \right] \right) \right) \right),
\end{equation}
i.e. the same two-spider initial diagram as used in the previous section, and applying only the variants of the Z- and X-spider identity rules with an input arity of 2, one obtains after two evolution steps the multiway states graphs (i.e. the variants of multiway evolution graphs in which cycles are permitted) shown in Figure \ref{fig:Figure46}, with the corresponding branchial graphs shown in Figure \ref{fig:Figure47}; the Z-spider identity rule case is shown on the left, and the X-spider identity rule case is shown on the right.

\begin{figure}[ht]
\centering
\includegraphics[width=0.695\textwidth]{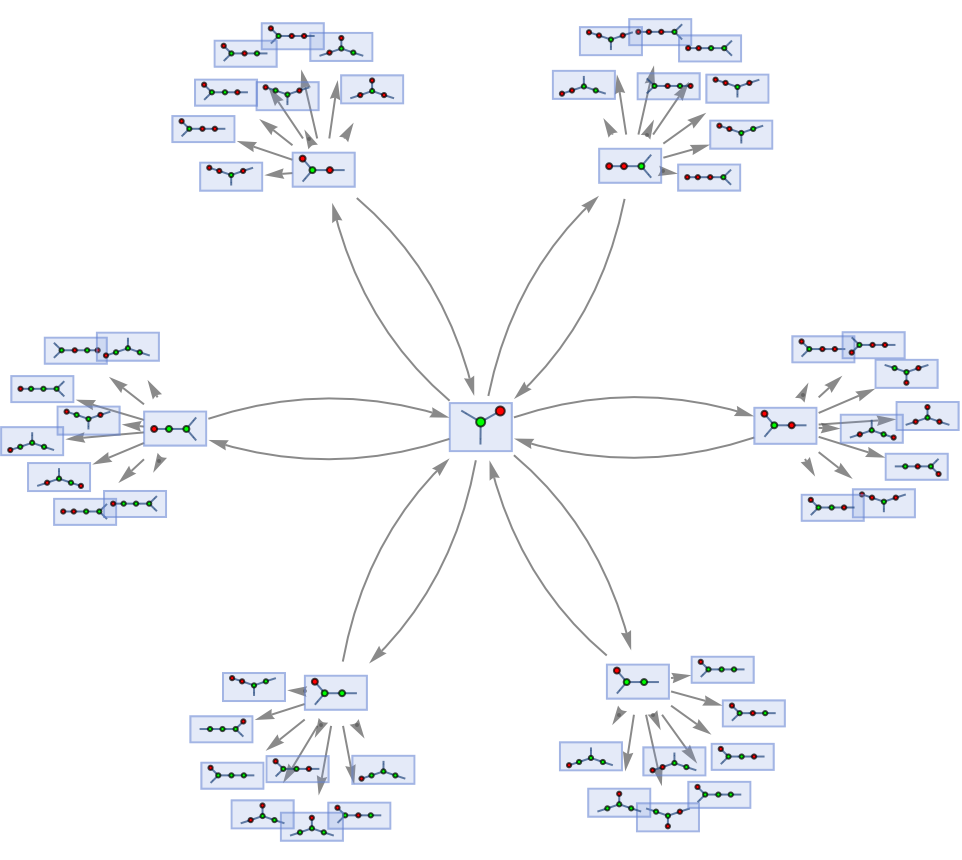}
\caption{The multiway states graph (i.e. the variant of the multiway evolution graph in which cycles are permitted) corresponding to the first two steps in the non-deterministic evolution of a composite ZX-calculus multiway operator system, starting from a simple two-spider initial diagram, using both of the (input arity 2 variants of the) Z- and X-spider identity rules.}
\label{fig:Figure48}
\end{figure}

\begin{figure}[ht]
\centering
\includegraphics[width=0.495\textwidth]{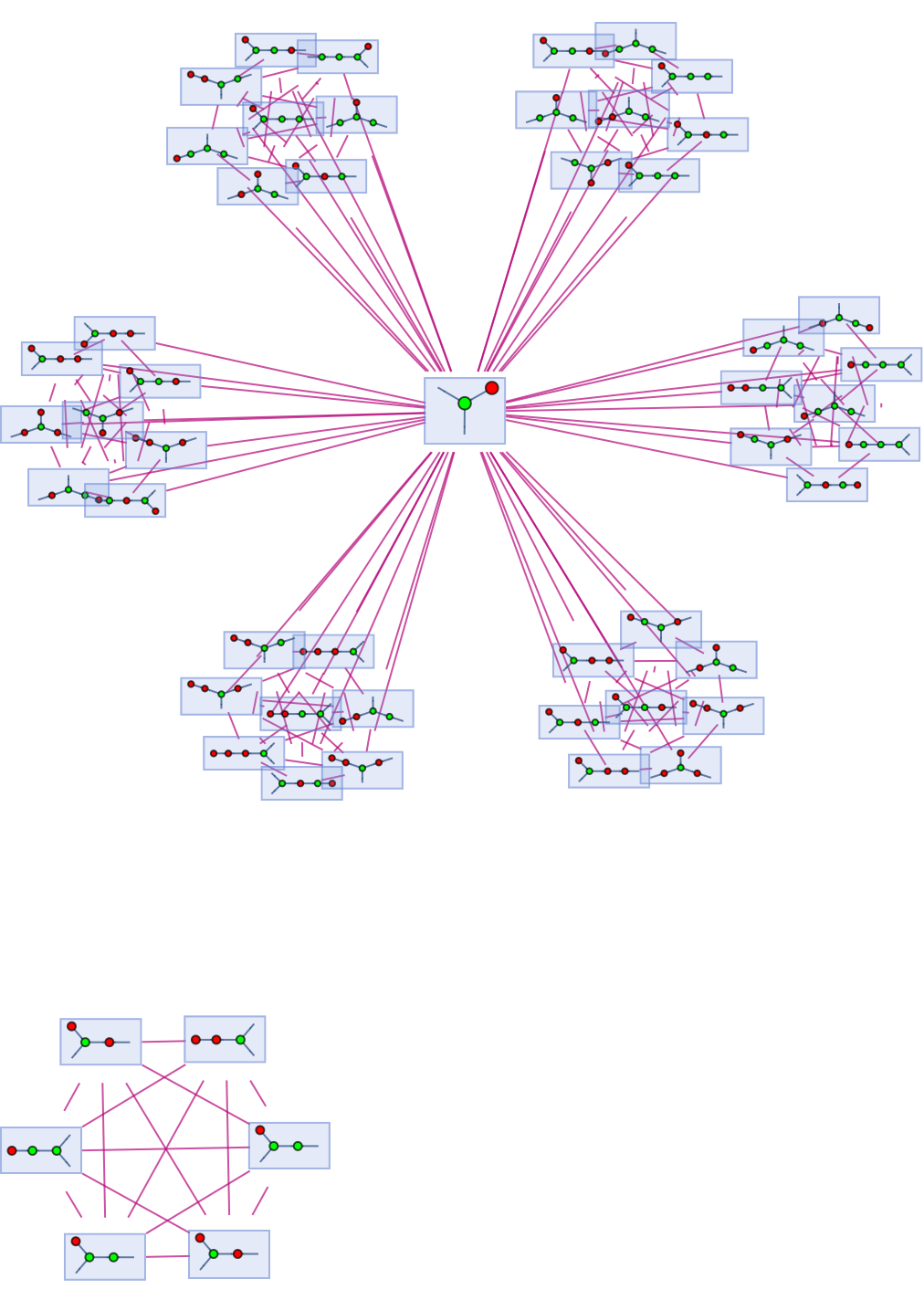}
\caption{The corresponding branchial graph (i.e. branchlike hypersurface), as witnessed within the default foliation of the multiway states graph for the first two steps in the non-deterministic evolution of a composite ZX-calculus multiway operator system, starting from a simple two-spider initial diagram, using both of the (input arity 2 variants of the) Z- and X-spider identity rules.}
\label{fig:Figure49}
\end{figure}

Parallel composition of the Z- and X-spider identity rules yields a composite multiway operator system in which either rule can be applied at any step. The resultant composite multiway states graph is shown in Figure \ref{fig:Figure48}, with the corresponding composite branchial graph shown in Figure \ref{fig:Figure49}.

\begin{figure}[ht]
\centering
\includegraphics[width=0.195\textwidth]{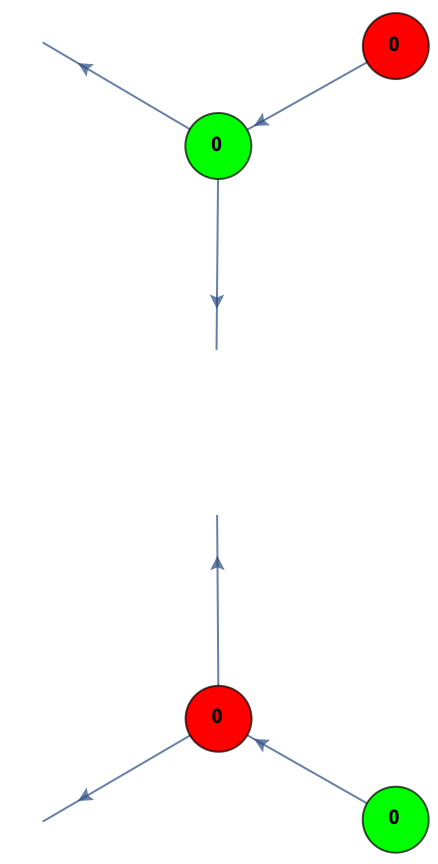}
\caption{A composite four-spider ZX-diagram, composed by stacking two (mutually color-inverted) two-spider diagrams on top of each other, and hence representing a monoidal product.}
\label{fig:Figure50}
\end{figure}

\begin{figure}[ht]
\centering
\includegraphics[width=0.695\textwidth]{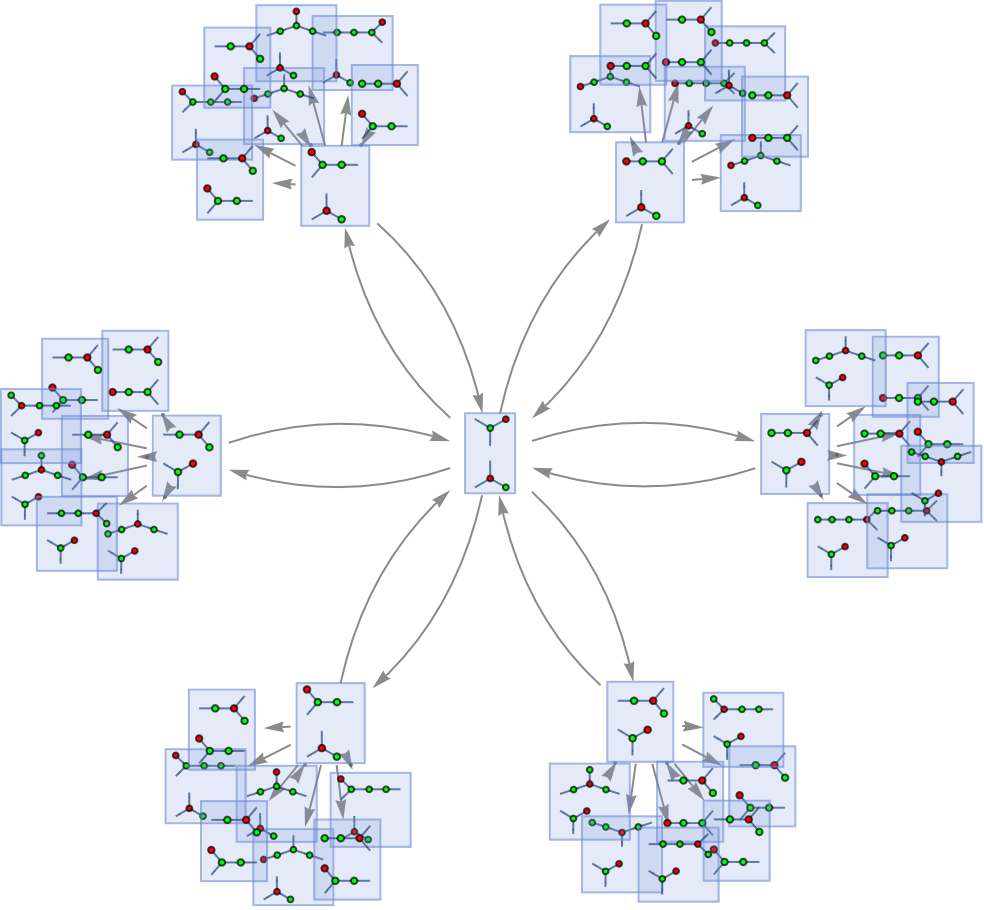}
\caption{The multiway states graph (i.e. the variant of the multiway evolution graph in which cycles are permitted) corresponding to the first two steps in the non-deterministic evolution of a ZX-calculus multiway operator system, starting from a composite four-spider initial diagram (obtained as the monoidal product of two two-spider ZX-diagrams), using only the (input arity 2 variant of the) Z-spider identity rule.}
\label{fig:Figure51}
\end{figure}

\begin{figure}[ht]
\centering
\includegraphics[width=0.495\textwidth]{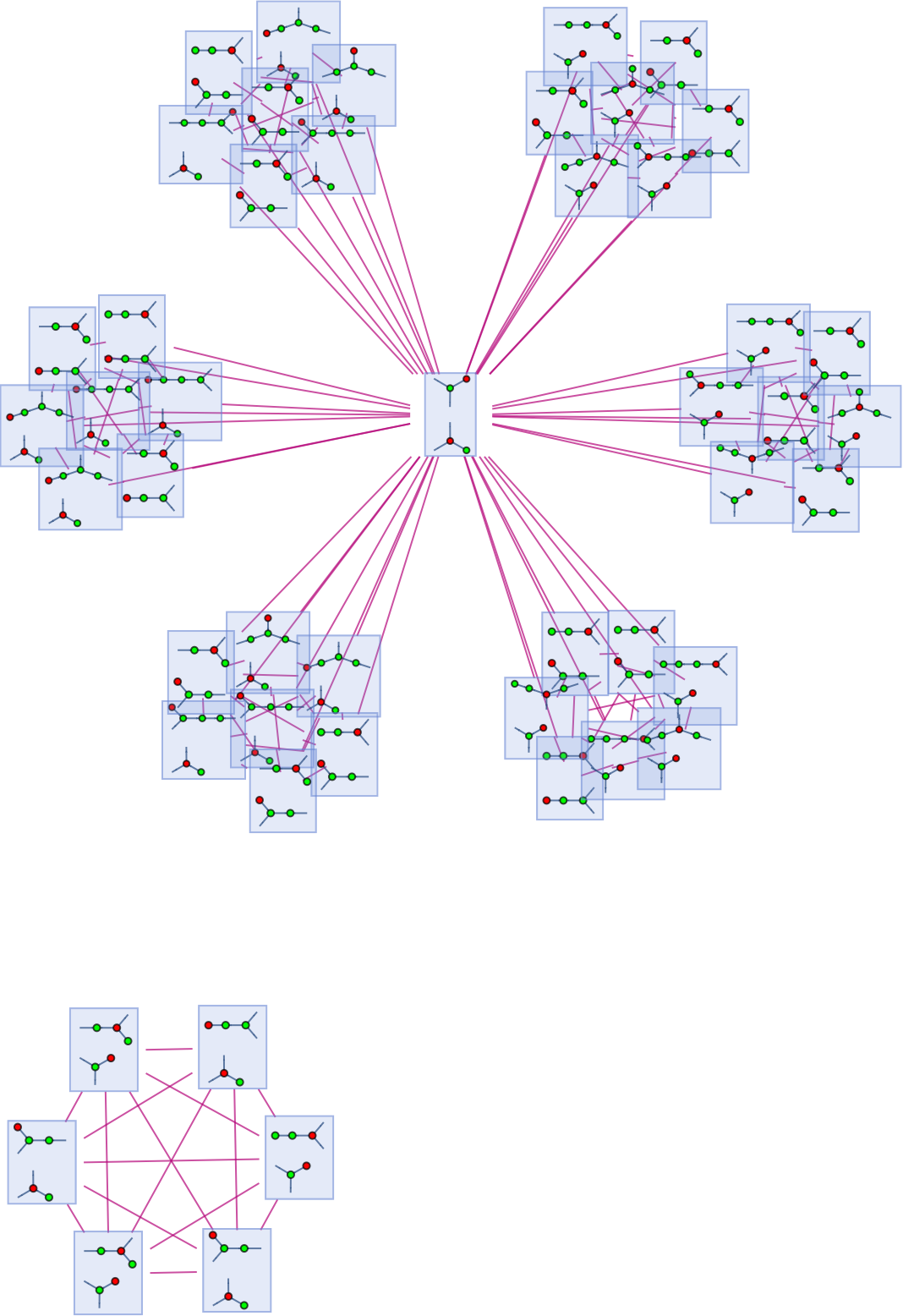}
\caption{The corresponding branchial graph (i.e. branchlike hypersurface), as witnessed within the default foliation of the multiway states graph for the first two steps in the non-deterministic evolution of a ZX-calculus multiway operator system, starting from a composite four-spider initial diagram (obtained as the monoidal product of two two-spider ZX-diagrams), using only the (input arity 2 variant of the) Z-spider identity rule.}
\label{fig:Figure52}
\end{figure}

On the other hand, monoidal products in the ZX-calculus can be taken simply by stacking diagrams vertically. Thus, consider now an initial expression of the form:

\begin{equation}
\begin{split}
X \left[ x_2, 0, 1, 0 \right] \otimes \left( Z \left[ z_2, 1, 2, 0 \right] \otimes \left( W \left[ x_2, z_2 \right] \otimes \left( W \left[ z_2, o_3 \right] \otimes \left( W \left[ z_2, o_4 \right] \otimes \left( Z \left[ z_1, 0, 1, 0 \right] \otimes \right. \right. \right. \right. \right.\\
\left. \left. \left. \left. \left. \left( X \left[ x_1, 1, 2, 0 \right] \otimes \left( W \left[ z_1, x_1 \right] \otimes \left( W \left[ x_1, o_1 \right] \otimes W \left[ x_1, o_2 \right] \right) \right) \right) \right) \right) \right) \right) \right)
\end{split},
\end{equation}
i.e. a composite four-spider initial diagram composed by stacking two (mutually color-inverted) two-spider diagrams on top of each other, as shown in Figure \ref{fig:Figure50}. Applying only the Z-spider identity rule (with an input arity of 2), one obtains after two evolution steps the multiway states graph (i.e. the variant of the multiway evolution graph in which cycles are permitted) shown in Figure \ref{fig:Figure51}, with the corresponding branchial graph shown in Figure \ref{fig:Figure52}. Modulo trivial state equivalences (in particular, the ``cusps'' in the branchial graphs in the monoidal product case each contain exactly one fewer state than the corresponding ``cusps'' in the rulial composition case, due to the presence of a trivial color-inversion symmetry in the diagram), one can see rather explicitly that the multiway and branchial structures obtained by the monoidal product of ZX-diagrams are isomorphic to the multiway and branchial structures obtained by rulial composition (i.e. by the disjoint union of rules).

This realization immediately lends credibility to the claim that rulial composition endows ${\mathbf{MuGraph}}$, the putative category of multiway systems, with a monoidal structure, and moreover indicates strongly that this monoidal structure is fully compatible between the branchial, multiway and rulial levels. Indeed, after performing the same isomorphism tests against 65 instances of ZX-diagrams with spider input/output arities up to 1, 235 instances of diagrams with input/output arities up to 2 and 1001 diagrams with input/output arities up to 3, the same compatibility between the monoidal structure of the ZX-diagrams and the compositional structure of the rulial multiway system was found to hold in all cases. To prove that this compatibility holds in general, we employ the techniques introduced by Dixon and Kissinger\cite{dixon} for constructing general monoidal theories using typed open graphs; we will first provide a brief overview of the Dixon-Kissinger construction, followed by a description of how it can then be extended for our particular purposes.

\clearpage

\subsection{The Rulial Multiway System of ZX-Diagrams and the Category of Directed Cospans of Selective Adhesive Rules}

We now proceed to prove that the rulial multiway system defined by applying all possible rewriting rules of the ZX-calculus to a given ZX-diagram forms a subcategory of the category of directed cospans of selective adhesive rules. Intuitively, any particular multiway evolution graph, associated with a particular choice of rewriting rule, can thus be thought of as corresponding to a particular fiber in the associated \textit{rulial space}. In this way, the multiway evolution graphs themselves inherit certain properties from this underlying space of rules, in such a way as to guarantee compatibility between the merging and splitting operations of the multiway evolution and the generalized tensor product and partial trace operations of the ZX-calculus (this can be shown by constructing an explicit functor from the rulial space to the multiway evolution graphs themselves, such that the monoidal structure reduces to a simple homomorphism of functors, although this construction lies outside the scope of the present article).

\begin{definition}
A ``functor category'', denoted ${\left[ \mathbf{C}, \mathbf{D} \right]}$ for categories ${\mathbf{C}}$ and ${\mathbf{D}}$, is a category whose class of objects ${\mathrm{ob} \left( \left[ \mathbf{C}, \mathbf{D} \right] \right)}$ is the class of functors:

\begin{equation}
F : \mathbf{C} \to \mathbf{D},
\end{equation}
and whose class of morphisms ${\hom \left( \left[ \mathbf{C}, \mathbf{D} \right] \right)}$ is the class of natural transformations:

\begin{equation}
\eta : F \to G,
\end{equation}
between these functors, where:

\begin{equation}
G : \mathbf{C} \to \mathbf{D},
\end{equation}
is assumed to be another object in ${\mathrm{ob} \left( \left[ \mathbf{C}, \mathbf{D} \right] \right)}$.
\end{definition}
Natural transformations can be composed, since if:

\begin{equation}
\mu \left( A \right) : F \left( A \right) \to G \left( A \right),
\end{equation}
is a natural transformation between the functors:

\begin{equation}
F : \mathbf{C} \to \mathbf{D}, \qquad \text{ and } \qquad G : \mathbf{C} \to \mathbf{D},
\end{equation}
and:

\begin{equation}
\eta \left( A \right) : G \left( A \right) \to H \left( A \right),
\end{equation}
is a natural transformation between the functors:

\begin{equation}
G : \mathbf{C} \to \mathbf{D}, \qquad \text{ and } \qquad H : \mathbf{C} \to \mathbf{D},
\end{equation}
then the composition:

\begin{equation}
\eta \left( A \right) \mu \left( A \right) : F \left( A \right) \to H \left( A \right)
\end{equation}
defines a natural transformation from $F$ to $H$, and therefore ${\left[ \mathbf{C}, \mathbf{D} \right]}$ does indeed satisfy the axioms of a category. Strictly speaking, in the definition above, one should restrict ${\mathbf{C}}$ to be a \textit{small category} (i.e. one in which the collections of objects ${\mathrm{ob} \left( \mathbf{C} \right)}$ and of morphisms ${\hom \left( \mathbf{C} \right)}$ form sets, as opposed to proper classes).

In the work of Dixon and Kissinger, an ``open graph'' is formalized as a generalization of a graph in which edges do not need to be attached to vertices on both ends; they can ``hang loose'' at one or both ends, and an edge can consequently even connect back to itself to form a closed loop. Note that, within our multiway operator formulation of the ZX-calculus, we are implicitly performing diagrammatic rewritings over open graphs (as opposed to ordinary graphs), since the input wires, output wires and loops in the ZX-diagrams form edges that are open at one or both ends, and therefore this formalization is of eminent relevance to the present argument. The category of open graphs is not itself an adhesive category, but through the Dixon-Kissinger construction of a \textit{selective adhesive functor}, the category can inherit sufficient adhesivity from the ambient category of typed graphs into which it is embedded so as to allow the techniques of double-pushout rewriting to be applied. We can formalize the category of open graphs as a functor category in the following way:

\begin{definition}
A ``partial graph morphism''\cite{lowe}, denoted:

\begin{equation}
f : G \to H,
\end{equation}
for graphs $G$ and $H$, is given by a pair of partial functions:

\begin{equation}
f_V : G_V \to H_V, \qquad \text{ and } \qquad f_E : G_E \to H_E,
\end{equation}
where ${G_V}$ and ${G_E}$ denote the vertex and edge sets of graph $G$, respectively, and likewise for graph $H$, satisfying certain properties. More specifically, if the function ${f_E}$ is defined for an edge ${e \in G_E}$, then ${f_V}$ should be defined for both ${s_G \left( e \right)}$ and ${t_G \left( e \right)}$ (where ${s_G}$ and ${t_G}$ denote the source and target functions for directed edges in graph $G$, respectively), and the following pair of composition rules should hold:

\begin{equation}
f_V \circ s_G = s_H \circ f_E, \qquad \text{ and } \qquad f_V \circ t_G = t_H \circ f_E.
\end{equation}
\end{definition}

\begin{definition}
A partial graph morphism $f$ becomes a ``total graph morphism'' if the functions ${f_V}$ and ${f_E}$ are both themselves total.
\end{definition}

\begin{definition}
A graph $G$ is a ``typed graph'' if there exists a distinct graph ${TG}$ known as the ``type graph'', along with a total graph morphism of the form:

\begin{equation}
type_{G} : G \to TG,
\end{equation}
known as the ``typing morphism''.
\end{definition}

\begin{definition}
The category of graphs, denoted ${\mathbf{Graph}}$, is the functor category ${\left[ \mathbb{G}, \mathbf{Set} \right]}$ where ${\mathbb{G}}$ is defined by:

\begin{equation}
\begin{tikzcd}
E \arrow[r, "s", bend right = 20] \arrow[r, "t", bend left = 20] & P
\end{tikzcd},
\end{equation}
where $E$ designates the edges of the graph, $P$ designates points on the graph and $s$ and $t$ are the source and target functions, respectively, such that the incoming edges to point $p$ are those edges $e$ for which ${t \left( e \right) = p}$, and the outgoing edges are those for which ${s \left( e \right) = p}$.
\end{definition}
The rationale for using the terminology of ``points'' as opposed to ``vertices'' arises from the fact that some points will be ``true'' vertices, whilst others will simply be ``dummy'' points that lie on the open ends of edges, which are nevertheless useful when formalizing replacement rules over subgraphs. Furthermore, we can use the following type graph, henceforth denoted ${2_{\mathcal{G}}}$:

\begin{equation}
\begin{tikzcd}
V \arrow[r, bend right = 20] & \epsilon \arrow[l, bend right = 20] \arrow[loop right]
\end{tikzcd},
\end{equation}
to distinguish between the two. If a graph $G$ is typed by ${2_{\mathcal{G}}}$, i.e. if there exists a typing morphism:

\begin{equation}
\tau : G \to 2_{\mathcal{G}},
\end{equation}
then those points $p$ in graph $G$ which map to $V$, i.e. such that ${\tau \left( p \right) = V}$, are the usual (``true'') vertices, whilst all others, for which ${\tau \left( p \right) = \epsilon}$, are the ``dummy'' points lying on the open ends of edges.

\begin{definition}
A ``comma category''\cite{lawvere}, denoted ${\left( S \downarrow T \right)}$, where $S$ and $T$ (``source'' and ``target'', respectively) are a pair of functors with the same codomain:

\begin{equation}
\begin{tikzcd}
\mathbf{A} \arrow[r, "S"] & \mathbf{C} & \mathbf{B} \arrow[l, "T"]
\end{tikzcd},
\end{equation}
for categories ${\mathbf{A}}$, ${\mathbf{B}}$ and ${\mathbf{C}}$, is a category whose class of objects ${\mathrm{ob} \left( \left( S \downarrow T \right) \right)}$ is the class of all triples ${\left( A, B, h \right)}$ for which $A$ is an object in ${\mathrm{ob} \left( \mathbf{A} \right)}$, $B$ is an object in ${\mathrm{ob} \left( \mathbf{B} \right)}$ and:

\begin{equation}
h : S \left( A \right) \to T \left( B \right),
\end{equation}
is a morphism in ${\hom \left( \mathbf{C} \right)}$. Moreover, the class of morphisms ${\hom \left( \left( S \downarrow T \right) \right)}$ between objects ${\left( A, B, h \right)}$ and ${\left( A^{\prime}, B^{\prime}, h^{\prime} \right)}$ in ${\mathrm{ob} \left( \left( S \downarrow T \right) \right)}$ is the class of all pairs ${\left( f, g \right)}$ of morphisms:

\begin{equation}
f : A \to A^{\prime}, \qquad \text{ and } \qquad g : B \to B^{\prime},
\end{equation}
in ${\hom \left( \mathbf{A} \right)}$ and ${\hom \left( \mathbf{B} \right)}$, respectively, for which the following diagram commutes:

\begin{equation}
\begin{tikzcd}
S \left( A \right) \arrow[r, "S \left( f \right)"] \arrow[d, "h"] & S \left( A^{\prime} \right) \arrow[d, "h^{\prime}"]\\
T \left( B \right) \arrow[r, "T \left( g \right)"] & T \left( B^{\prime} \right)
\end{tikzcd}.
\end{equation}
Morphisms compose according to the equational rule:

\begin{equation}
\left( f^{\prime}, g^{\prime} \right) \circ \left( f, g \right) = \left( f^{\prime} \circ f, g^{\prime} \circ g \right),
\end{equation}
assuming that the right-hand expression is defined, and the identity morphism on object ${\left( A, B, h \right)}$ is assumed to be given by ${\left( id_A, id_B \right)}$.
\end{definition}
Comma categories thus provide a different intuition for the basic structure of a category, in which morphisms (rather than the objects themselves) become the fundamental items of study. As a special case, one can consider:

\begin{definition}
If one now considers a comma category for which ${\mathbf{C} = \mathbf{A}}$, $S$ is the identity functor and ${\mathbf{B} = \mathbf{1}}$ (i.e. the category containing a single object, denoted ${*}$, equipped with a single identity morphism), then one obtains a ``slice category'', denoted ${\left( \mathbf{A} \downarrow A_{*} \right)}$, where ${T \left( * \right) = A_{*}}$, for some object ${A_{*}}$ in ${\mathrm{ob} \left( \mathbf{A} \right)}$. In other words, one has:

\begin{equation}
\begin{tikzcd}
\mathbf{A} \arrow[r, "id_{\mathbf{A}}"] & \mathbf{A} & \mathbf{1} \arrow[l, "A_{*}"]
\end{tikzcd}.
\end{equation}
\end{definition}
The objects in ${\mathrm{ob} \left( \left( \mathbf{A} \downarrow A_{*} \right) \right)}$, namely ${\left( A, *, h \right)}$, can therefore be reduced to objects of the form ${\left(A, h \right)}$, where:

\begin{equation}
h : A \to A_{*},
\end{equation}
where, in order to make the reduction more explicit, we choose instead to refer to $h$ as ${\pi_A}$. Thus, we can simplify the morphisms:

\begin{equation}
\left( f, id_{*} \right) : \left( A, \pi_A \right) \to \left( A^{\prime}, \pi_{A^{\prime}} \right),
\end{equation}
in ${\hom \left( \left( \mathbf{A} \downarrow A_{*} \right) \right)}$ down to morphisms of the form:

\begin{equation}
f : A \to A^{\prime},
\end{equation}
for which the following diagram commutes:

\begin{equation}
\begin{tikzcd}
A \arrow[rr, "f"] \arrow[dr, "\pi_A"] & & A^{\prime} \arrow[dl, "\pi_{A^{\prime}}"]\\
& A_{*} &
\end{tikzcd}.
\end{equation}

\begin{definition}
The category of open graphs, denoted ${\mathbf{OGraph}}$, is a subcategory of the slice category ${\left( \mathbf{Graph} \downarrow 2_{\mathcal{G}} \right)}$ (i.e. a category whose objects are objects in ${\mathrm{ob} \left( \left( \mathbf{Graph} \downarrow 2_{\mathcal{G}} \right) \right)}$ and whose morphisms are morphisms in ${\hom \left( \left( \mathbf{Graph} \downarrow 2_{\mathcal{G}} \right) \right)}$, with the same identities and compositions of morphisms). More specifically, the class of objects ${\mathrm{ob} \left( \mathbf{OGraph} \right)}$ is given by those objects in ${\mathrm{ob} \left( \left( \mathbf{Graph} \downarrow 2_{\mathcal{G}} \right) \right)}$ for which every ``dummy'' point has no more than one incoming edge and one outgoing edge. The class of morphisms ${\hom \left( \mathbf{OGraph} \right)}$ is given by those morphisms in ${\hom \left( \left( \mathbf{Graph} \downarrow 2_{\mathcal{G}} \right) \right)}$ which are ``true'' vertices, i.e. any edge that is adjacent to a vertex ${f \left( v \right)}$ must also be in the codomain of $f$.
\end{definition}

\begin{definition}
An ``input'' is a ``dummy'' point $p$ that has no incoming edges; the set of all inputs of an open graph $G$ is denoted ${\mathrm{In} \left( G \right)}$.
\end{definition}

\begin{definition}
An ``output'' is a ``dummy'' point $p$ that has no outgoing edges; the set of all outputs of an open graph $G$ is denoted ${\mathrm{Out} \left( G \right)}$.
\end{definition}

\begin{definition}
An ``isolated point'' is a ``dummy'' point $p$ that has no incoming or outgoing edges (i.e. it is both an input and an output).
\end{definition}

\begin{definition}
A ``graphical signature'' is a function:

\begin{equation}
T : A \to O^{*} \times O^{*},
\end{equation}
for sets $A$ and $O$, where ${O^{*}}$ denotes the set of finite lists from $O$.
\end{definition}
An intuition for graphical signatures is that they assign each element of $A$ with an input and an output type, similar to the tensor schemes introduced by Joyal and Street\cite{joyal2}. Any graphical signature $T$ allows us to construct an associated type graph ${T_{\mathcal{G}}}$ whose vertex set is given by ${O \cup A}$, and in which every vertex ${o \in O}$ has a self-loop. For each vertex ${a \in A}$, ${T \left( a \right)}$ yields a pair of words $D$ and $C$, which define the domain and codomain of $a$, respectively (corresponding to the input and output types of $a$). For every element $d$ of word $D$, there exists a directed edge ${d \to a}$ in the type graph ${T_{\mathcal{G}}}$, and likewise there exists a directed edge ${a \to c}$ for every element $c$ in word $C$.

\begin{definition}
The edge neighborhood, denoted ${N \left( v \right)}$, of a given vertex $v$ in graph $G$ is the set of edges that are adjacent to $v$.
\end{definition}

\begin{definition}
A ``local isomorphism'' between graphs $G$ and $H$ is a map:

\begin{equation}
f : G \to H,
\end{equation}
such that, for every vertex $v$ in graph $G$, the edge function of $f$ restricts to the bijection:

\begin{equation}
f^{v} : N \left( v \right) \to N \left( f \left( v \right) \right).
\end{equation}
\end{definition}
In what follows, let ${\left( \mathbf{Graph} \downarrow T_{\mathcal{G}} \right)_{\cong}}$ denote the subcategory of the slice category ${\left( \mathbf{Graph} \downarrow T_{\mathcal{G}} \right)}$ whose class of objects ${\mathrm{ob} \left( \left( \mathbf{Graph} \downarrow T_{\mathcal{G}} \right)_{\cong} \right)}$ consists of all pairs of the form:

\begin{equation}
\left( G, \tau : G \to T_{\mathcal{G}} \right),
\end{equation}
where ${\tau}$ designates a local isomorphism, and whose class of morphisms ${\hom \left( \left( \mathbf{Graph} \downarrow T_{\mathcal{G}} \right)_{\cong} \right)}$ consists of all local isomorphisms. This forms a \textit{full} subcategory of ${\left( \mathbf{Graph} \downarrow T_{\mathcal{G}} \right)}$, i.e. one in which, for every pair of objects $A$ and $B$ in ${\mathrm{ob} \left( \left( \mathbf{Graph} \downarrow T_{\mathcal{G}} \right)_{\cong} \right)}$, one has:

\begin{equation}
\hom_{\left( \mathbf{Graph} \downarrow T_{\mathcal{G}} \right)_{\cong}} \left( A, B \right) = \hom_{\left( \mathbf{Graph} \downarrow T_{\mathcal{G}} \right)} \left( A, B \right).
\end{equation}
For every type graph ${T_{\mathcal{G}}}$, there exists a graph homomorphism:

\begin{equation}
\kappa : T_{\mathcal{G}} \to 2_{\mathcal{G}},
\end{equation}
that sends every point in the set $A$ to $V$ and every point in the set $O$ to ${\epsilon}$, such that composing every object in ${\mathrm{ob} \left( \left( \mathbf{Graph} \downarrow T_{\mathcal{G}} \right) \right)}$ on the left with the homomorphism ${\kappa}$ yields a \textit{forgetful} functor (i.e. a functor that partially ``forgets'' the structure of its input) of the form:

\begin{equation}
U_{\kappa} : \left( \mathbf{Graph} \downarrow T_{\mathcal{G}} \right) \to \left( \mathbf{Graph} \downarrow 2_{\mathcal{G}} \right),
\end{equation}
that sends objects of the form:

\begin{equation}
\tau : G \to T_{\mathcal{G}},
\end{equation}
in ${\mathrm{ob} \left( \left( \mathbf{Graph} \downarrow T_{\mathcal{G}} \right) \right)}$ to objects of the form ${\kappa \circ \tau}$ in ${\mathrm{ob} \left( \left( \mathbf{Graph} \downarrow 2_{\mathcal{G}} \right) \right)}$.

\begin{definition}
An ``open ${T_{\mathcal{G}}}$-graph'' is a ${T_{\mathcal{G}}}$-graph such that the object ${U_{\kappa} \left( \mathbf{G} \right)}$ in ${\mathrm{ob} \left( \left( \mathbf{Graph} \downarrow 2_{\mathcal{G}} \right) \right)}$ is itself an open graph.
\end{definition}

\begin{definition}
The category of open ${T_{\mathcal{G}}}$-graphs, denoted ${\mathbf{OGraph}_{T_{\mathcal{G}}}}$, is the full subcategory of ${\left( \mathbf{Graph} \downarrow T_{\mathcal{G}} \right)_{\cong}}$ whose objects are themselves open graphs.
\end{definition}

\begin{definition}
A ``cospan'' is any diagram that consists of two maps with a common domain:

\begin{equation}
B \rightarrow A \leftarrow C,
\end{equation}
and so generalizes the binary relation between two objects in a category ${\mathbf{C}}$, by considering instead three objects, $A$, $B$ and $C$ in ${\mathrm{ob} \left( \mathbf{C} \right)}$, and the pair of morphisms:

\begin{equation}
f : B \to A, \qquad \text{ and } \qquad f : C \to A,
\end{equation}
in ${\hom \left( \mathbf{C} \right)}$.
\end{definition}

\begin{definition}
The category of ``directed cospans'' of ${\mathbf{OGraph}_{T_{\mathcal{G}}}}$, denoted ${\mathbf{DCsp} \left( \mathbf{OGraph}_{T_{\mathcal{G}}} \right)}$, assuming a graphical signature of the form:

\begin{equation}
T : A \to O^{*} \times O^{*},
\end{equation}
is the category whose class of objects ${\mathrm{ob} \left( \mathbf{DCsp} \left( \mathbf{OGraph}_{T_{\mathcal{G}}} \right) \right)}$ is the class of words in ${O^{*}}$, i.e. the objects are the point graphs in ${\mathbf{OGraph}_{T_{\mathcal{G}}}}$, where points are endowed with a total order. Each morphism:

\begin{equation}
G : X \to Y,
\end{equation}
is a cospan of the form:

\begin{equation}
\begin{tikzcd}
Y \arrow[r, "c"] & G & X \arrow[l, "d"]
\end{tikzcd},
\end{equation}
in which $d$ denotes the inclusion map of ${\mathrm{In} \left( G \right) \cong X}$, $c$ denotes the inclusion map of ${\mathrm{Out} \left( G \right) \cong Y}$ and the open graph $G$ contains no isolated points.
\end{definition}
The crucial realization is that ${\mathbf{DCsp} \left( \mathbf{OGraph}_{T_{\mathcal{G}}} \right)}$ forms a symmetric monoidal category (strictly speaking, a 2-category) in the following, rather natural, way:

\begin{definition}
The ``coproduct'' of two objects ${A_1}$ and ${A_2}$ in ${\mathrm{ob} \left( \mathbf{C} \right)}$ for some category ${\mathbf{C}}$, denoted ${A_1 \amalg A_2}$, is an object for which there exists a pair of morphisms:

\begin{equation}
i_1 : A_1 \to A_1 \amalg A_2, \qquad \text{ and } \qquad i_2 : A_2 \to A_1 \amalg A_2,
\end{equation}
in ${\hom \left( \mathbf{C} \right)}$, such that, for any object $B$ in ${\mathrm{ob} \left( \mathbf{C} \right)}$ and any pair of morphisms:

\begin{equation}
f_1 : A_1 \to B, \qquad \text{ and } \qquad f_2 : A_2 \to B,
\end{equation}
in ${\hom \left( \mathbf{C} \right)}$, there exists a unique morphism:

\begin{equation}
f : A_1 \amalg A_2 \to B,
\end{equation}
in ${\hom \left( \mathbf{C} \right)}$ such that the following compositional equations are satisfied:

\begin{equation}
f_1 = f \circ i_1, \qquad \text{ and } \qquad f_2 = f \circ i_2.
\end{equation}
\end{definition}
These compositional equations for the universal property are equivalent to stating that the following diagram commutes:

\begin{equation}
\begin{tikzcd}
& B &\\
A_1 \arrow[ur, "f_1"] \arrow[r, "i_1"] & A_1 \amalg A_2 \arrow[u, "f", dashed] & A_2 \arrow[l, "i_2"] \arrow[ul, "f_2"]
\end{tikzcd}.
\end{equation}
The morphisms ${i_1}$ and ${i_2}$ are commonly known as ``canonical injections''. Now, the coproducts in ${\mathbf{OGraph}_{T_{\mathcal{G}}}}$ endow the category of cospans ${\mathbf{DCsp} \left( \mathbf{OGraph}_{T_{\mathcal{G}}} \right)}$ with a monoidal structure, such that, for the pair of cospans:

\begin{equation}
G : A \to B, \qquad \text{ and } \qquad H : C \to D,
\end{equation}
the monoidal product ${G \otimes H}$ is the cospan:

\begin{equation}
\begin{tikzcd}
B \amalg D \arrow[r, "o"] & G \amalg H & A \amalg C \arrow[l, "i"]
\end{tikzcd},
\end{equation}
where $i$ and $o$ denote the two induced maps of the coproducts. The symmetry isomorphisms:

\begin{equation}
\sigma_{A, B} : A \otimes B \to B \otimes A,
\end{equation}
are constructed using the induced swap map:

\begin{equation}
\sigma = \left[ i_2, i_1 \right],
\end{equation}
where ${i_1}$ and ${i_2}$ denote the canonical injections of the coproduct ${A \amalg B}$:

\begin{equation}
\begin{tikzcd}
B \amalg A \arrow[r, "id"] & B \amalg A & A \amalg B \arrow[l, "\sigma"]
\end{tikzcd}.
\end{equation}
Moreover, pairs of morphisms:

\begin{equation}
G : A \to B, \qquad \text{ and } H : B \to C,
\end{equation}
in ${\hom \left( \mathbf{DCsp} \left( \mathbf{OGraph}_{T_{\mathcal{G}}} \right) \right)}$ compose via pushouts, and, for any point graph $A$, the identity morphism of $A$ in ${\hom \left( \mathbf{DCsp} \left( \mathbf{OGraph}_{T_{\mathcal{G}}} \right) \right)}$ is the cospan given by the identity of graph $A$ in ${\hom \left( \mathbf{OGraph}_{T_{\mathcal{G}}} \right)}$, namely:

\begin{equation}
\begin{tikzcd}
A \arrow[r, "id"] & A & A \arrow[l, "id"]
\end{tikzcd}.
\end{equation}

The final step in the argument returns to our previous definition of a hypergraph transformation rule as a span of monomorphisms of the form:

\begin{equation}
\rho = \left( l : K \to L, r : K \to R \right),
\end{equation}
which one can then extend to the case of open graphs by considering \textit{selective adhesive functors}:

\begin{definition}
A ``selective adhesive functor'', denoted $S$, is a functor:

\begin{equation}
S : \mathbf{C} \to \mathbf{A},
\end{equation}
between a category ${\mathbf{C}}$ and an adhesive category ${\mathbf{A}}$ that preserves monomorphisms, creates isomorphisms, reflects pushouts and is ``faithful'' in the sense that it induces an injective function of the form:

\begin{equation}
S_{X, Y} : \hom_{\mathbf{C}} \left( X, Y \right) \to \hom_{\mathbf{A}} \left( S \left( X \right), S \left( Y \right) \right),
\end{equation}
for every pair of objects $X$ and $Y$ in ${\mathrm{ob} \left( \mathbf{C} \right)}$.
\end{definition}

\begin{definition}
A ``selective adhesive span'' for a selective adhesive functor of the form:

\begin{equation}
S : \mathbf{C} \to \mathbf{A},
\end{equation}
is a span of the form:

\begin{equation}
\begin{tikzcd}
A & B \arrow[l, "f"] \arrow[r, "g"] & C
\end{tikzcd},
\end{equation}
that has a pushout that is preserved by the functor $S$ (this pushout is known as a ``selective adhesive pushout'').
\end{definition}

\begin{definition}
The ``pushout complement'' of a pair of morphisms:

\begin{equation}
m : C \to A, \qquad \text{ and } \qquad g : A \to D,
\end{equation}
in ${\hom \left( \mathbf{C} \right)}$ for some category ${\mathbf{C}}$, is a pair of morphisms:

\begin{equation}
f : C \to B, \qquad \text{ and } \qquad n : B \to D,
\end{equation}
such that the following square commutes and is a pushout:

\begin{equation}
\begin{tikzcd}
C \arrow[r] \arrow[d] & A \arrow[d]\\
B \arrow[r] & D
\end{tikzcd}.
\end{equation}
\end{definition}
In this way, pushout complements are essentially ``completions'' of pairs of morphisms to form pushout squares (and are thus highly analogous to the Knuth-Bendix completions\cite{cohn}\cite{knuth}\cite{huet2} of branch pairs described in the conjectural formulation of quantum measurement in the Wolfram model).

\begin{definition}
A ``selective adhesive pushout complement'' for a pair of morphisms:

\begin{equation}
m : C \to A, \qquad \text{ and } \qquad g : A \to D,
\end{equation}
in ${\hom \left( \mathbf{C} \right)}$ for some category ${\mathbf{C}}$, is a pair of morphisms:

\begin{equation}
f : C \to B, \qquad \text{ and } \qquad n : B \to D,
\end{equation}
such that the following diagram is a selective adhesive pushout:

\begin{equation}
\begin{tikzcd}
B \arrow[r] \arrow[d] & A \arrow[d]\\
B \arrow[r] & D
\end{tikzcd}.
\end{equation}
\end{definition}

\begin{definition}
A ``selective rule match'' for a transformation rule:

\begin{equation}
\rho = \left( l : K \to L, r : K \to R \right),
\end{equation}
within an object $G$ is a monomorphism of the form:

\begin{equation}
m : L \to G,
\end{equation}
such that the pair of morphisms:

\begin{equation}
\begin{tikzcd}
K \arrow[r, "l"] & L \arrow[r, "m"] & G
\end{tikzcd},
\end{equation}
has a selective adhesive pushout complement.
\end{definition}

\begin{definition}
A ``selective adhesive rewrite'' for a transformation rule:

\begin{equation}
\rho = \left( l : K \to L, r : K \to R \right),
\end{equation}
and a selective adhesive rule match:

\begin{equation}
m : L \to G,
\end{equation}
is a diagram of the following form:

\begin{equation}
\begin{tikzcd}
L \arrow[d, "m"] & K \arrow[l, "l"] \arrow[d, "f"] \arrow[r, "r"] & R \arrow[d]\\
G & G^{\prime} \arrow[l] \arrow[r] & G \left[ \rho \right]_{m}
\end{tikzcd},
\end{equation}
where ${G^{\prime}}$ denotes the selective adhesive pushout complement of the pair of morphisms:

\begin{equation}
\begin{tikzcd}
K \arrow[r, "l"] & L \arrow[r, "m"] & G
\end{tikzcd},
\end{equation}
such that the right-hand pushout is selective adhesive.
\end{definition}
Thus, by constructing the category of selective adhesive rules, tentatively denoted ${\mathbf{SARule}}$, and considering its category of directed cospans ${\mathbf{DCsp} \left( \mathbf{SARule} \right)}$, one immediately obtains a lifting of the monoidal structure from ${\mathbf{DCsp} \left( \mathbf{OGraph}_{T_{\mathcal{G}}} \right)}$ to ${\mathbf{DCsp} \left( \mathbf{SARule} \right)}$. More specifically, for any pair of cospans:

\begin{equation}
G : A \to B, \qquad \text{ and } \qquad H : C \to D,
\end{equation}
in ${\mathbf{DCsp} \left( \mathbf{SARule} \right)}$, the monoidal product ${G \otimes H}$ is simply given by the cospan:

\begin{equation}
\begin{tikzcd}
B \amalg D \arrow[r, "o"] & G \amalg H & A \amalg C \arrow[l, "i"]
\end{tikzcd},
\end{equation}
where $i$ and $o$ are the induced maps of the coproducts in ${\mathbf{SARule}}$. This demonstrates that the rulial multiway system defined by applying all possible rules of the ZX-calculus to a given ZX-diagram does indeed form a subcategory of the category of directed cospans of selective adhesive rules.

It is worth nothing that, in the above argument, we have considered only rewriting rules over typed open graphs, since this is the particular case that pertains to the multiway operator systems of the ZX-calculus. More generally, one may extend the above construction of the category of directed cospans to string substitution systems, set substitution systems, or any other abstract rewriting system within which appropriate adhesive functors can be defined. 

\clearpage

\section{Concluding Remarks}
\label{sec:section4}

It has become increasingly clear over the past several years that the theory of diagrammatic rewriting systems over combinatorial structures has the potential to yield significant insights in regards to the foundations of physics; the Wolfram model and the ZX-calculus (and categorical quantum mechanics more generally) offer distinct yet complementary approaches to studying such systems. It is hoped that the present article will begin a fruitful process of establishing a more rigorous mathematical correspondence between the two, in the hopes that both approaches may benefit from the methods and intuitions yielded by the other. In this article, we have made a first step towards formally establishing this correspondence, in that we have:

\begin{enumerate}
\item
Demonstrated that the diagrammatic rewriting formalism of the ZX-calculus can indeed be embedded and realized within the more general formalism of Wolfram model multiway operator systems, using a novel reformulation of the Wolfram model in terms of double-pushout rewriting systems and adhesive categories.

\item
Shown explicitly (and subsequently proved formally) that the rulial evolution graphs, multiway evolution graphs and branchial graphs described by the Wolfram model are indeed endowed with a mutually compatible monoidal structure, that is furthermore naturally compatible with the monoidal product of ZX-diagrams.

\item
Introduced new Wolfram Language tools for generating, manipulating, computing and displaying ZX-diagrams, and simulating their diagrammatic rewritings as Wolfram model multiway systems.
\end{enumerate}

In other words, the categories of branchial graphs, multiway evolution graphs and rulial evolution graphs are therefore indeed symmetric monoidal categories of exactly the kind studied in the context of categorical quantum mechanics, suggesting that these structures may be interpreted as categorical quantum systems based on hypergraph rewritings. Assuming moreover that they are endowed with an involutive dagger structure (which we conjecture is given by the inversion of multiway evolution edges, although we have not yet proved formally that this operation is compatible with the induced monoidal product - this remains a subject for future work), plus the requisite compact structure (which would allow us to compute generalized duals of structures like branchial graphs), this will help to establish a complete and rigorous mathematical equivalence between the global multiway and categorical approaches to quantum mechanics. More specifically, since the category ${\mathbf{FdHilb}}$ of finite-dimensional Hilbert spaces is itself a dagger compact category, it would therefore guarantee that, under appropriate foliation of the rulial multiway system, the structure that one obtains is isomorphic to that of a tensor product of finite-dimensional Hilbert spaces, exactly as our standard mathematical formulation of the Wolfram model predicts\cite{gorard2}. It is important to note that, whereas the standard formulation of categorical quantum mechanics builds only upon the monoidal categorical structure of ZX-diagrams themselves, we have considered here the extension of this construction to the category of general \textit{rewriting systems} over ZX-diagrams.

More practically, there is also an immediate and rather exciting potential for application of the generalized Wolfram model/multiway system approaches developed within this article to the development of circuit optimization and automated diagrammatic reasoning algorithms over quantum circuits, and over string diagrams more generally, using multiway-based equational theorem-proving techniques; a follow-up to the present article addressing this particular topic is currently in preparation. Indeed, there even exists the possibility of using the multiway operator system formalism presented here to investigate alternative (and perhaps provably minimal) equational axiomatizations of categorical quantum mechanics with equivalent expressive power to the ZX-calculus. To the best of our knowledge it is currently unknown, but would be exceedingly interesting to find out, how the structure of multiway systems is affected by the use of alternative diagrammatic languages and axiomatizations, such as the recently-proposed ZW- \cite{coecke9}, \cite{hadzihasanovic2} and ZH-calculi \cite{backens3} for describing W-state and classically non-linear computation, respectively.

Finally, another future research direction following from the present project involves an attempt to make a formal correspondence between the Wolfram model higher category-theoretic structures appearing in the context of homotopy type theory and the univalent foundations program \cite{univalent}. The objective of such a direction would be to forge a connection between the kinds of combinatorial structures (such as multiway systems, branchial graphs, causal networks, etc.) studied in the context of the Wolfram model, and the kinds of spatial structures investigated in the context of Shulman's formulation of cohesive homotopy type theory \cite{shulman}. The existence of such a connection would likely hinge upon an interpretation of the rulial multiway system as an ${\infty}$-groupoid, which thus inherits the structure of a formal homotopic space via Grothendieck's hypothesis\cite{arsiwalla}. It is therefore hoped that the formal identification of dagger compact closed categories within the Wolfram model itself will provide a solid foundation for the construction of such higher category-theoretic generalizations.

\clearpage

\section*{Appendix}

\appendix

The first section of this appendix provides a glossary of relevant concepts commonly encountered in the formalism of the Wolfram model, whereas the second section aims to present a general overview of the mathematical formalism of monoidal categories as they are commonly employed in category-theoretic approaches to quantum mechanics in general, and in the ZX-calculus approach to quantum information theory in particular.

\section{Glossary of Wolfram Model Terminology}
\label{sec:section5}

\begin{itemize}
\item \textbf{Branchial Graph}: The graph whose vertex set is the set of states in a particular layer (or \textit{slice}) of the multiway evolution graph, and in which states are connected by directed edges if and only if they share a common ancestor in the evolution graph. Otherwise known as a \textit{branchlike hypersurface}, by analogy to spacelike hypersurfaces in causal networks. Branchial graphs are used to represent instantaneous superpositions between pure states.

\item \textbf{Branchial Space}: The spatial structure defined by a branchial graph, much like how physical space is the spatial structure defined by a hypergraph. In this way, branchial space has the same relationship to the multiway evolution graph as physical space has to an ordinary causal network. The default metric on branchial space (i.e. the Fubini-Study metric) is defined in such a way that entangled states are nearby.

\item \textbf{Branchtime}: The spatial structure defined by a multiway evolution graph, i.e. the time-extended version of branchial space. In this way, branchtime has the same relationship to the multiway evolution graph as spacetime has to an ordinary causal network.

\item \textbf{Causal Network}: The network of causal relationships between updating events. Specifically, the vertex set of the causal network is the set of updating events, and a directed edge exists between events $A$ and $B$ if and only if the input for event $B$ makes use of hyperedges that were produced by the output of event $A$, such that event $B$ could not be applied unless event $A$ had previously been applied. The flux of edges through particular hypersurfaces in the causal network is related to certain projections of the energy-momentum tensor.

\item \textbf{Causal Invariance}: A property of multiway systems whereby all possible evolution paths yield causal networks that are (eventually) isomorphic as directed acyclic graphs. Since the notion of confluence in the theory of abstract rewriting systems is a necessary (though not sufficient) condition for causal invariance, it follows that whenever causal invariance exists, every branch in the multiway evolution graph must eventually merge. For the particular case of a terminating (strongly normalizing) rewriting system, causal invariance therefore implies that all evolution paths yield the same eventual state. For a causal network representing the causal structure of a Lorentzian manifold, causal invariance implies Lorentz symmetry.

\item \textbf{Completion}: An additional rule or collection of rules introduced into a multiway system that brings it closer to causal invariance (some multiway systems can be made causal invariant by adding only a finite number of completions). Completions, specifically Knuth-Bendix completions, are commonly used in automated theorem-proving algorithms, as a means of forcing confluence within equational rewriting systems. Knuth-Bendix completions also constitute a conjectural approach to representing the process of projective quantum measurement within the framework of the Wolfram model, by allowing one to ``collapse'' superpositions of states in branchial space.

\item \textbf{Foliation}: A method for defining a universal time function over the vertices of a directed acyclic graph (i.e a function mapping vertices to integers), in such a way that the level sets of that function, known as \textit{slices}, cover the entire graph without intersecting. Foliations of a causal network yield successive configurations of hypergraphs, representing spacelike hypersurfaces. Foliations of a multiway evolution graph yield successive configurations of branchial graphs, representing instantaneous superpositions between pure states (\textit{branchlike hypersurfaces}).

\item \textbf{Hypergraph}: The basic structure used for representing space in the Wolfram model. Hypergraphs are generalizations of ordinary graphs in which \textit{hyperedges} can connect any arbitrary non-empty subset of vertices (such that ordinary graphs correspond to the special case in which all hyperedges are of arity 2). In this way, a hypergraph can be represented purely formally as a collection of abstract relations between elements.

\item \textbf{Multiway Evolution Causal Graph}: A composition of a multiway evolution graph and a causal network, i.e. a directed acyclic graph containing both global states and updating events, and in which there exist two fundamentally different kinds of edges - \textit{evolution edges}, which connect different states in the multiway evolution, and \textit{causal edges}, which show the causal relationships between the updating events. In this way, the multiway evolution graph contains information regarding the causal relationships not only on a single branch of multiway evolution (as in a standard causal network), but also \textit{between} different branches of multiway evolution. In a causal invariant system, the multiway evolution causal graph can be effectively \textit{factored} into many isomorphic spacetime causal networks.

\item \textbf{Multiway Evolution Graph}: A directed acyclic graph whose vertex set is the set of possible states of a given multiway system, and a directed edge exists between states $A$ and $B$ if and only if there exists an updating event (i.e. a rule application) that transforms state $A$ to state $B$. The limiting behavior of geodesics in the multiway evolution graph is governed by a path integral.

\item \textbf{Multiway System}: An abstract rewriting system that has been enriched with additional causal structure between rewriting events (see \textit{Multiway Evolution Graph} and \textit{Multiway Evolution Causal Graph} for further details).

\item \textbf{Observer}: Any ordered sequence of non-intersecting level surfaces of a universal time function, defined over a directed acyclic graph. In the case of a causal network, this corresponds to a foliation of spacetime (and therefore to a relativistic observer, embedded in a particular reference frame). In the case of a multiway evolution graph, this corresponds to a foliation of \textit{branchtime} (and therefore to a quantum observer).

\item \textbf{Rulial Multiway System}: The multiway system constructed by applying all possible rules of a given class (e.g. hypergraph tansformation rules, string substitution rules, Turing machine rules, etc.) to states of a given system. For instance, the rulial multiway system for Turing machines is obtained by allowing all possible non-deterministic transitions between Turing machine states. See \textit{Rulial Space} for further details.

\item \textbf{Rulial Space}: The space defined by allowing all possible rules of a given class (e.g. hypergraph transformation rules, string substitution rules, Turing machine rules, etc.) to be followed between states of a system. In other words, it is the spatial structure associated with the evolution of a rulial multiway system. For instance, the rulial space for Turing machines is obtained by allowing all possible non-deterministic transitions between Turing machine states. Rulial space can be formulated as a category of rewriting rules, and it functorially acquires the properties of an adhesive category. Consequently, the monoidal structure of both multiway evolution graphs and branchial graphs (which are obtained by foliations of multiway evolution graphs) is inherited from the composition of rules in rulial space.

\item \textbf{Spacelike Hypersurface}: A subset of spacetime that contains only spacelike-separated events, i.e. events that form antichains in the associated causal network, such that a time-ordered sequence of spacelike hypersurfaces defines a foliation of spacetime. The events on a spacelike hypersurface may be considered by a relativistic observer to be ``simultaneous''. In the Wolfram model, the flux of causal edges through spacelike hypersurfaces is associated with energy.

\item \textbf{Timelike Hypersurface}: A subset of spacetime that contains only timelike-separated events, i.e. events that are connected by edges in the associated causal network, and which are therefore orthogonal to spacelike hypersurfaces. In the Wolfram model, the flux of causal edges through timelike hypersurfaces is associated with momentum.

\item \textbf{Updating Event}: A single application of a rule (i.e. an application of the rewrite relation ${\to}$) in a multiway system. Updating events form the set of vertices in the causal network and the set of edges in the multiway evolution graph.

\end{itemize}

\section{Overview of Monoidal Categories and Categorical Quantum Mechanics}
\label{sec:section6}

\begin{definition}
A ``category''\cite{maclane}, denoted ${\mathbf{C}}$, consists of a class of ``objects'', denoted ${\mathrm{ob} \left( \mathbf{C} \right)}$, as well as a class of ``morphisms'', denoted ${\hom \left( \mathbf{C} \right)}$, between objects, where the composition of morphisms satisfies axioms of both identity and associativity. More specifically, each morphism $f$ in ${\hom \left( \mathbf{C} \right)}$ is of the general form:

\begin{equation}
f : A \to B,
\end{equation}
where $A$ and $B$ are both objects in ${\mathrm{ob} \left( \mathbf{C} \right)}$, such that the ``hom-class'' of the pair ${\left( A, B \right)}$, denoted ${\hom_{\mathbf{C}} \left( A, B \right)}$, represents the class of all morphisms from $A$ to $B$. For every triple of objects $A$, $B$ and $C$ in ${\mathrm{ob} \left( \mathbf{C} \right)}$, there exists a binary operation:

\begin{equation}
\hom \left( A, B \right) \times \hom \left( B, C \right) \to \hom \left( A, C \right),
\end{equation}

known as the ``composition of morphisms'', denoted ${g \circ f}$, where:

\begin{equation}
f : A \to B, \qquad \text{ and } \qquad g : B \to C,
\end{equation}
which satisfies the properties of identity and associativity, such that for all objects $X$ in ${\mathrm{ob} \left( \mathbf{C} \right)}$, there exists an ``identity morphism'', denoted:

\begin{equation}
id_X : X \to X,
\end{equation}
such that every pair of morphisms:

\begin{equation}
f : A \to X, \qquad g : X \to B,
\end{equation}
one has:

\begin{equation}
id_X \circ f = f, \qquad \text{ and } \qquad g \circ id_X = g,
\end{equation}
and moreover that, whenever:

\begin{equation}
f : A \to B, \qquad g : B \to C, \qquad \text{ and } \qquad h : C \to D,
\end{equation}
one has:

\begin{equation}
h \circ \left( g \circ f \right) = \left( h \circ g \right) \circ f.
\end{equation}
\end{definition} 

\begin{definition}
A (covariant) ``functor'', denoted $F$, is a map between categories ${\mathbf{C}}$ and ${\mathbf{D}}$ that preserves both the identity morphisms and the composition of morphisms in ${\hom \left( \mathbf{C} \right)}$. More specifically, $F$ associates each object $X$ in ${\mathrm{ob} \left( \mathbf{C} \right)}$ with an object ${F \left( X \right)}$ in ${\mathrm{ob} \left( \mathbf{D} \right)}$, as well as each morphism:

\begin{equation}
f : A \to B
\end{equation}
in ${\hom \left( \mathbf{C} \right)}$ with a morphism:

\begin{equation}
F \left( f \right) : F \left( A \right) \to F \left( B \right)
\end{equation}
in ${\hom \left( \mathbf{D} \right)}$, such that, for all objects $X$ in ${\mathrm{ob} \left( \mathbf{C} \right)}$, one has the associated identity morphism:

\begin{equation}
F \left( id_X \right) = id_{F \left( X \right)},
\end{equation}
and for all pairs of morphisms:

\begin{equation}
f: A \to B , \qquad \text{ and } \qquad g : B \to C,
\end{equation}
in ${\hom \left( \mathbf{C} \right)}$, one has the composition of morphisms:

\begin{equation}
F \left( g \circ f \right) = F \left( g \right) \circ F \left( f \right).
\end{equation}
\end{definition}

A functor can also be contravariant, in which case it simply reverses the direction of morphisms (and hence also the direction of composition):

\begin{definition}
A ``contravariant'' functor, denoted $F$, is a map between categories ${\mathbf{C}}$ and ${\mathbf{D}}$ that preserves the identity morphisms and reverses the composition of morphisms in ${\hom \left( \mathbf{C} \right)}$. More specifically, $F$ associates each object $X$ in ${\mathrm{ob} \left( \mathbf{C} \right)}$ with an object ${F \left( X \right)}$ in ${\mathbf{D}}$, as well as each morphism:

\begin{equation}
f : A \to B,
\end{equation}
in ${\hom \left( \mathbf{C} \right)}$ with a (reversed) morphism:

\begin{equation}
F \left( f \right) = F \left( B \right) \to F \left( A \right),
\end{equation}
in ${\hom \left( \mathbf{D} \right)}$, such that, for all objects $X$ in ${\mathrm{ob} \left( \mathbf{C} \right)}$, one has the associated identity morphism:

\begin{equation}
F \left( id_X \right) = id_{F \left( X \right)},
\end{equation}
and for all pairs of morphisms:

\begin{equation}
f : A  \to B, \qquad \text{ and } \qquad g : B \to C,
\end{equation}
in ${\hom \left ( \mathbf{C} \right)}$, one has the (reversed) composition of morphisms:

\begin{equation}
F \left( g \circ f \right) = F \left( f \right) \circ F \left( g \right).
\end{equation}
\end{definition}

\begin{definition}
A ``diagram'', denoted ${\mathbf{D}}$, in a category ${\mathbf{C}}$ is a (covariant) functor from an ``index category'', denoted ${\mathbf{J}}$, to ${\mathbf{C}}$:

\begin{equation}
D : \mathbf{J} \to \mathbf{C}.
\end{equation}
\end{definition}

\begin{definition}
A ``commutative diagram'' is a diagram in which all directed paths with the same start and endpoints yield the same result.
\end{definition}

\begin{definition}
The ``product category'', denoted ${\mathbf{C} \times \mathbf{D}}$, of two categories ${\mathbf{C}}$ and ${\mathbf{D}}$, generalizes the notion of a Cartesian product of sets, such that the class of objects ${\mathrm{ob} \left( \mathbf{C} \times \mathbf{D} \right)}$ consists of pairs of objects ${\left( A, B \right)}$ where $A$ is an object in ${\mathrm{ob} \left( \mathbf{C} \right)}$ and $B$ is an object in ${\mathrm{ob} \left( \mathbf{D} \right)}$, the class of morphisms ${\hom \left( \mathbf{C} \times \mathbf{D} \right)}$ consists of pairs of morphisms ${\left( f, g \right)}$:

\begin{equation}
\left( f, g \right) : \left( A_1, B_1 \right) \to \left( A_2, B_2 \right),
\end{equation}
where:

\begin{equation}
f : A_1 \to A_2,
\end{equation}
is a morphism in ${\hom \left( \mathbf{C} \right)}$, and:

\begin{equation}
g : B_1 \to B_2,
\end{equation}
is a morphism in ${\hom \left( \mathbf{D} \right)}$. Moreover, the composition of morphisms is given by component-wise composition within each of the two constituent categories:

\begin{equation}
\left( f_2, g_2 \right) \circ \left( f_1, g_1 \right) = \left( f_2 \circ f_1, g_2 \circ g_1 \right),
\end{equation}
where ${f_1, f_2}$ are morphisms in ${\hom \left( \mathbf{C} \right)}$, and ${g_1, g_2}$ are morphisms in ${\hom \left( \mathbf{D} \right)}$, and the identity morphisms are given by pairs of identity morphisms taken from each of the constituent categories:

\begin{equation}
id_{\left( A, B \right)} = \left( id_{A}, id_{B} \right).
\end{equation}
\end{definition}

\begin{definition}
A ``bifunctor'' is any functor whose domain is a product category (and which therefore acts like a functor in two arguments).
\end{definition}

\begin{definition}
A ``natural transformation'', denoted ${\eta}$, is a map between (covariant) functors $F$ and $G$ (which are themselves maps between categories ${\mathbf{C}}$ and ${\mathbf{D}}$) that preserves the composition of morphisms in both ${\mathbf{C}}$ and ${\mathbf{D}}$. More specifically, ${\eta}$ associates each object $A$ in ${\mathrm{ob} \left( \mathbf{C} \right)}$ with a morphism:

\begin{equation}
\eta_A : F \left( A \right) \to G \left( A \right),
\end{equation}
in ${\hom \left( \mathbf{D} \right)}$, known as the ``component'' of ${\eta}$ at object $A$, such that, for all morphisms:

\begin{equation}
f : A \to B,
\end{equation}
in ${\hom \left( \mathbf{C} \right)}$, one has:

\begin{equation}
\eta_Y \circ F \left( f \right) = G \left( f \right) \circ \eta_A.
\end{equation}
\end{definition}

This last condition is equivalent to stating that the following diagram commutes:

\begin{equation}
\begin{tikzcd}
A \arrow[d, "f"] & F \left( A \right) \arrow[d, "F \left( f \right)"] \arrow[r, "\eta_{A}"] & G \left( A \right) \arrow[d, "G \left( f \right)"]\\
B & F \left( B \right) \arrow[r, "\eta_{B}"] & G \left( B \right)
\end{tikzcd}
\end{equation}
If both $F$ and $G$ are contravariant functors, then the vertical arrows in this diagram are reversed.

\begin{definition}
In a category ${\mathbf{C}}$, an ``isomorphism'' between objects $A$ and $B$ in ${\mathrm{ob} \left( \mathbf{C} \right)}$, denoted ${A \cong B}$, is any morphism:

\begin{equation}
f : A \to B,
\end{equation}
that is also equipped with an inverse morphism:

\begin{equation}
g : B \to A,
\end{equation}
i.e. such that:

\begin{equation}
f \circ g = id_{B}, \qquad \text{ and } \qquad g \circ f = id_{A}.
\end{equation}
\end{definition}

\begin{definition}
A ``natural isomorphism'' is a natural transformation ${\eta}$ with the property that, for every object $X$ in ${\mathrm{ob} \left( \mathbf{C} \right)}$, ${\eta_X}$ is an isomorphism in ${\mathbf{D}}$.
\end{definition}

\begin{definition}
A ``coherence condition'' is a requirement that certain compositions of morphisms must be equal.
\end{definition}

\begin{definition}
A ``monoidal category''\cite{kelly}\cite{kelly2}, denoted ${\left( \mathbf{C}, \otimes, I \right)}$, is a category that is equipped with a ``monoidal structure'' (or a ``monoidal product''), i.e. a bifunctor, denoted ${\otimes}$:

\begin{equation}
\otimes : \mathbf{C} \times \mathbf{C} \to \mathbf{C},
\end{equation}
which satisfies an associativity axiom (up to natural isomorphism) and which generalizes the notion of a tensor product of vector spaces\cite{baez2}\cite{joyal}, as well as an object $I$, known as the ``identity object'' of the monoidal structure, that acts as both a left and right identity for the monoidal product (again, up to natural isomorphisms). More specifically, for each triple of objects $A$, $B$ and $C$ in ${\mathrm{ob} \left( \mathbf{C} \right)}$, there exists a natural isomorphism ${\alpha}$, known as the ``associator'', whose components are given by:

\begin{equation}
\alpha_{A, B, C} : A \otimes \left( B \otimes C \right) \cong \left( A \otimes B \right) \otimes C,
\end{equation}
and, for each single object $A$ in ${\mathrm{ob} \left( \mathbf{C} \right)}$, there exists a pair of natural isomorphisms ${\lambda}$ and ${\rho}$, known as the ``left unitor'' and the ``right unitor'', respectively, and whose components are given by:

\begin{equation}
\lambda_A : I \otimes A \cong A, \qquad \text{ and } \qquad \rho_A : A \otimes I \cong A.
\end{equation}
\end{definition}

These two foundational requirements of associativity and left/right identity for the monoidal structure can be represented using the following coherence conditions, expressed here as the requirement that the following two diagrams commute:

\begin{equation}
\begin{tikzcd}
A \otimes \left( B \otimes \left( C \otimes D \right) \right) \arrow[d, "id_{A} \otimes \alpha_{B, C, D}"] \arrow[r, "\alpha_{A, B, C \otimes D}"] & \left( A \otimes B \right) \otimes \left( C \otimes D \right) \arrow[r, "\alpha_{A \otimes B, C, D}"] & \left( \left( A \otimes B \right) \otimes C \right) \otimes D\\
A \otimes \left( \left( B \otimes C \right) \otimes D \right) \arrow[rr, "\alpha_{A, B \otimes C, D}"] & & \left( A \otimes \left( B \otimes C \right) \right) \otimes D \arrow[u, "\alpha_{A, B, C \otimes id_{D}}"]
\end{tikzcd},
\end{equation}
for every tuple of objects $A$, $B$, $C$ and $D$ in ${\mathrm{ob} \left( \mathbf{C} \right)}$, and:

\begin{equation}
\begin{tikzcd}
A \otimes \left( I \otimes B \right) \arrow[rr, "\alpha_{A, I, B}"] \arrow[dr, "id_{A} \otimes \lambda_{B}"] & &  \left( A \otimes I \right) \otimes B \arrow[dl, "\rho_{A} \otimes id_{B}"]\\
& A \otimes B &
\end{tikzcd},
\end{equation}
for all pairs of objects $A$ and $B$ in ${\mathrm{ob} \left( \mathbf{C} \right)}$.

\begin{definition}
A monoidal category ${\left( \mathbf{C}, \otimes, I \right)}$ is known as a ``symmetric monoidal category'' if, for every pair of objects $A$ and $B$ in ${\mathrm{ob} \left( \mathbf{C} \right)}$, there exists a natural isomorphism ${\sigma}$ whose components are given by:

\begin{equation}
\sigma_{A, B} : A \otimes B \cong B \otimes A,
\end{equation}
such that ${\sigma}$ is coherent with the associator and the left and right unitors, as well as obeying an inverse law. More specifically, the following three diagrams all commute:

\begin{equation}
\begin{tikzcd}
\left( A \otimes B \right) \otimes C \arrow[r, "\sigma_{A, B} \otimes id_{C}"] \arrow[d, "\alpha_{A, B, C}"] & \left( B \otimes A \right) \otimes C \arrow[d, "\alpha_{B, A, C}"]\\
A \otimes \left( B \otimes C \right) \arrow[d, "\sigma_{A, B \otimes C}"] & B \otimes \left( A \otimes C \right) \arrow[d, "id_{B} \otimes \sigma_{A, C}"]\\
\left( B \otimes C \right) \otimes A \arrow[r, "\alpha_{B, C, A}"] & B \otimes \left( C \otimes A \right)
\end{tikzcd},
\end{equation}
for all triples of objects $A$, $B$ and $C$ in ${\mathrm{ob} \left( \mathbf{C} \right)}$,

\begin{equation}
\begin{tikzcd}
A \otimes I \arrow[rr, "\sigma_{A, I}"] \arrow[dr, "\rho_{A}"] & & I \otimes A \arrow[dl, "\lambda_{A}"]\\
& A &
\end{tikzcd},
\end{equation}
for all single objects $A$ in ${\mathrm{ob} \left( \mathbf{C} \right)}$, and:

\begin{equation}
\begin{tikzcd}
& B \otimes A \arrow[dr, "\sigma_{B, A}"] &\\
A \otimes B \arrow[ur, "\sigma_{A, B}"] \arrow[rr, equal, "id_{A \otimes B}"] & & A \otimes B
\end{tikzcd},
\end{equation}
for all pairs of objects $A$ and $B$ in ${\mathrm{ob} \left( \mathbf{C} \right)}$, where, in the above, ${\alpha}$, ${\lambda}$ and ${\rho}$ denote the associator isomorphism and the left and right unitor isomorphisms of the monoidal category, respectively.
\end{definition}

Loosely speaking, a symmetric monoidal category captures the intuitive notion of the monoidal structure being ``as commutative as possible''.

\begin{definition}
The ``opposite category'', denoted ${\mathbf{C}^{op}}$, of a given category ${\mathbf{C}}$, is the category obtained by reversing the direction of all morphisms in ${\hom \left( \mathbf{C} \right)}$.
\end{definition}

\begin{definition}
A ``dagger category''\cite{burgin}\cite{lambek}, denoted ${\left( \mathbf{C}, \dag \right)}$, is a category that is equipped with a ``dagger structure'', namely an involutive functor (i.e. a functor that is also its own inverse), denoted ${\dag}$:

\begin{equation}
\dag : \mathbf{C}^{op} \to \mathbf{C},
\end{equation}
which naturally generalizes the notion of a Hermitian adjoint of a linear operator on Hilbert space, where ${\mathbf{C}^{op}}$ designates the opposite category of ${\mathbf{C}}$. More specifically, the dagger structure associates every morphism:

\begin{equation}
f : A \to B,
\end{equation}
in ${\hom \left( \mathbf{C} \right)}$ with its adjoint morphism:

\begin{equation}
f^{\dag} : B \to A,
\end{equation}
such that, for every object $A$ in ${\mathrm{ob} \left( \mathbf{C} \right)}$, the adjoint of the identity morphism ${id_{A}}$ is itself:

\begin{equation}
id_{A} = id_{A}^{\dag} : A \to A,
\end{equation}
for every morphism:

\begin{equation}
f : A \to B,
\end{equation}
the dagger structure is involutive:

\begin{equation}
f^{\dag \dag} = f : A \to B,
\end{equation}
and for every pair of morphisms:

\begin{equation}
f : A \to B, \qquad \text{ and } \qquad g : B \to C,
\end{equation}
the dagger structure reverses the order of composition:

\begin{equation}
\left( g \circ f \right)^{\dag} = f^{\dag} \circ g^{\dag} : C \to A.
\end{equation}
\end{definition}

\begin{definition}
A ``dagger symmetric monoidal category'', denoted ${\left( \mathbf{C}, \otimes, I, \dag \right)}$, is a symmetric monoidal category ${\left( \mathbf{C}, \otimes, I \right)}$ that is also equipped with a dagger structure that is compatible with the underlying monoidal structure. More specifically, for every pair of morphisms:

\begin{equation}
f : A \to B, \qquad \text{ and } \qquad g : C \to D,
\end{equation}
in ${\hom \left( \mathbf{C} \right)}$, one has:

\begin{equation}
\left( f \otimes g \right)^{\dag} = f^{\dag} \otimes g^{\dag} : B \otimes D \to A \otimes C,
\end{equation}
if ${\alpha}$ denotes the associator isomorphism, then for every triple of objects $A$, $B$ and $C$ in ${\mathrm{ob} \left( \mathbf{C} \right)}$, one has:

\begin{equation}
\alpha_{A, B, C}^{\dag} = \alpha_{A, B, C}^{-1} : A \otimes \left( B \otimes C \right) \cong \left( A \otimes B \right) \otimes C,
\end{equation}
if ${\lambda}$ and ${\rho}$ denote the left and right unitor isomorphisms, respectively, then for every single object $A$ in ${\mathrm{ob} \left( \mathbf{C} \right)}$, one has:

\begin{equation}
\lambda_{A}^{\dag} = \lambda_{A}^{-1} : A \cong I \otimes A, \qquad \text{ and } \qquad \rho_{A}^{\dag} = \rho_{A}^{-1} : A \cong A \otimes I,
\end{equation}
and if ${\sigma}$ denotes the monoidal symmetry isomorphism, then for every pair of objects $A$ and $B$ in ${\mathrm{ob} \left( \mathbf{C} \right)}$, one has:

\begin{equation}
\sigma_{A, B}^{\dag} = \sigma_{A, B}^{-1} : B \otimes A \cong A \otimes B.
\end{equation}
\end{definition}

\begin{definition}
A symmetric monoidal category ${\left( \mathbf{C}, \otimes, I \right)}$ is designated as ``compact closed''\cite{kelly3} if every object $A$ in ${\mathrm{ob} \left( \mathbf{C} \right)}$ has an associated dual object, denoted ${A^{*}}$, which is unique up to a canonical (i.e. unique) isomorphism and which generalizes the notion of a dual vector space. More specifically, the dual ${A^{*}}$ is an object that is equipped with a pair of morphisms known as the ``unit'':

\begin{equation}
\eta_{A} : I \to A^{*} \otimes A,
\end{equation}
and the ``counit'':

\begin{equation}
\epsilon_{A} : A \otimes A^{*} \to I,
\end{equation}
which satisfy the following pair of equations regarding their compositions:

\begin{equation}
\lambda_{A} \circ \left( \epsilon_{A} \otimes A \right) \circ \alpha_{A, A^{*}, A}^{-1} \circ \left( A \otimes \eta_{A} \right) \circ \rho_{A}^{-1} = id_{A},
\end{equation}
and:

\begin{equation}
\rho_{A^{*}} \circ \left( A^{*} \otimes \epsilon_{A} \right) \circ \alpha_{A^{*}, A, A^{*}} \circ \left( \eta_{A} \otimes A^{*} \right) \circ \lambda_{A^{*}}^{-1} = id_{A^{*}},
\end{equation}
where, as usual, ${\alpha}$, ${\lambda}$ and ${\rho}$ denote the associator, left unitor and right unitor isomorphisms, respectively.
\end{definition}

These two equational requirements for ${\left( \mathbf{C}, \otimes, I \right)}$ to be compact closed can be restated diagrammatically as the assertion that, for every object $A$ in ${\mathrm{ob} \left( \mathbf{C} \right)}$, the following compositions of morphisms:

\begin{equation}
\begin{tikzcd}
A \arrow[r, "\cong"] & A \otimes I \arrow[r, "A \otimes \eta"] & A \otimes \left( A^{*} \otimes A \right) \arrow[r, "\cong"] & \left( A \otimes A^{*} \right) \otimes A \arrow[r, "\epsilon \otimes A"] & I \otimes A \arrow[r, "\cong"] & A
\end{tikzcd},
\end{equation}
and:

\begin{equation}
\begin{tikzcd}
A^{*} \arrow[r, "\cong"] & I \otimes A^{*} \arrow[r, "\eta \otimes A^{*}"] & \left( A^{*} \otimes A \right) \otimes A^{*} \arrow[r, "\cong"] & A^{*} \otimes \left( A \otimes A^{*} \right) \arrow[r, "A^{*} \otimes \epsilon"] & A^{*} \otimes I \arrow[r, "\cong"] & A^{*},
\end{tikzcd}
\end{equation}
must equal ${id_{A}}$ and ${id_{A^{*}}}$, respectively.

\begin{definition}
A ``dagger compact closed category''\cite{doplicher}\cite{baez3}, denoted ${\left( \mathbf{C}, \otimes, I, \dag \right)}$, is a dagger symmetric monoidal category that is also compact closed, obeying a suitable compatibility condition between the dagger structure and the compact structure. More specifically, for every object $A$ in ${\mathrm{ob} \left( \mathbf{C} \right)}$, the unit ${\eta_{A}}$ and counit ${\epsilon_{A}}$ are related by the dagger structure in such a way that the following diagram commutes:

\begin{equation}
\begin{tikzcd}
I \arrow[r, "\epsilon_{A}^{\dag}"] \arrow[dr, "\eta_{A}"] & A \otimes A^{*} \arrow[d, "\sigma_{A  \otimes A^{*}}"]\\
& A^{*} \otimes A
\end{tikzcd}.
\end{equation}
\end{definition}

The relationship between category theory and the foundations of quantum mechanics arises from the fact that the category ${\mathbf{FdHilb}}$ of finite-dimensional vector spaces and linear maps naturally forms a dagger compact closed category\cite{selinger2}\cite{hasegawa}, with the class of morphisms being the class of linear maps between Hilbert spaces, the monoidal structure reducing to the regular tensor product of Hilbert spaces and the dagger structure reducing to the regular Hermitian adjoint of linear operators. The category of infinite-dimensional Hilbert spaces (which we shall not consider here) is not dagger compact, but nevertheless still forms a dagger symmetric monoidal category in much the same way.

Of greater relevance for the present article is the fact that the collection of all possible diagrams in the ZX-calculus forms a dagger compact closed category ${\mathbf{C}}$, in which the class of objects ${\mathrm{ob} \left( \mathbf{C} \right)}$ is simply the class of natural numbers ${\mathbb{N}}$, and the monoidal structure ${\otimes}$ is given by natural number addition ${+}$. Every morphism:

\begin{equation}
f : n \to m,
\end{equation}
in ${\hom \left( \mathbf{C} \right)}$ is a ZX-diagram, to be interpreted as a linear map acting on $n$ qubits as input and yielding $m$ qubits as output. The composition of morphisms in ${\hom \left( \mathbf{C} \right)}$ thus yields the horizontal composition of ZX-diagrams (in which one takes the output wires of a diagram on the left and connects them to the input wires of a diagram on the right, or vice versa), with the monoidal product of morphisms yielding the vertical ``stacking'' of ZX-diagrams, such that the former operation captures the essence of sequential application of linear maps, whilst the latter captures the essence of parallel application. By taking the free category generated by a finite set of ZX-generators under composition and monoidal product, and then modding out by both the compact structure and the intrinsic equational diagrammatic rewriting rules that define the ZX-calculus (to be defined subsequently), one thus obtains the category of all possible ZX-diagrams.

Note that the ZX-generators themselves are known as ``spiders'', and are directly analogous to the shapes that appear within the Penrose graphical notation\cite{penrose}\cite{cvitanovic} for multilinear functions; in this way, each ZX-diagram may simply be interpreted as a tensor network representing an arbitrary multilinear map. Specifically, each tensor is represented as a ``spider'', and summations over pairs of indices are represented as wires; moreover, free indices are represented as ``dangling'' wires, with sums represented by loops, and finally the identity tensor, i.e. the Dirac delta ${\delta_{i}^{j}}$, is also represented as a single wire.

\section*{Acknowledgments}

The authors would like to thank Stephen Wolfram for his enthusiastic encouragement, as well as Nik Murzin for his diligent proofreading of an early draft of this manuscript.

\end{document}